\documentclass[amsmath,amssymb,onecolumn,superscriptaddress]{revtex4}

\usepackage{graphicx}% 
\usepackage{dcolumn}% 
\def\be{\begin{equation}}
\def\ee{\end{equation}}
\def\beq{\begin{eqnarray}}

\usepackage{bm}
\usepackage{graphicx}
\usepackage{dcolumn}
\usepackage{amsmath}
\usepackage[latin1]{inputenc}
\usepackage{graphicx, psfrag}
\usepackage{amssymb}
\usepackage[colorlinks=true, citecolor=blue, urlcolor = blue, linkcolor= red, bookmarks=true]{hyperref}
\usepackage{float}
\usepackage{mathtools}
\usepackage{amsmath}
\usepackage{amsfonts}
\usepackage{dcolumn}
\usepackage{natbib}
\RequirePackage{doi}
\usepackage{hyperref}
\usepackage{subfigure}
\usepackage{pgfplots}
\usepackage{epstopdf}
\usepackage{booktabs}
\usepackage{orcidlink}
\usepackage{xcolor}

\begin{document}
\title{Charged analog of anisotropic dark energy star in Rastall gravity}

\author{Pramit Rej \orcidlink{0000-0001-5359-0655} \footnote{Corresponding author}}
\email[Email:]{pramitrej@gmail.com, pramitr@sccollegednk.ac.in, pramit.rej@associates.iucaa.in}
 \affiliation{Department of Mathematics, Sarat Centenary College, Dhaniakhali, Hooghly, West Bengal 712 302, India}

\begin{abstract}\noindent
Dark energy is one of the potential strategies for preventing compact objects from gravitationally collapsing into singularities. Because it is the cause of the accelerated expansion of our universe, it has the greatest impact on the cosmos. Thus, it is plausible that dark energy will interact with any stellar object that is compact in the universe [\textit{Phys. Rev. D} \textbf{103}, 084042 (2021)]. Our main goal in this work is to create a simplified model, in the Rastall gravitational framework, of a charged strange star coupled to anisotropic dark energy in Krori-Barua spacetime [\textit{J. Phys. A, Math. Gen.} \textbf{8}:508, 1975]. Here, we consider a specific strange star object, Her X-1, with observed values of mass $=(0.85 \pm 0.15)M_{\odot}$ and radius $= 8.1_{-0.41}^{+0.41}$ km., so that we can develop our model. In this context, we began by modeling dark energy using the equation of state (EoS), in which the dark energy density is proportional to the isotropic perfect fluid matter-energy density. The Darmois-Israel condition has been used to calculate the unknown constants that are present in the metric. We perform a detailed analysis of the model's physical properties, including the mass-radius relation, pressure, density, metric function, and dark energy parameters for different Rastall coupling parameter values. We also examine the stability and force equilibrium of our proposed stellar configuration. Following a comprehensive theoretical analysis, we discovered that our suggested model is both singularity-free and meets all stability requirements needed to be a stable, physically reasonable stellar model.

\end{abstract}

\maketitle

\section{Introduction}
Albert Einstein's general theory of relativity (GTR) is a crucial gravitational tool for comprehending the dynamics of cosmic bodies and other phenomena, as well as the structure of space-time. Since it is difficult to find precise analytical solutions for Einstein field equations that describe compact configurations such as strange stars, theoretical modeling has made significant progress. Specifically, fluid-sphere models with anisotropic matter distributions have been designed. Anisotropies contribute significantly to the stability and equilibrium of the stellar structure \cite{Maurya:2019zyc}. The effect of local anisotropy on the global properties of relativistic compact objects has been extensively studied by several researchers \cite{tamta2017new,bondi1992anisotropic, DEB2017239, dev2002anisotropic,di1997cracking, krori1984some,mak2003anisotropic, maurya2016new}.\par
According to cosmology, the apparent matter of the universe only accounts for $5\%$ of gravity, while the remaining $26\%$ and $69\%$ are explained by dark matter and dark energy, respectively. Dark energy is thought to be a spatially uniform cosmic fluid. However, it can be expanded to a non-homogenous spacetime by assuming negative radial pressure in the EoS and transverse pressure by the field equations \cite{Lobo:2005uf}. Because dark energy is evenly distributed throughout the universe, both in space and time, its significance does not diminish as it expands. However, despite not being visible through a telescope and not being able to discharge, absorb, or reflect electromagnetic radiation, dark matter is still fascinating. Despite this, it has been established that dark matter plays a significant role in the structure and evolution of the universe and is widespread throughout it. Following observations of Type-Ia supernovae ("one-A") in 1998 \cite{SupernovaCosmologyProject:1998vns}, the High-z Supernova Search Team and the Lawrence Berkeley National Laboratory's Supernova Cosmology Project postulated an accelerated expansion of the universe. For dark energy, these Type-Ia supernovae offer the most conclusive evidence. Compact astrophysical objects whose interior pressure $p$ and energy density $\rho$ obey a typical dark energy EoS, such as $p = -\rho$, have been the subject of numerous studies.\par
These objects have been known by a number of different names in the literature. We call them "dark energy stars" to make it easier to use \cite{beltracchi2019formation, Chapline:2004jfp}. Perhaps the origins of dark energy are unknown, but by observing how quickly the universe expands and how large-scale structures such as galaxies and clusters of galaxies form as a result of gravitational instabilities, one can determine the existence of dark energy. Although dark energy stars have the potential to exhibit spacetime singularities, non-singular dark energy stars are more often investigated. Astrophysical measurements taken recently have revealed that the universe is expanding more quickly than before. Using the Hubble Space Telescope, Riess and colleagues discovered that the universe is expanding $9 \%$ faster than expected \cite{Riess:2019cxk}. Despite working against gravity, dark energy accelerates the expansion of the universe while slowing the development of large-scale structures, resulting in a violation of the strong-energy condition (SEC). Due to dark energy's fundamental importance in cosmology, we usually look for local astrophysical appearances of the phenomenon. The EoS of dark energy can be characterized as $p = \omega\rho$, where $\omega < -\frac{1}{3}$ \cite{sushkov2005wormholes,bibi2016solution}. In this regard, Feng et al. conducted a statistical joint analysis using three distinct observational data sets to investigate various phenomenological interaction models for dark energy and dark matter \cite{feng2008observational}.\par
\textcolor{blue}{In order to address these concerns, many research groups have modified the general Einstein-Hilbert (EH) action during last few decades, which corresponds to different versions of modified gravitational theories \cite{Capozziello:2002rd, Nojiri:2006ri, Nojiri:2010wj, Bamba:2012cp, Rastall:1972swe, Rastall:1976uh, Harko:2011kv, Odintsov:2013iba, Haghani:2013oma, Ayuso:2014jda}. Several researchers have conducted pioneering astrophysical investigations using these modified gravity theories described in the references \cite{Shirafuji:1996im, Nashed:2001im, Nashed:2001cp, Nashed:2010ocg, ElHanafy:2014efn, Nashed:2015pga, Nashed:2018piz, Nashed:2018qag, Nashed:2019yto, Nashed:2020mnp, Nashed:2022zyi}. Rastall gravitational theory \cite{Rastall:1972swe} introduced by P. Rastall in 1972 is one of those variants by proposing a modification to the Einstein field equations. Rastall introduced a non-minimal coupling between the material inside and the geometry. Rastall's research is consistent with a number of observational data sets, theoretical predictions, and cosmological phenomena \cite{Darabi:2017coc, Moradpour:2016fur, Li:2019jkv}. The curvature-matter theory of gravity \cite{Koivisto:2005yk, Bertolami:2007gv, Harko:2014gwa} is another hypothesis. It is similar to the Rastall theory in that it holds that the standard energy-momentum conservation law is not valid because matter and geometry are related in a non-minimal way. It is demonstrated that even under this Rastall gravity, if we suppose that the universe is filled with cosmic fluids with a constant EoS \cite{capone2010possibility, Batista:2011nu}, a dark energy-like source is still required to characterize the current accelerating universe. A new study considers a universe in which other cosmic fluids have constant EoS and a constant performs the role of vacuum energy (such as the dark energy candidate) to address its conflicts and agreements with findings (Rastall-$\Lambda$CDM model) \cite{Akarsu:2020yqa}. In accordance with the observational data, Rastall modified the GR by adding a constant coupling parameter known as the Rastall parameter. This constant coupling parameter assists to couple the geometry non-minimally with the matter field. Another interesting characteristic of this theory is that the resulting set of equations is simpler than that of previous modified variants of GR, making them easier to inspect.} In vacuum, this modified gravity theory aligns with Einstein's gravity and follows Mach's principle \cite{Majernik:2006jg}. Researchers find Rastall gravity (RG) to be incredibly fascinating because of its numerous astrophysical and cosmological applications. Several cosmological models with significant outcomes have been explored Rastall gravity \cite{Fabris:2012hw, Fabris:2011wz}. This RG framework explores the phenomenological impacts of cosmology, such as on accelerated cosmic expansion and inflation \cite{Batista:2011nu, Batista:2012hv, Campos:2012ez, Carames:2014twa, Salako:2016ihq, Moradpour:2015ymo, Moradpour:2017shy, Moradpour:2017ycq, Das:2018dzp, Ghaffari:2020nnk, Saleem:2020bwa, Singh:2020akk} and on particle creation mechanisms \cite{Bishi:2023mwv}.
The particle creation procedure \cite{Gibbons:1977mu, Parker:1971pt, Ford:1986sy, Pereira:2009kv} and the RG theory share similarities in the cosmic evolutionary process as both don't follow the conservation of the energy-momentum tensor (i.e., $\nabla_\mu T^{\mu \nu}= 0$). Rastall proposed a remarkable change in GR to improve conservation law as a non-conservation, as shown below.:
\begin{equation} \label{rt1}
\nabla_\mu T^{\mu \nu}= \alpha R^{; \nu}
\end{equation}
i.e., the covariant derivative of the energy-momentum tensor doesn't vanish but instead is proportional to the divergence of the Ricci scalar $R$. The proportionality factor $\alpha$ is defined as the Rastall coupling parameter. The Rastall coupling parameter ($\alpha$) quantifies the variation from GR. The conservation law can only be verified in Minkowski spacetime, which is because the correlation between matter field and geometry was selected. Even with non-zero spacetime curvature, energy-momentum tensor conservation can still be achieved. When $\alpha$ reaches its limiting value, this modification reverts to GR since the particle creation process is not impacted in
the energy-momentum tensor conservation process. The following modified field equations-generally termed Rastall field equations-are derived from the above modified conservation law:
\begin{equation} \label{rt2}
G_{\mu \nu} + \kappa' \alpha g_{\mu \nu} R= \kappa' T_{\mu \nu}
\end{equation}
where $\kappa'$ is the usual gravitational coupling constant of RG. \par
The above equation can be rewritten as 
\begin{equation} \label{rt3}
G_{\mu \nu} = \kappa' S_{\mu \nu}
\end{equation}
where $S_{\mu \nu}$ being the effective energy-momentum tensor defined as $$S_{\mu \nu}= T_{\mu \nu} - \frac{\kappa' \alpha T}{4\kappa' \alpha -1} g_{\mu \nu}$$
\textcolor{blue}{Now taking trace on both sides of Eq.~(\ref{rt2}), we may construct the following relationship: 
\begin{equation} \label{rt4}
R(4\kappa' \alpha -1)= T
\end{equation}
This implies that $\kappa' \alpha = \frac{1}{4}$ is not allowed in RG.}

\textcolor{blue}{Several research works have already been done within this RG framework. Abbas and Shahzad \cite{Abbas:2019evn} investigated compact anisotropic objects in the RG domain by employing the Krori-Barua metric. They further studied the corresponding charged analog within this framework \cite{Shahzad:2019bqf} and then also proposed a novel approach to solve the field equations in RG for isotropic matter content with a quintessence field \cite{Shahzad:2020gjj}. Mustafa et al. \cite{Mustafa:2021fzk} investigated stellar structures by applying static spherically symmetric spacetime in RG via an embedding approach. Hansraj and Banerjee \cite{Hansraj:2020clg} examined the astrophysical implications of stellar structures in equilibrium. Other noteworthy research works on astrophysical objects in RG that should be acknowledged here are given in the Refs. \cite{Moradpour:2016ubd, Shahzad:2020bwr, Saleem:2023ukn, ashraf2024possible, ElHanafy:2022kjl, waseem2024isotropic, Ghosh:2021byh, Majeed:2022uyz, Abbas:2020kju}.}\par
Thus, inspired by the fascinating details about the fictitious dark energy and earlier works in RG theory mentioned above, \textcolor{blue}{the Krori-Barua metric has been utilized here to model a charged strange star} coupled to inhomogeneous anisotropic dark energy within the RG framework assuming that the density of isotropic perfect fluid matter is proportional to the radial pressure exerted by dark energy on the entire system. In the Einstein-Maxwell field equation, there are eight unknown functions (e.g., $\rho$, $\rho^{DE}$, $p$, $p^{DE}_r$, $p^{DE}_t$, $\nu$, $\lambda$, $E$) as well as four equations. As a result, in order to solve this system, additional information or conditions must be provided, as mentioned in this work. We provide the fundamental field equations for a charged strange star model with inhomogeneous anisotropic dark energy in the RG framework. The field equations' solutions are analyzed as well as the smooth correspondence between internal and external spacetime. A thorough discussion has been made concerning the physical study of this present model. \textcolor{blue}{Some significant findings have been summarized on the construction of a dense, stable} stellar structure with dark energy's repulsive scalar field in this RG framework.\par
The current article is designed as follows. It is divided into five major sections. In section \ref{interior}, \textcolor{blue}{the popular Krori-Barua metric has been employed to deal with Einstein-Maxwell field equations} for charged static and spherically symmetric matter distribution coupled with dark energy. Then fix the constants of this model using junction conditions for a particular compact star candidate Her X - 1. In the following section \ref{phy}, \textcolor{blue}{the physical characteristics of this model have been analyzed}. \textcolor{blue}{In Section \ref{stable}, the stability of the current model has been analyzed from several perspectives and ensure the equilibrium of the force.} The final section \ref{con} discusses the important findings of this current investigation and provides some concluding remarks.

\section{Solutions of Einstein-Maxwell Field Equations}\label{interior}

To describe the interior space-time of a static spherically symmetric object, we consider the interior line element in a static spherically symmetric $4-\mathcal{D}$ space-time in the Schwarzchild coordinate system $(x^i=t,\,r,\,\theta,\,\phi)$ as follows,
\begin{equation} \label{line1}
ds^{2}=e^{\nu (r)}dt^{2}-e^{\lambda (r)}dr^{2}-r^{2}(d\theta^{2}+\sin^{2}\theta d\phi^{2}),
\end{equation}
Assume that the energy-momentum tensor for an anisotropic charged with two fluids is made up of matter with matter-energy density $\rho$, pressure $p$ of the corresponding ordinary baryonic matter (OBM), electric field intensity $E$, and 'dark' energy density $\rho^{DE}$, corresponding radial pressure $p_r^{DE}$, and tangential pressure $p_t^{DE}$ respectively. The sub-index "DE" represents dark energy throughout the manuscript. The dark energy density can be described in terms of the (varying) cosmological constant ($\Lambda$) as $\rho^{\text{DE}}=\frac{\Lambda}{8\pi}$ \cite{Ghezzi:2009ct}.\par
The corresponding energy-momentum tensor for an anisotropic charged system with two fluids can be expressed \cite{Ghezzi:2009ct} as,
\begin{equation} \label{t1}
  \begin{rcases}
    \begin{aligned}
      T_0^0 &=\rho^{\text{Tot}}+E^2= (\rho+\rho^{DE}) +E^2, \\
      T_1^1 &=-p_r^{\text{Tot}}+E^2=-(p+p_r^{DE}) +E^2, \\
      T_2^2 &=T_3^3=-p_t^{\text{Tot}}-E^2=-(p+p_t^{DE}) -E^2 ,\\
      T_0^1 &=T_1^0=0.
    \end{aligned}
  \end{rcases} \text{Energy-Momentum Tensor}
\end{equation}
where, $\rho^{\text{Tot}}=(\rho+\rho^{DE}$), $p_r^{\text{Tot}}=(p+p_r^{DE}$), and $p_t^{\text{Tot}}=(p+p_t^{DE}$) being the total matter-energy density and total pressure components, respectively.\par

Now, assuming a geometrized unit system ($G = c = 1$), we have the following Einstein-Maxwell field equations:
\begin{eqnarray}
T_0^0:~\kappa(\rho+\rho^{DE}) +E^2 &=& e^{-\lambda}\left(\frac{\lambda'}{r} + \frac{e^{\lambda}}{r^{2}}-\frac{1}{r^{2}} \right)+ 8\pi \alpha e^{-\lambda} \Bigg[\nu'\bigg(\nu'-\lambda'+\frac{2}{r}\bigg) + \nu'' -\frac{2\lambda'}{r}-\frac{2e^{\lambda}}{r^{2}} + \frac{2}{r^{2}}\Bigg], \label{fe1}\\
T_1^1:~\kappa (p+p_r^{DE}) -E^2  &=& e^{-\lambda}\left(\frac{\nu'}{r} - \frac{e^{\lambda}}{r^{2}}+\frac{1}{r^{2}} \right)- 8\pi \alpha e^{-\lambda} \Bigg[\nu'\bigg(\nu'-\lambda'+\frac{2}{r}\bigg) + \nu'' -\frac{2\lambda'}{r}-\frac{2e^{\lambda}}{r^{2}} + \frac{2}{r^{2}}\Bigg], \label{fe2}\\
T_2^2=T_3^3:~\kappa (p+p_t^{DE}) +E^2 &=& \frac{1}{4}e^{-\lambda}\Bigg[\nu'(\nu' r-\lambda')+ 2\nu''+2\frac{\nu'-\lambda'}{r}\Bigg]- 8\pi \alpha e^{-\lambda} \Bigg[\nu'\bigg(\nu'-\lambda'+\frac{2}{r}\bigg) + \nu'' \nonumber\\ && -\frac{2\lambda'}{r}-\frac{2e^{\lambda}}{r^{2}} + \frac{2}{r^{2}}\Bigg]\label{fe3}
\end{eqnarray}
where $\kappa=8\pi$ and primes $(^{\prime})$ indicate the derivative with respect to the radial coordinate `r'. 
Here $E$ is the electric field intensity is given by,
\begin{eqnarray}
E(r) &=& \frac{\kappa}{2r^2} \int_0^r \sigma r^2 \sqrt{e^{\lambda }} = \frac{q(r)}{r^2},\label{fe5}
\end{eqnarray}
where $q(r)$ indicates the total charge within the sphere of radius $r$ under consideration.\par
Thus the charge density $\sigma=\sigma(r)$ can be written in terms of electric field intensity as,
\begin{eqnarray}
\sigma &=& \frac{2}{\kappa r^2 \sqrt{e^{\lambda }}}  (E r^2)',\label{fe4}
\end{eqnarray}
In this paper, we use the well-known metric potentials  that Krori-Barua \cite{Tolman:1939jz,osti_4507306} proposed (usually known as KB {\em ansatz}) given by,
\begin{equation}\label{kb}
  \begin{rcases}
    \begin{aligned}
    \lambda(r)= A r^2 ,\\
    \nu(r)= Br^2+ C
\end{aligned}
  \end{rcases} \text{Krori-Barua Metric}
\end{equation}
where $A$, $B$ are constants with units km$^{-2}$ and $C$ is a dimensionless quantity. We will compute the numerical values of these constants through a smooth matching of interior and exterior spacetimes. The metric potentials constitute a non-singular stellar model, which will be discussed in the next sections.\par
Using the metric expressions given in (\ref{kb}), the field equations (\ref{fe1})-(\ref{fe3}) have the following form :
\begin{eqnarray}
\kappa(\rho+\rho^{DE}) +E^2 &=&\frac{e^{-A r^2} }{r^2}\Bigg[e^{A r^2} (1 - 16 \alpha \pi) + 2 A r^2 - 16 \alpha \pi \Big(2 (A - B) r^2 -1\Big) (1 + B r^2)-1\Bigg],   \label{f1}\\
\kappa (p+p_r^{DE}) - E^2 &=& \frac{e^{-A r^2} }{r^2}\Bigg[e^{A r^2} (16 \alpha \pi -1) + 2 B r^2 + 16 \alpha \pi \Big(2 (A - B) r^2 -1\Big) (1 + B r^2)+1\Bigg], \label{f2}\\
\kappa (p+p_t^{DE}) +E^2 &=&  \frac{e^{-A r^2} }{r^2}\Bigg[16 \alpha \pi \Big[e^{A r^2} + \Big(2 (A - B) r^2 -1\Big) (1 + B r^2)\Big] + 
 r^2 \Big[B (2 + B r^3) -A (1 + B r^2)\Big]\Bigg] .\label{f3}
\end{eqnarray}

\subsection{Assumption of Equation of State due to the dark energy}\label{eosde}

The nonlinearity of the aforementioned field equations (\ref{f1})-(\ref{f3}) makes it excessively difficult to obtain explicit solutions, even though the system of equations is mathematically well defined. 
\textcolor{blue}{To overcome this issue, three assumptions have been made, as suggested earlier by the researchers mentioned in Refs.}\cite{Ghezzi:2005iy,Ghezzi:2009ct,Barreto:2006cr} given as follows: \\

(i) The radial dark pressure component ($p_r^{DE}$) and the dark energy density ($\rho^{DE}$) are connected by the following relation,
\begin{eqnarray}\label{as1}
p_r^{DE} + \rho^{DE} = 0,
\end{eqnarray}

(ii) The dark energy density ($\rho^{DE}$) is linearly proportional to the OBM density, i.e.,
\begin{eqnarray}\label{as2}
\rho^{DE}=\beta \rho,
\end{eqnarray}
where $\beta(>0)$ is a proportionality factor defined as the "dark energy coupling parameter". \\

(iii) Finally, following the prior research in Refs. \cite{Barreto:2006cr, das2015anisotropic}, it is preferable to assume the difference between radial and tangential pressure components due to dark energy is proportional to the square of the electric field intensity $E$, given by the following relation,
\begin{eqnarray}\label{as3}
 p_t^{DE} - p_r^{DE} = E^2 /4\pi
\end{eqnarray}

\subsection{Proposed model of dark energy star}

Solving equations (\ref{f1})-(\ref{f3}) with the above mentioned assumptions (\ref{as1})-(\ref{as3}) yields the explicit expressions for $\rho$, $p$, and $E^2$ as follows:
\begin{eqnarray}
 \rho &=& \frac{e^{-A r^2}}{32 (1 + \beta) \pi r^2}\Bigg[  e^{A r^2} (3 - 64 \alpha \pi) + 9 A r^2 + B r^4 (A - B r) - 
 64 \alpha \pi \Big( 2 (A - B) r^2 -1\Big) (1 + B r^2)  -3\Bigg] ,   \label{f11}\\
 p &=& \frac{e^{-A r^2}}{32 (1 + \beta) \pi r^2}\Bigg[3 + e^{A r^2} ( 64 \alpha \pi -3) + \Big[ 8 B + 
    8 (A + B) \beta -A\Big] r^2 + B r^4 (  B r -A) + 64 \alpha \pi \Big[ 2 (A - B) r^2 \nonumber\\ && -1\Big](1 + B r^2) \Bigg]   ,\label{f12}\\
E^2 &=&  \frac{e^{-A r^2} \Big[  e^{A r^2} + B^2 r^5 - A (r^2 + B r^4) -1\Big]}{4 r^2} .\label{f13}
\end{eqnarray} \\

The analytical expressions for matter-energy density, radial and tangential pressure arising from dark energy are as follows:
\begin{eqnarray} 
\rho^{DE} &=&  \frac{\beta e^{-A r^2} \Big[ e^{A r^2} (3 - 64 \alpha \pi) + 9 A r^2 + B r^4 (A - B r) - 64 \alpha \pi \Big( 2 (A - B) r^2 -1\Big) (1 + B r^2) -3\Big]}{32 (1 + \beta) \pi r^2},    \label{de1}\\
p_r^{DE} &=& - \frac{\beta e^{-A r^2} \Big[ e^{A r^2} (3 - 64 \alpha \pi) + 9 A r^2 + B r^4 (A - B r) - 64 \alpha \pi \Big( 2 (A - B) r^2 -1\Big) (1 + B r^2) -3\Big]}{32 (1 + \beta) \pi r^2},  \label{de2}\\
p_t^{DE}  &=&  \frac{e^{-A r^2}}{32 (1 + \beta) \pi r^2} \Bigg[ \beta \Big[1 + e^{A r^2} (64 \alpha \pi -1 ) - 11 A r^2 + 3 B r^4 ( B r -A) + 
    64 \alpha \pi \Big( 2 (A - B) r^2 -1\Big) (1 + B r^2)\Big] +\nonumber\\ && \ 
 2 \Big[ e^{A r^2} + B^2 r^5 - A (r^2 + B r^4) -1 \Big]\Bigg].  \label{de3}
\end{eqnarray}

The analytical expressions for total matter-energy density, total radial and tangential pressure for this present model are as follows:

\begin{eqnarray} 
\rho^{\text{Tot}} &=& \rho + \rho^{DE}   \nonumber\\ 
&=& \frac{e^{-A r^2} \Big[ e^{A r^2} (3 - 64 \alpha \pi) + 9 A r^2 + 
   B r^4 (A - B r) - 64 \alpha \pi \Big( 2 (A - B) r^2 -1\Big) (1 + B r^2) -3\Big]}{32 \pi r^2}, \label{ef1}\\ \nonumber\\
p_r^{\text{Tot}} &=& p + p_r^{DE}  \nonumber\\ 
&=& \frac{e^{-A r^2} \Big[ e^{A r^2} (-3 + 64 \alpha \pi) -(A - 8 B) r^2 + 
   B r^4 (-A + B r) + 64 \alpha \pi \Big( 2 (A - B) r^2 -1\Big) (1 + B r^2) +3\Big]}{32 \pi r^2},\label{ef2}\\ \nonumber\\
p_t^{\text{Tot}} &=& p + p_t^{DE}  \nonumber\\ 
&=& \frac{e^{-A r^2} \Big[ e^{A r^2} ( 64 \alpha \pi -1) +\Big( B (8 + 3 B r^3) -3 A (1 + B r^2)\Big) r^2 + 
    64 \alpha \pi \Big( 2 (A - B) r^2 -1\Big) (1 + B r^2) +1\Big]}{32 \pi r^2}.  \label{ef3}
\end{eqnarray}

\subsection{Determination of constants using Junction or Matching Conditions}\label{match}

To generate the model parameter profiles, the values of $A$, $B$, and $C$ must be fixed. In order to examine the significant values of unknown constants, we smoothly match our interior space-time to the external space-time described by the Reissner-Nordstr\"om metric \cite{reissner1916eigengravitation,nordstrom1918energy} given by

\begin{eqnarray}
ds_{\Sigma}^{2} &=& -\left(1 - \frac{2\mathcal{M}}{r} + \frac {\mathcal{Q}^2}{r^2}\right)dt^2 + \left(1 - \frac{2\mathcal{M}}{r} + \frac {\mathcal{Q}^2}{r^2}\right)^{-1}dr^2
 + r^2(d\theta^2+\sin^2\theta d\phi^2), \label{eq22}
\end{eqnarray}
where $\mathcal{Q}$ represents the total charge contained within the surface $r = \mathcal{R}$. The continuity of the metric coefficients $g_{tt}$, $g_{rr}$, and $\frac{\partial g_{tt}}{\partial r}$ over the boundary surface $r= \mathcal{R}$ between the interior and exterior regions yields the following set of relations:

\begin{eqnarray}
1 - 2\tilde{\mathbb{X}}+ \tilde{\mathbb{Y}} = e^{\nu (\mathcal{R})} &=& e^{B\mathcal{R}^2+C},\label{eq23}\\
1 - 2\tilde{\mathbb{X}} + \tilde{\mathbb{Y}} = e^{-\lambda (\mathcal{R})} &=&  e^{-A\mathcal{R}^2},\label{eq24}\\
\tilde{\mathbb{X}} - \tilde{\mathbb{Y}} = R\Big[ \frac{\partial}{\partial r} e^{\nu (r)} \Big]_{r=\mathcal{R}} &=& B \mathcal{R} e^{B\mathcal{R}^2+C}.\label{eq25}
\end{eqnarray}
where, $\tilde{\mathbb{X}}=\frac{\mathcal{M}}{\mathcal{R}}$, $\tilde{\mathbb{Y}}=\frac {\mathcal{Q}^2}{\mathcal{R}^2} $ and both $\tilde{\mathbb{X}},\,\tilde{\mathbb{Y}}$ are dimensionless quantity.\\

Now simultaneously solving the Eqs.~(\ref{eq23})-(\ref{eq25}) along with the  condition $p(r=\mathcal{R})=0$, we determine the values of the constants $A$, $B$, $C$ and $\beta$ as,
\begin{eqnarray}
A &=& - \frac{1}{\mathcal{R}^2} \ln \left( 1 - \frac{2\mathcal{M}}{\mathcal{R}} + \frac {\mathcal{Q}^2}{\mathcal{R}^2}
\right),\label{eq26}\\ 
B &=&   \frac{1}{\mathcal{R}^2} \left[\frac{\mathcal{M}}{\mathcal{R}} - \frac {\mathcal{Q}^2}{\mathcal{R}^2}\right] \left[1 - \frac{2\mathcal{M}}{\mathcal{R}} + \frac {\mathcal{Q}^2}{\mathcal{R}^2}
\right]^{-1}, \label{eq27}\\ 
C &=&  \ln \left[ 1 - \frac{2\mathcal{M}}{\mathcal{R}} +
\frac {\mathcal{Q}^2}{\mathcal{R}^2} \right]- \frac{ \frac{\mathcal{M}}{\mathcal{R}} - \frac {\mathcal{Q}^2}{\mathcal{R}^2}}{1 - \frac{2\mathcal{M}}{\mathcal{R}} + \frac {\mathcal{Q}^2}{\mathcal{R}^2} }  \label{eq28}\\
\beta &=& \frac{e^{A \mathcal{R}^2} (3 - 64 \alpha \pi) + (A - 8 B) \mathcal{R}^2 + 
 B \mathcal{R}^4 (A - B \mathcal{R}) - 
 64 \alpha \pi \Big(2 (A - B) \mathcal{R}^2 -1\Big) (1 + B \mathcal{R}^2) -3}{8 (A + B) \mathcal{R}^2}\label{eq29}
\end{eqnarray} \\

\textcolor{blue}{Finally, the values of every concerned constant in the KB metric have been determined in terms of $\mathcal{M}, \mathcal{R}$, and $\mathcal{Q}$.} From (\ref{eq29}) we see that the dark energy coupling parameter $\beta$ is a function of the Rastall coupling parameter $\alpha$. \textcolor{blue}{To examine the physical characteristics of this current model, the compact star Her X-1 have been specifically examined}  with observed mass and radius $\mathcal{M} = 0.85 \pm 0.15~M_{\odot},\, \mathcal{R} = 8.1_{-0.41}^{+0.41}$ km \cite{Abubekerov:2008inw}. Furthermore, it has been additionally assumed that $\mathcal{Q} = 1.31$ and $\kappa'=0.02$ for the simplicity of constructing this stellar model.

\section{Exploration of celestial physical attributes}\label{phy}

\subsection{Consistency in metric potentials}\label{mp}

\begin{figure}[H]
    \centering
        \includegraphics[scale=.55]{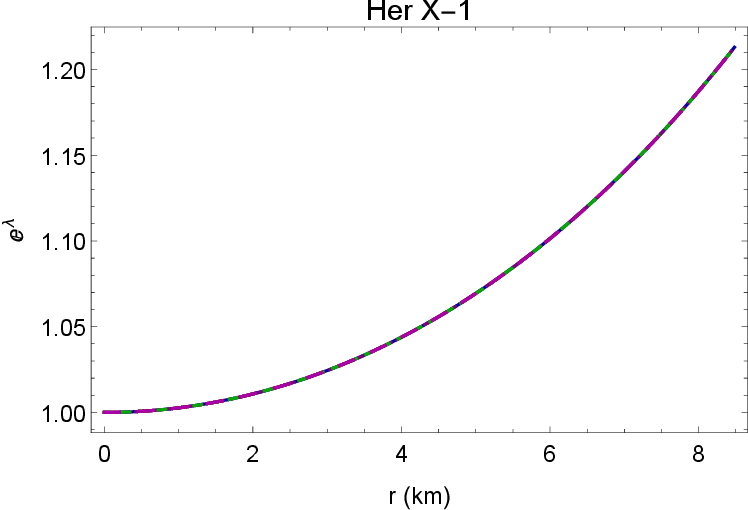}
         \includegraphics[scale=.55]{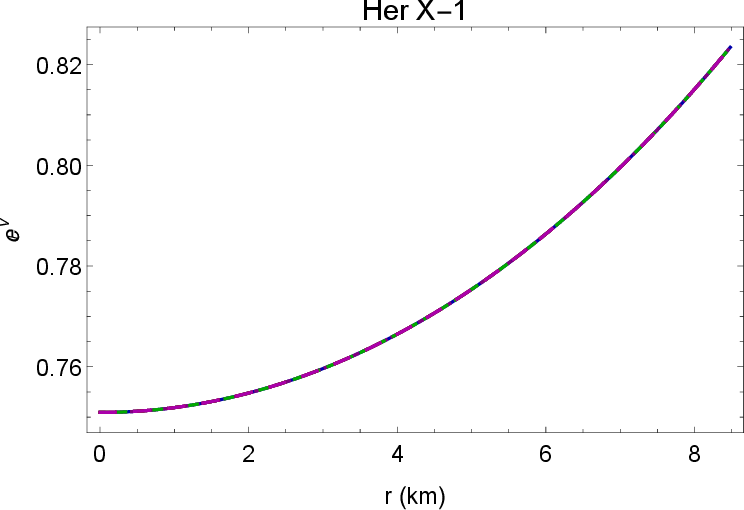}
        \caption{Variation of metric functions $e^{\lambda(r)}$ and $e^{\nu(r)}$ with respect to `r' with required magnified inset.}\label{metric}
\end{figure}

The temporal components $e^{\nu(r)}$ and the spatial components $e^{\lambda(r)}$ have been examined. It is simple for us to verify that, $[{e^{\nu(r)}}]_{r = 0}= e^C > 0$, a non-zero constant, and $[{e^{\lambda(r)}}]_{r=0} = 1$ and which result in the fact that both metric potential components being finite at the centre and have regularity throughout $r < R$ \cite{Delgaty:1998uy, Pant:2010iub}. 
Moreover, $\Big[\frac{d(e^{\nu(r)})}{dr}\Big]_{r=0} = \Big[2Br{e^{Br^2 +C}}\Big]_{r=0} =0$ \text{and}\\ $\Big[\frac{d(e^{\lambda(r)})}{dr}\Big]_{r=0} = \Big[2Ar{e^{Ar^2}}\Big]_{r=0} =0$,\\
that means the derivatives of the metric potential components diminish at the star's center. The radial profiles of the components of the metric coefficients in Fig.~(\ref{metric}) demonstrate their positive and consistent patterns within the star's interior. The interior space-time smoothly matches the asymptotically flat exterior space-time at the surface ($r = \mathcal{R}$), satisfying the Darmois-Israel condition \cite{Chu:2021uec, darmois1927equations, Israel:1966rt}. By matching these, we may determine the values of the constant parameters that define this model. As a result, we confirm that the components of the metric potential behave properly in the range $(0, \mathcal{R})$.

\subsection{Stellar patterns of OBM energy density, pressure, and electric field}

\begin{figure}[H]
    \centering
        \includegraphics[scale=.47]{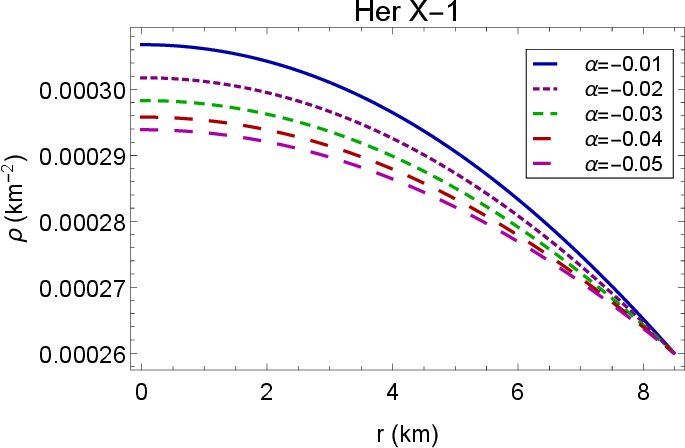}
         \includegraphics[scale=.47]{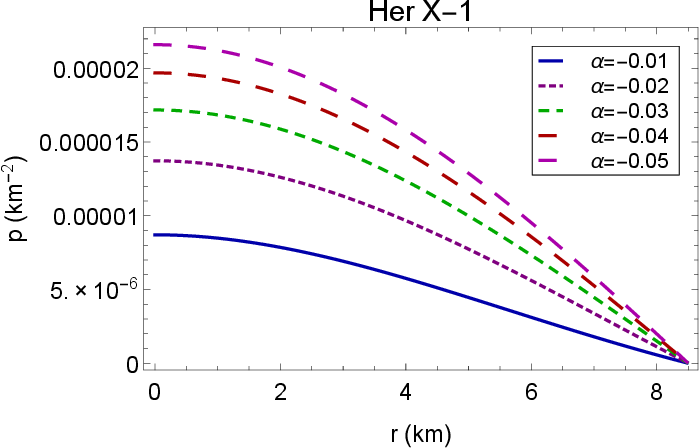}
         \includegraphics[scale=.47]{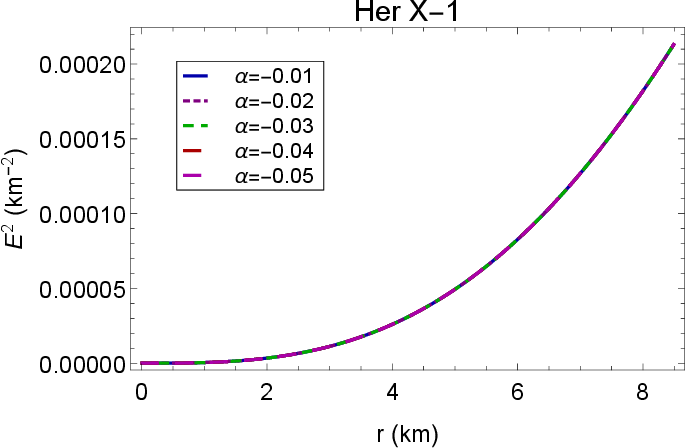}
        \caption{Profiles of the OBM matter-energy density, pressure, and electric field intensity ($E^2$) in relation to to `r'.}\label{rho}
\end{figure}

Similar to how pressure is important for defining the stellar boundary and general stability, the matter-energy density of the confined matter is an essential component in determining stellar stability against gravitational collapse \cite{chandrasekhar1984stars}. The central values of $\rho~ \text{and}~ p$ under RG are obtained from equations (\ref{f11}) and (\ref{f12}) as follows:
$\rho_c= [\rho]_{r=0} = \frac{3 A - 48 A \alpha \pi + 48 \alpha B \pi}{8 \pi (1 + \beta)} > 0$,
 $p_c= [p]_{r=0} =  \frac{2 B (1 + \beta - 24 \alpha \pi) + A (2 \beta + 48 \alpha \pi -1)}{8 \pi (1 + \beta)} >0$.  \\
 Obviously, $\rho_c$ and $p_c$ are finite.\\
 According to Fig.~\ref{rho}, matter-energy density and pressure decrease monotonically with radius $r$ and peak in the star's center. As observed in Fig.~\ref{rho}, $[E^2]_{r=0} = 0,$ indicating that the electric field intensity is zero at the center and progressively increases as $r$ increases. Finally, it is seen that $\rho$, $p$, and $E^2$ within the star are not negative. For various values of the Rastall coupling parameter $\alpha$, we provided numerical values of central density ($\rho_c$), surface density ($\rho_s$), central pressure ($p_c$) and central pressure-density ratio $p_c/\rho_c$ in Table~\ref{tab1}.\\
 
\begin{figure}[H]
    \centering
        \includegraphics[scale=.55]{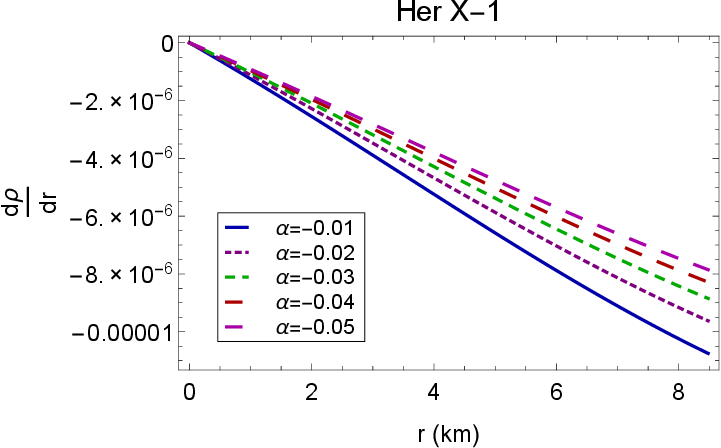}
         \includegraphics[scale=.55]{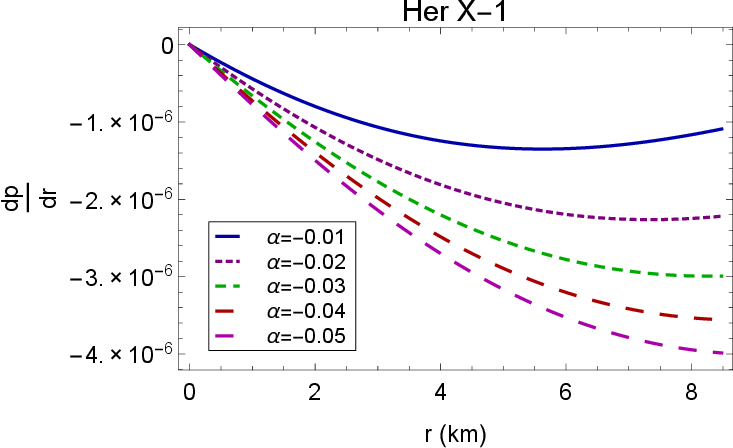}
        \caption{Gradients of matter density and pressure are plotted with respect to `r'.}\label{grad}
\end{figure}

Additionally, the OBM's pressure and density gradients for this current model are obtained as,
\begin{eqnarray}
  \frac{d\rho}{dr} &=& \frac{e^{-A r^2}}{32 (1 + \beta) \pi r^3}\Bigg[6 + 2 e^{A r^2} ( 64 \alpha \pi -3) + 6 A r^2 +  128 \alpha \pi \Big[(A - 2 B) (2 A - B) r^4  \nonumber\\ && + 2 A (A - B) B r^6 -1 - A r^2\Big] +r^4 \Big[ 2 A B (1 + B r^3) -3 B^2 r - 2 A^2 (9 + B r^2)\Big]\Bigg], \\
 \frac{dp}{dr} &=& \frac{e^{-A r^2}}{32 (1 + \beta) \pi r^3}\Bigg[ e^{A r^2} (6 - 128 \alpha \pi) - 6 (1 + A r^2) - 
 128 \alpha \pi \Big[(A - 2 B) (2 A - B) r^4 + 
    2 A (A - B) B r^6 \nonumber\\ && -1 - A r^2\Big] +r^4 \Big[3 B^2 r + 2 A^2 (1 - 8 \beta + B r^2) - 2 A B (9 + 8 \beta + B r^3)\Big]\Bigg]  
\end{eqnarray}
The density and pressure gradients remain negative throughout the fluid sphere and vanish at its core, as shown in Fig.~\ref{grad}.
We see that,
\begin{equation}
\frac{d\rho}{dr} \Bigg\rvert_{r=0} = 0, \frac{dp}{dr} \Bigg\rvert_{r=0} = 0,
\end{equation}
Now, differentiating two times of $\rho, p$ with respect to $r$ we get,
$$\Bigg[\frac{d^2\rho}{dr^2}\Bigg]_{r=0} = \frac{A (-21 A + 2 B) + 64 \alpha \pi (5 A^2 - 10 A B + 4 B^2) }{4\kappa (1 + \beta)},$$\\
$$\Bigg[\frac{d^2 p}{dr^2}\Bigg]_{r=0} = \frac{A \Big(5 A - 18 B - 16\beta (A + B) \Big) - 64 \alpha \pi (5 A^2 - 10 A B + 4 B^2) }{4\kappa (1 + \beta)}.$$
Numerically, we can easily verify that the second derivatives assume negative values at the center for the chosen Rastall coupling parameter range, $\alpha \in [-0.05, -0.01]$ considered in this work. As a result, we can conclude that the density and pressure attain maximum values at the stellar core.

\textcolor{blue}{Now, we all are interested in evaluating the allowed range of the Rastall coupling parameter $\alpha$ as proposed in some earlier investigations \cite{Moradpour:2016fur, ElHanafy:2022kjl, Afshar:2023uyw}.\\ 
We know from the assumption of dark energy EoS that $\beta>0$.\\
Now $p_c>0 \implies \alpha > \frac{A - 2 B - 2 (A + B) \beta}{48 (A - B) \pi}$,\\
and $\rho_c > 0 \implies \alpha < \frac{A}{16 (A - B) \pi}$.\\
Also since, $p_c/\rho_c <1 \implies \alpha < \frac{-2 A + B + (A + B) \beta}{48 (-A + B) \pi}.$\\
So we finally obtain the desired range of the Rastall coupling parameter $\alpha$ as,
\begin{equation}
\frac{A - 2 B - 2 (A + B) \beta}{48 (A - B) \pi} < \chi < \text{Min}\Bigg\{ \frac{A}{16 (A - B) \pi}, \frac{-2 A + B + (A + B) \beta}{48 (-A + B) \pi}\Bigg\}.
\end{equation}}

\subsection{Dark energy density and dark pressure}

According to physical cosmology, dark energy is an imaginary form of energy that permeates space and has the propensity to speed up the expansion of the universe \cite{Peebles:2002gy}. Two hypothesized types of dark energy are the cosmological constant, which has a constant energy density and fills space uniformly, and scalar fields such as quintessence or moduli, which have a dynamic energy density that varies in time and space. The reason why it is called "dark" is that it has no electric charge and does not react to electromagnetic radiation, like light. It allows us to compare our observations to an idea known as the cosmological constant, which Albert Einstein proposed in 1917 to mitigate the effects of gravity in his equations to construct a universe that was neither expanding nor shrinking. Einstein eventually removed it from his computations. Dark energy has a significant negative pressure, which leads to the expansion of the universe. When an object pushes outward against its surroundings, it exerts positive pressure. For fluids, this is the typical scenario. When an object pulls on its surroundings instead, a negative pressure, also known as tension, arises. The Friedmann-Lemaitre-Robertson-Walker (FLRW) metric, a cosmological application of GR, states that, like mass density, an object's internal pressure affects its gravitational attraction towards other objects. Negative pressure produces gravitational repulsion.\par
It is still mysterious what this dark energy is exactly. It is generally referred to be very homogeneous, not particularly dense, and to not interact with any of the fundamental forces other than gravity. Its low density (approximately $10^{-29}$ gm / cm$^3$) makes laboratory detection very challenging. Dark energy presently accounts for more than three-quarters of the universe's total mass-energy, according to the standard cosmological model. Dark energy is being used to build a cyclic model of the cosmos \cite{Baum:2006ee}.\par
\begin{figure}[H]
    \centering
        \includegraphics[scale=.46]{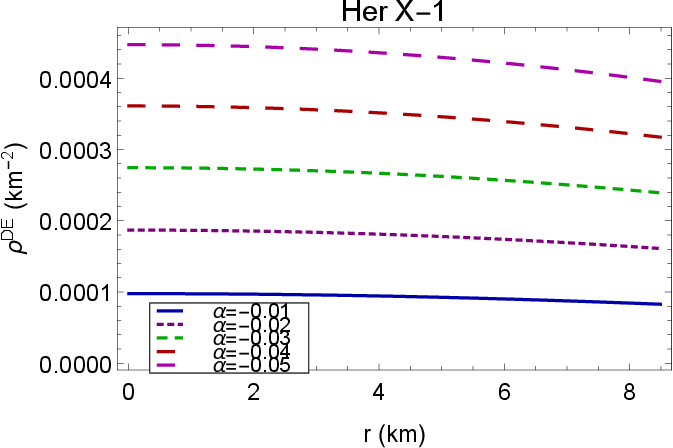}
        \includegraphics[scale=.46]{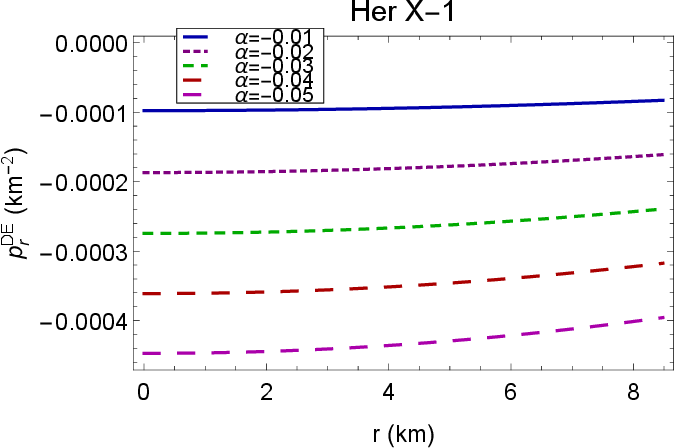}
         \includegraphics[scale=.46]{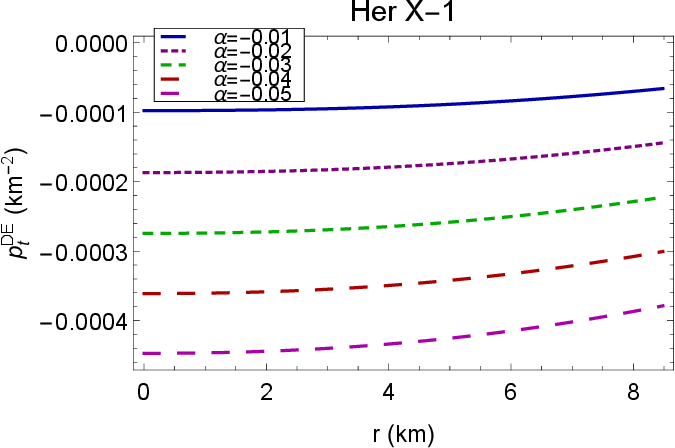}
        \caption{The variation of dark energy density and dark pressure with respect to `r'.}\label{dark}
\end{figure}

This present study simulates the repulsive nature of dark energy, with positive energy density ($\rho^{DE}>0$) and negative radial dark pressure ($p_r^{DE}<0$). Figure ~\ref{dark} illustrates dark energy density, radial dark pressure, and transverse dark pressure. Based on the seed assumption ~(\ref{as3}), Fig.~\ref{dark} shows that the transverse dark pressure $p_t^{DE}$ is negative throughout the region.

\subsection{Mass-radius profile and compactness}

\begin{figure}[H]
    \centering
        \includegraphics[scale=.5]{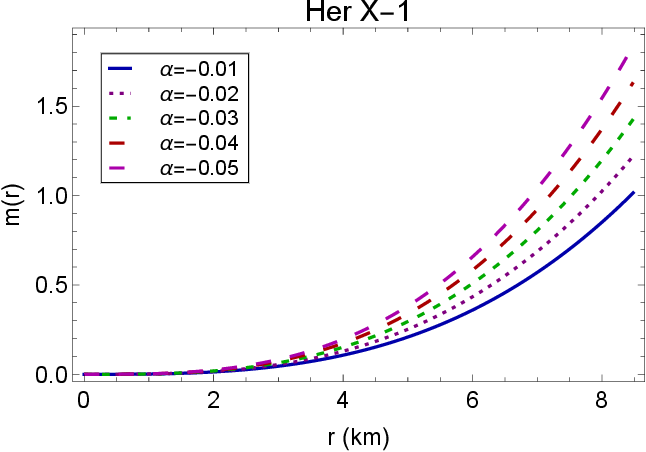}
        \includegraphics[scale=.5]{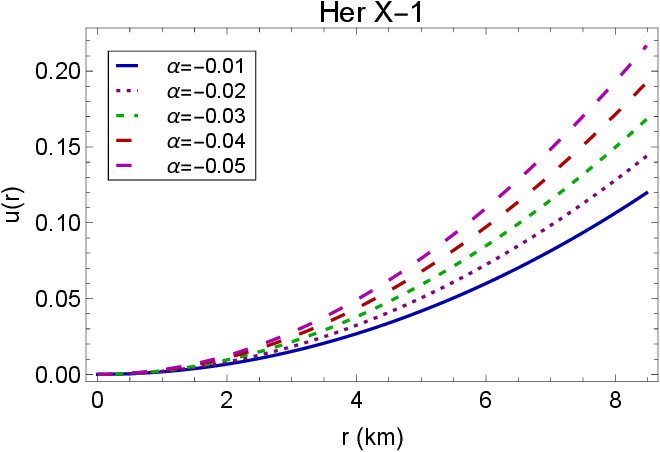}
       \caption{The mass function and compactness factor are plotted against `r'.}\label{mass}
\end{figure}

Since the active mass of a stellar system is gravitationally confined to a finite spatial extent $r= \mathcal{R}$, it is clear that the stellar mass depends on the density profile and increases with the confining radius \cite{buchdahl1959general, glendenning2012compact}. By evaluating the integral corresponding directly to the energy density (\ref{f11}) and electric field intensity (\ref{f13}) along with the following expression \cite{florides1983complete, kumar2022isotropic}, we can simply estimate the profile of mass function $m(r)$ for an electrically charged fluid sphere in the context of Rastall gravity as,
\begin{eqnarray}
 m(r) &=& 4\pi\int^r_0{\Big(\rho^{\text{Tot}} r^2+ r q \sigma  \sqrt{e^{\lambda}}\Big)\,dr} \nonumber\\  
 &=& 4\pi\int^r_0{\Big[\big(\rho + \rho^{DE}\big) r^2+  \sigma  E r^3 \sqrt{e^{\lambda}}\Big]\,dr} \nonumber\\  
  &=& \frac{1}{8} \Bigg[(5 - 64 \alpha \pi) r - \frac{r e^{-A r^2}
      \Big[96 \alpha B^2 \pi + 64 A \alpha B^2 \pi r^2 + A^3 (r^2 + B r^4) - 
      A^2 \big(-5 + B^2 r^5 + 64 \alpha (\pi + B \pi r^2)\big)\Big]}{A^2} \nonumber\\ && + \frac{48 \alpha B^2 \pi^{3/2} erf(\sqrt{A} r)}{A^{5/2}} \Bigg],~~~\label{mm}
 \end{eqnarray}
where $erf(x)$ is the Gaussian error function.\\
It should be noted that the mass function $m(r)$ is a function of $r$ such that $m(r = 0) = 0$, while $m(r= \mathcal{R}) = \mathcal{M}$. In Figure~\ref{mass}, the evolutionary nature of the mass function (\ref{mm}) versus the radial coordinate $r$ for a spherically symmetric dark energy stellar model has been examined \textcolor{blue}{versus the Rastall parameter $\alpha$.}
\par
We can easily check that the mass function is regular (or non-singular) and non-negative at every point within the stellar structure. In addition, it is precisely proportional to the radial distance $r$, and the maximum mass is achieved on the surface $r=\mathcal{R}$, as shown in Fig.~\ref{mass}.
The compactification factor ($u(r)$), which is the ratio of the maximum allowable mass to the radius, ensures the stability of a compact star model. Buchdahl \cite{buchdahl1959general} in one of his pioneering research works derived an upper bound for the ratio, i.e. $\frac{2m(r)}{r}< \frac{8}{9}$, for a static spherically symmetric perfect fluid sphere. Mak et al. \cite{Mak:2001ie} later extended it to charged fluid spheres. Jotania and Tikekar \cite{Jotania:2006zq} classified the celestial compact objects into several groups based on the ratio of $\frac{m(r)}{r}$. It is obvious that the parameter $u(r)$ is dimensionless. \\
To determine the compactness of this current stellar model, the ratio can be obtained as
\begin{eqnarray}
 u(r) &=& \frac{m(r)}{r} \nonumber\\  
      &=& \frac{1}{8} \Bigg[(5 - 64 \alpha \pi) - \frac{e^{-A r^2}
      \Big[96 \alpha B^2 \pi + 64 A \alpha B^2 \pi r^2 + A^3 (r^2 + B r^4) - 
      A^2 \big(-5 + B^2 r^5 + 64 \alpha (\pi + B \pi r^2)\big)\Big]}{A^2} \nonumber\\ && + \frac{48 \alpha B^2 \pi^{3/2} erf(\sqrt{A} r)}{A^{5/2}r} \Bigg]~~~\label{ur}
 \end{eqnarray}

Figure~\ref{mass} (right panel) illustrates the evolution of the compactness ratio $u(r)$ of this stellar model, which increases with $r$ for several values of Rastall parameter $\alpha$. From the figure, we can easily verify that $u(r)$ is completely positive and regular within the stellar region.

\subsection{Surface redshift and Gravitational redshift}
\begin{figure}[H]
    \centering
        \includegraphics[scale=.52]{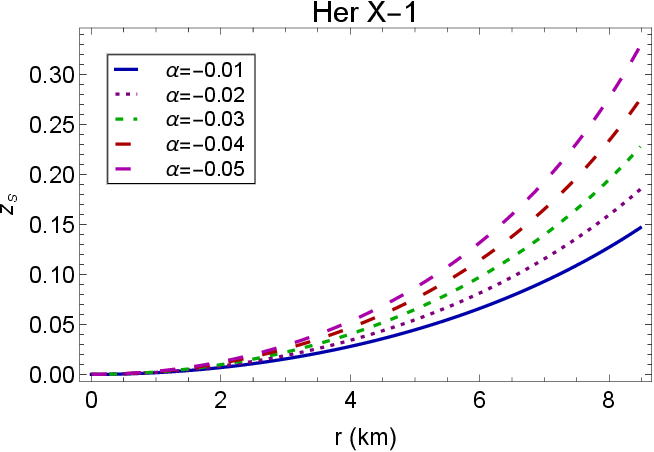}
        \includegraphics[scale=.46]{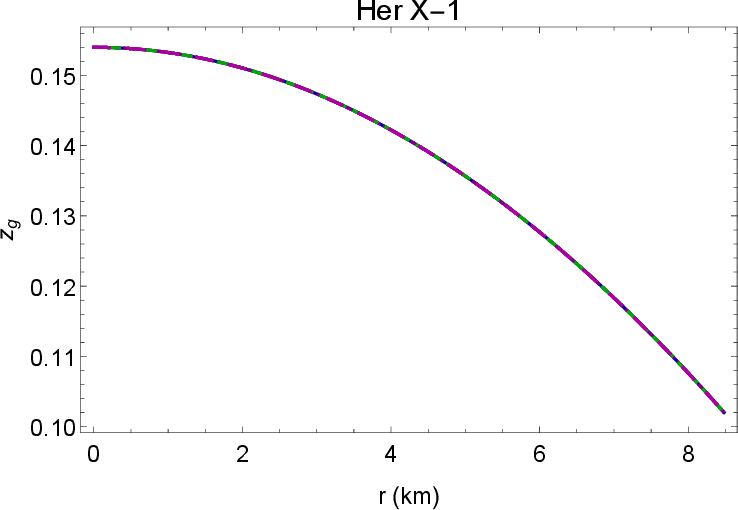}
         \caption{Surface redshift and gravitational redshift are plotted against `r'.}\label{red}
\end{figure}

The surface redshift $z_s(r)$ of a stellar model can be obtained in terms of the compactness ratio $u(r)$ as 
\begin{eqnarray}
z_s(r)= \Big[1-2u(r)\Big]^{-\frac{1}{2}}-1~~~\label{zs}
\end{eqnarray}
Barraco and Hamity \cite{Barraco:2002ds} demonstrated that for an isotropic star without a cosmological constant, $z_s < 2$. Bohmer and Harko \cite{Boehmer:2006ye} later suggested that the surface redshift of an anisotropic star might reach a maximum value of $z_s \leq 5$ in the presence of the cosmological constant. Later, Ivanov \cite{Ivanov:2002xf} refined the maximum acceptable limit of $z_s$ as $5.211$. However, for a quasiblack hole, the redshift at the stellar surface is infinitely large-that is, it has a numerical value of the order of 100 \cite{Arbanil:2014usa}.\\
Whereas, the gravitational redshift (or inner redshift) $z_g(r)$ within a static line element can be represented as 
\begin{eqnarray}
z_g(r)=e^{-\frac{\nu(r)}{2}}-1. 
\end{eqnarray}
\textcolor{blue}{In Fig.~(\ref{red}), both variations of $z_s(r)$ and $z_g(r)$ have been displayed for this model from center to surface.} These plots make evident how the surface redshift and gravitational redshift exhibit opposing trends throughout the fluid sphere. It is evident that $z_g(r)$ starts from its maximum in the center and then progressively decreases as the radius increases, reaching its minimum close to the surface. 

 Table~\ref{tab1} displays the numerically computed results for the mass (in $M_{\odot})$), compactification factor ($u(R)$), and surface redshift ($z_s(R)$) for a range of values of the Rastall coupling parameter $\alpha$.

\begin{table}[H]
\begin{center}
\caption{\label{tab1} Numerically computed values of $\beta$, central density $(\rho_c)$, surface density $(\rho_s)$, central pressure $(p_c)$, $u(\mathcal{R})$, $z_s(\mathcal{R})$, $p_c/\rho_c$, mass ($M_{\odot}$) correspond to the CS candidate Her X-1 for different values of $\alpha$.}
\begin{tabular}{ccccccccc}
\hline
\hline
          & \multicolumn{8}{c}{Her X-1}  \\
\cline{1-9}
$\alpha$  & $\beta$  &$\rho_c$  &   $\rho_s$  &   $p_c$  &       $u(\mathcal{R})$      &    $z_s(\mathcal{R})$  &   $p_c/\rho_c$ & $M (~M_{\odot})$ \\
  &  & $\text{gm}~\text{cm}^{-3}$  &   $\text{gm}~\text{cm}^{-3}$  &$\text{dyne}~\text{cm}^{-2}$  &     &     &    &  \\  
\hline
     -0.01 & 0.318396 &$4.13957 \times 10^{14}$  & $3.50679 \times 10^{14}$  & $1.17527 \times 10^{13}$ & 0.120224 & 0.147416 & 0.0283911  & 1.0219 \\

     -0.02 & 0.619168 &$4.07183 \times 10^{14}$  & $3.50679 \times 10^{14}$  & $1.85268 \times 10^{13}$ & 0.144606 & 0.186123 & 0.0455 & 1.22915\\

     -0.03 & 0.919941 &$4.02531 \times 10^{14}$  & $3.50679 \times 10^{14}$  & $2.31785 \times 10^{13}$ & 0.168988 & 0.229032 & 0.0575819 & 1.4364\\

     -0.04 & 1.22071 &$3.99139 \times 10^{14}$  & $3.50679 \times 10^{14}$  & $2.65701 \times 10^{13}$ & 0.19337 & 0.276961 & 0.0665686  & 1.64365\\

     -0.05 & 1.52149 &$3.96557 \times 10^{14}$  & $3.50679 \times 10^{14}$  & $2.91526 \times 10^{13}$ & 0.217752 & 0.330975 & 0.0735144 & 1.85089\\

\hline
\hline
\end{tabular}
\end{center}
\end{table}

\section{Stability analysis of our model and equilibrium of forces}\label{stable}

\subsection{Causality condition}
In this part, we will look into another crucial "physical acceptability condition" for realistic models: the causality condition via sound velocity. The system will be stable in the context of sound speed if it meets both Herrera's cracking concept \cite{Herrera:1992lwz, Abreu:2007ew} and the causality criterion. According to this requirement for a physically viable model, the square of the sound velocity $V^2$ =$\frac{dp}{d\rho}$ must be less than unity. This indicates that sound travels at a far slower speed than light. 

\begin{figure}[H]
    \centering
        \includegraphics[scale=.52]{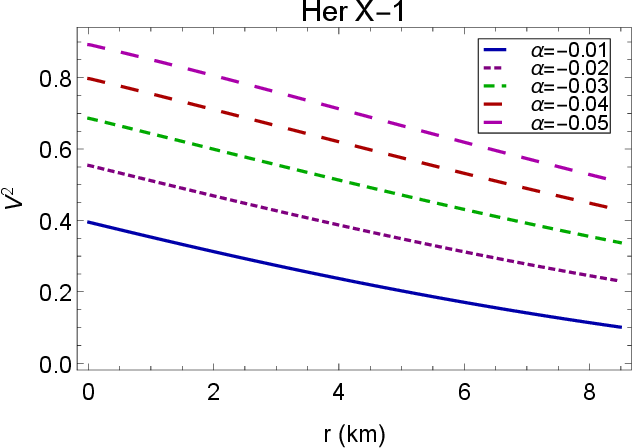}
       \caption{Square of sound velocity is plotted against `r'.}\label{sv}
\end{figure}

Fig.~(\ref{sv}) illustrates the visual analysis of this, and we can easily check that it falls within our desired range $(0, 1)$ within the stellar object. Hence, this suggested charged dark energy star model satisfies the causality condition. 

\subsection{Adiabatic index}

When examining the stability state of a fluid sphere model, the adiabatic index is an essential feature to consider. The "adiabatic index" is a ratio between two specific heats \cite{1976A&A....53..283H} that determines the stiffness of the EoS for a certain density, indicated by $\Gamma$. It can investigate the stability of both relativistic and non-relativistic fluid spheres. Earlier Chandrasekhar \cite{Chandrasekhar:1964zz} suggested that the dynamic stability of a stellar model can be assessed against an infinitesimal radial adiabatic perturbation. It has been suggested that the adiabatic index should be greater than $4/3$ to meet the stability. The adiabatic index ($\Gamma$) is defined as

\begin{eqnarray} \label{ad}
\Gamma=\frac{\rho+p}{p}V^2. 
\end{eqnarray}

\begin{figure}[H]
    \centering
        \includegraphics[scale=.52]{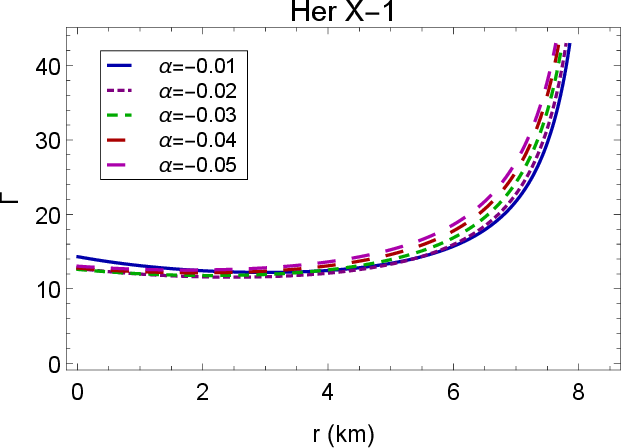}
       \caption{Evolution of adiabatic index $\Gamma$ w.r.t. radial distance `r'.}\label{gam}
\end{figure}
Fig.~(\ref{gam}) displays graphical representations of the adiabatic index. It is evident that the adiabatic index is greater than $4/3$ across the whole stellar distribution, indicating that the model depicts a stable configuration.

\subsection{Energy conditions}
\textcolor{blue}{Pointwise energy conditions were initially proposed as physically feasible constraints on matter within the context of mathematical relativity. For example, they try to express the attraction of gravity or the positive of mass. More importantly, they have been used as presumptions in mathematical relativity to argue against the possibility of wormholes and other exotic phenomena, as well as singularity theorems. Energy conditions (ECs) have a considerable impact in the context of exotic characteristics. These constraints restrict the contraction of the stress tensor at each point of the space-time. This part examines the ECs associated with an anisotropic dark energy connected to a charged-fluid sphere model.
The ECs are local inequalities defined in the context of GR that provide a relationship that adheres to specific boundaries between matter-energy density and pressure. Assuming these conditions are positive means that matter should flow along a time-like or null-world line. To rewrite these ECs in the Einstein equation, we can substitute the stress tensor with the Ricci curvature tensor, resulting in a geometric form compared to the original physical form. These ECs depend on the energy-momentum tensor $T_{\mu\nu}$. These ECs could be extended to modified gravity. So for this particular RG
case the ECs should also be modified depending on the Rastall coupling parameter \cite{Moradpour:2016ubd, Li:2019jkv, Zubair:2021dyg, ElHanafy:2022kjl, Mustafa:2022jso}.} \par

The six standard, model independent, pointwise ECs that are known as (i) Null energy condition (NEC), (ii) Weak energy condition (WEC), (iii) Strong energy condition (SEC), (iv) Weak dominating energy condition (WDEC), (v) Strong dominating energy condition (SDEC), and (vi) Trace energy condition (TEC) \cite{bondi1947spherically, witten1981new, visser1997energy, Andreasson:2008xw, garcia2011energy} should be satisfied by an anisotropically charged fluid sphere. \textcolor{blue}{Since SEC has a geometrical aspect, it assumes different forms in RG compared to GR.} 
For this current model, the corresponding ECs must hold simultaneously across the stellar model, as stated by the subsequent inequalities:

\begin{itemize}
\item \textbf{NEC}:~$ \rho + \frac{E^2}{8\pi}  \geq 0$, 
\item \textbf{WDEC}:~$\rho - |p| + \frac{E^2}{4\pi} \geq 0$, 
\item \textbf{SDEC}:~$\rho - 3p + \frac{E^2}{4\pi} \geq 0$, 
\item \textbf{SEC}:~$\rho + 3p + \frac{2\kappa' \alpha}{4\kappa' \alpha -1} (\rho - p) + \frac{E^2}{4\pi} \geq 0$, 
\item \textbf{WEC}:~$\rho + p + \frac{E^2}{4\pi} \geq 0$,
\item \textbf{TEC}:~$\rho - 3p \geq 0$.
\end{itemize}

\begin{figure}[H]
    \centering
        \includegraphics[scale=.47]{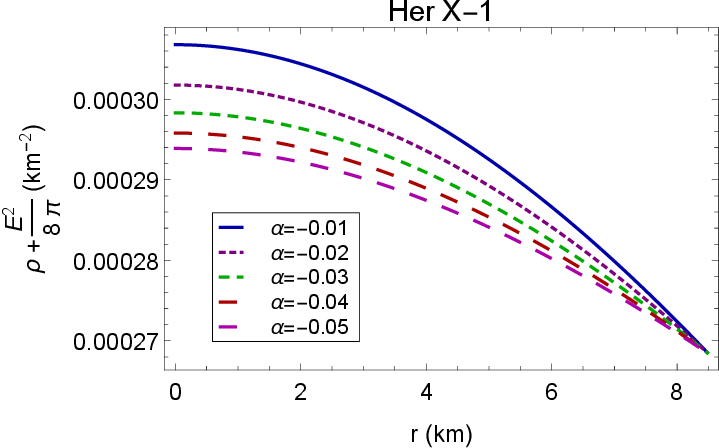}
        \includegraphics[scale=.47]{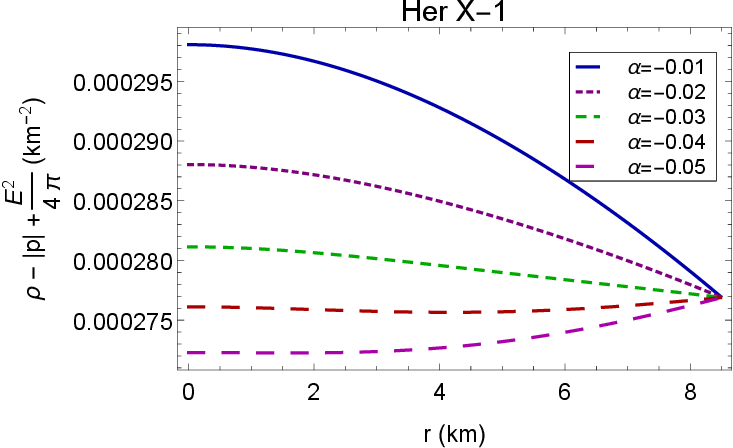}
        \includegraphics[scale=.47]{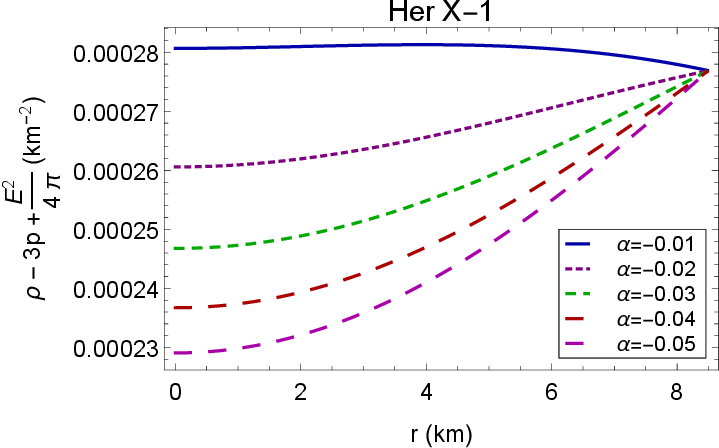}
        \includegraphics[scale=.47]{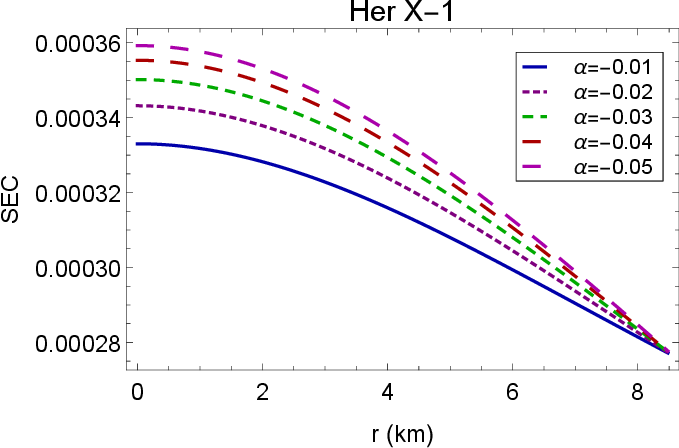}
        \includegraphics[scale=.47]{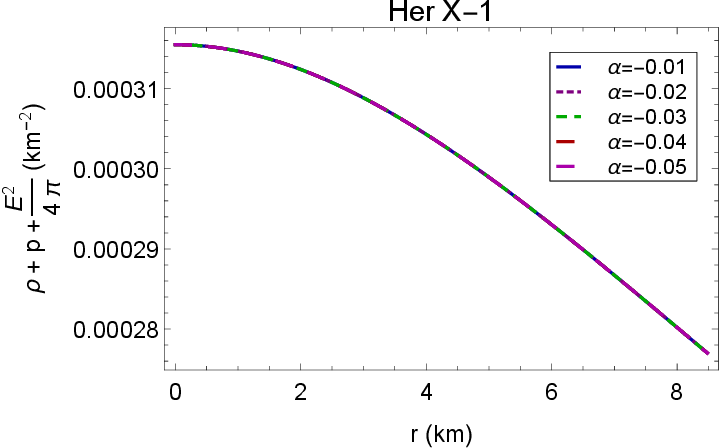}
        \includegraphics[scale=.47]{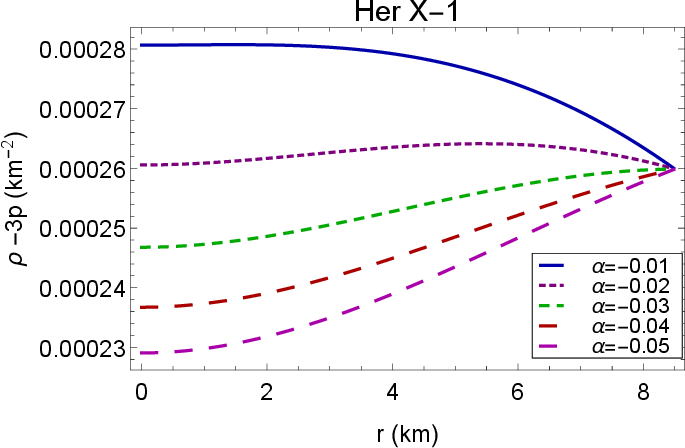}
         \caption{Variation of different energy conditions against radius `r'.}\label{ec}
\end{figure}

Figure (\ref{ec}) shows that this proposed dark energy star model meets all the energy criteria for all $r$. Hence, this model is physically reasonable.

\subsection{Hydrostatic Equilibrium via Modified TOV Equation}

Hydrostatic equilibrium is another fundamental feature that can be used to verify the stability of the model that is being presented in the presence of dark energy. Now in order to ensure the physical sustainability of this model's potential equilibrium state considering various forces acting upon it,\textcolor{blue}{ the Tolman-Oppenheimer-Volkoff (TOV) equation \cite{Oppenheimer:1939ne, ponce1987general, ponce1993limiting} has been utilized, which is governed by the following expression}

\begin{equation}\label{tov1}
-\frac{M_G(r)}{r^2} (\rho+p) e^{(\lambda-\nu)/2}-\frac{d}{dr}(p+p_r^{DE}) +\frac{2}{r}(p_t^{DE}-p_r^{DE}) +\frac{8\pi \alpha}{(32\pi \alpha -1)} \frac{d}{dr} (\rho - 3p + 4 \rho^{DE}) + E (r)\sigma(r) e^{\frac{\lambda(r)}{2}}=0,
\end{equation}
recommended by Tolman, Oppenheimer, and Volkoff and hence called the {\em TOV} equation.\\

Here, $M_G(r)$ is the gravitational mass at radius $r$, as determined by the Einstein field equations and the Tolman-Whittaker formula \cite{PhysRev.35.875}, is defined as
\begin{equation}\label{tov2}
M_G(r)=\frac{1}{2}r^2 \nu' e^{(\nu - \lambda)/2}.
\end{equation}

Now inserting the $M_G(r)$ expression in equation \eqref{tov1}, we obtain,
\begin{equation}\label{tov1a}
-\frac{\nu'}{2}(\rho+p)-\frac{d}{dr}(p+p_r^{DE})+\frac{2}{r}(p_t^{DE}-p_r^{DE}) +\frac{8\pi \alpha}{(32\pi \alpha -1)} \frac{d}{dr} (\rho - 3p + 4 \rho^{DE}) + E (r)\sigma(r) e^{\frac{\lambda(r)}{2}}=0.
\end{equation}

So the preceding equation (\ref{tov2}) can be rewritten as,
\begin{equation}
F_g + F_h + F_d + F_R + F_e=0,
\end{equation}
Here $F_g, F_h$ are $F_e$ are the gravitational, hydrodynamic, and electrical forces respectively along with a force due to dark energy $F_d$. An extra force $F_R$ arises due to the Rastall theory of gravity. These forces are given as follows:
\begin{eqnarray}
\text{Gravitational force:}~ F_g&=&-\frac{\nu'}{2}(\rho+p) \\
\text{Hydrodynamic force:}~ F_h&=& -\frac{d}{dr}(p+p_r^{DE})\\
\text{Dark energy force:}~ F_d&=& \frac{2}{r}(p_t^{DE}-p_r^{DE})\\
\text{Rastall force:}~ F_R &=& \frac{8\pi \alpha}{(32\pi \alpha -1)} \frac{d}{dr} (\rho - 3p + 4 \rho^{DE}) \\
\text{Electrical force:}~ F_e&=& E (r)\sigma(r) e^{\frac{\lambda(r)}{2}}.
\end{eqnarray}

\begin{figure}[H]
    \centering
        \includegraphics[scale=.465]{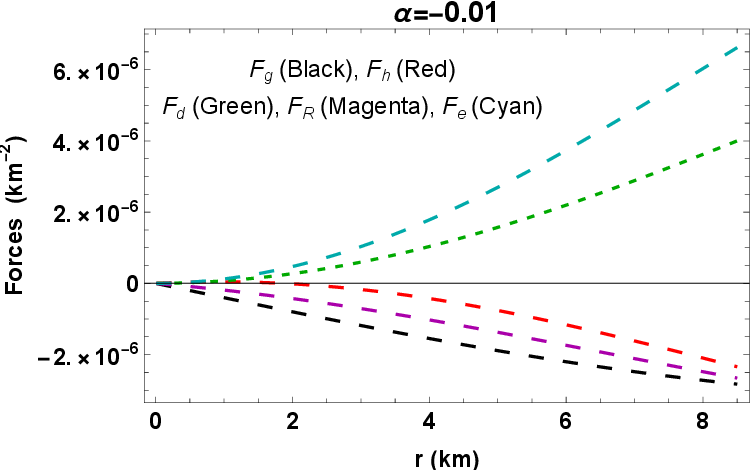}
        \includegraphics[scale=.465]{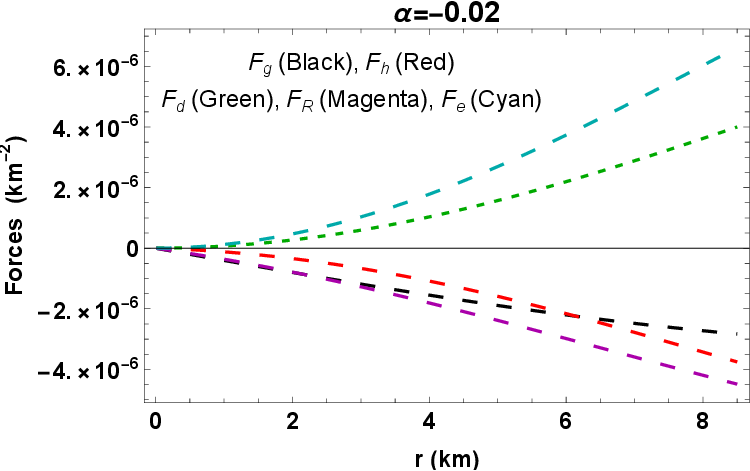}
        \includegraphics[scale=.465]{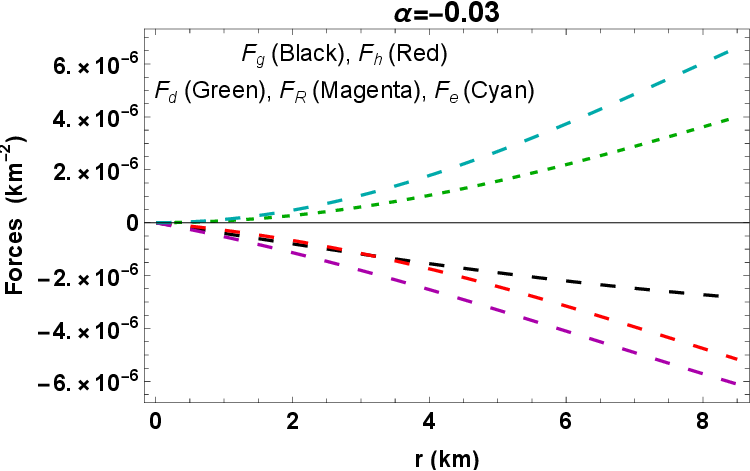}
        \includegraphics[scale=.465]{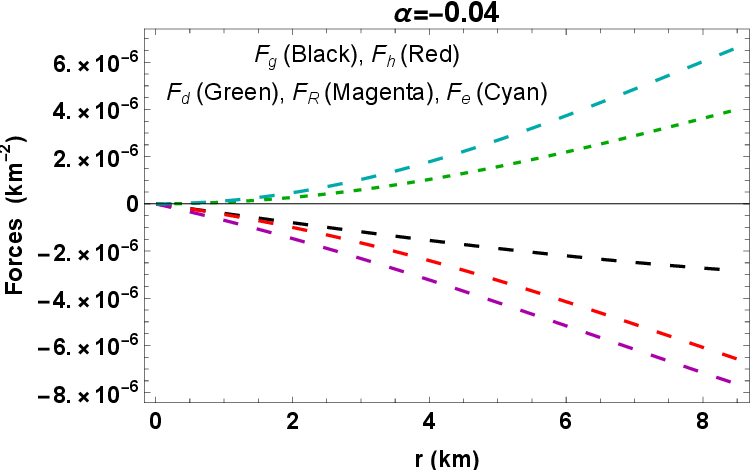}
        \includegraphics[scale=.465]{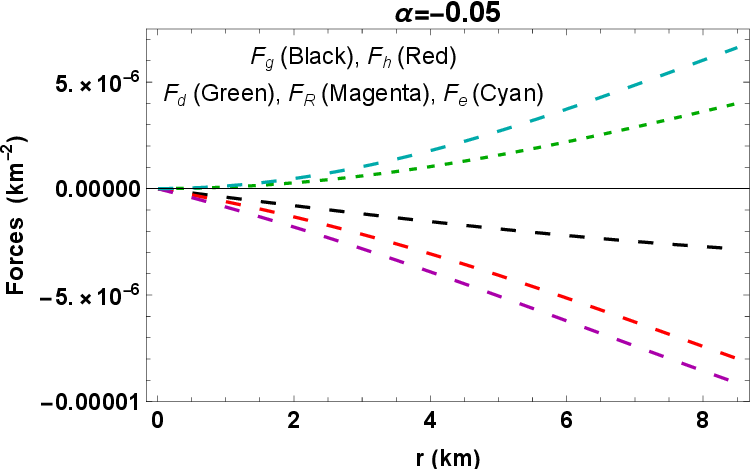}
        \caption{Variation of different forces w.r.t. radial coordinate $r$ for different values of $\alpha$.}
    \label{tov}
\end{figure}
To demonstrate the equilibrium situation of the stellar system, the balancing effect of these above forces is crucial. This above mechanism prevents the stellar system from collapsing to a point singularity. For the dark energy star model, we examine the combined effect of all five forces in Fig.~\ref{tov} \textcolor{blue}{versus the Rastall parameter $\alpha$.} From the figure, one can check that all five forces counterbalance the whole system. This implies that this model maintains a static equilibrium at all interior points of the stellar configuration. 

\subsection{Equation of State (EoS) and Zel'dovich criterion}

\begin{figure}[H]
    \centering
        \includegraphics[scale=.5]{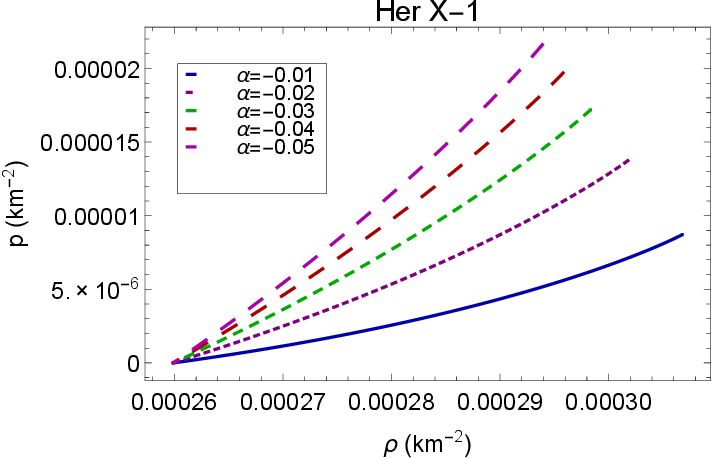}
        \includegraphics[scale=.52]{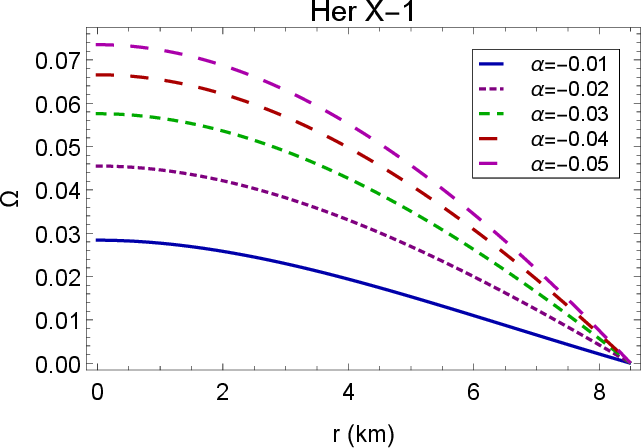}
        \caption{(i)Variation of pressure $p$ with respect to density $\rho$, (ii) the ratio $\Omega$ are shown inside the stellar interior with respect to $\alpha$.}\label{eos}
\end{figure}

The Zel'dovich criterion for stellar stability \cite{shapiro2008black, zeldovich1971relativistic, zel1972relativistic, l1962equation} states that the pressure-density ratio must be less than unity throughout the stellar interior. Assuming the ratios as: $\Omega(r) = \frac{p(r)}{\rho(r)}$, then it should be $\Omega(r)<1$. For simplicity, the fluid distribution has been assumed to be spatially homogeneous. However, several studies \cite{Zhuravlev:2000wg, Peebles:2002gy, Usmani:2008ce} suggest that $\Omega$ may vary over time and space. It is evident from Fig.~(\ref{eos}) that, for every $r < \mathcal{R}$ and for all Rastall coupling parameters, $\alpha \in [-0.05, -0.01 ]$, this model satisfies the Zel'dovich ratio $\Omega(r)<1$. Specifically, $\Omega$ lies within $(0, \frac{1}{3})$ indicating that the matter distribution in this model is non-exotic \cite{Shee:2015kqa} and montonically decreasing towards the stellar surface. It is worth noting that $\Omega$ has a significant impact on categorizing stellar structures \cite{Biswas:2019doe, Staykov:2014mwa, Ngubelanga:2015jne}.
In addition, a parametric plot of  $p$ vs $\rho$ has been given with Fig.~(\ref{eos}). This $\Omega$ is crucial to determine the star formations, since it clarifies the concept of the EoS parameters. This condition confirms physical viability and potential equilibrium state of the stellar structure.

\section{Discussion and Concluding Remarks}\label{con}

A novel relativistic dark energy model may be a feasible testing ground for simulating equilibrium versus gravitational collapse to a black hole. In the context of Rastall gravity, we presented the new configuration of an astrophysical stellar compact object as a charged dark energy star with inhomogeneous anisotropic dark energy. The well-known Krori-Barua {\em ansatz} has been employed to solve the Einstein-Maxwell field equations for a static symmetric perfect fluid sphere. Additionally, here we select a certain type of dark energy EoS to obtain different physical parameters. The numerical values of various physical quantities have been computed and presented in tabular form, as well as their graphical variations with respect to the radial coordinate $r$ by applying the observational data of compact star Her X-1, which are theoretically and potentially stable to model a new dark energy stellar model in the context of Rastall theory of gravity. Several studies have previously been conducted on dark energy stars consisting of ordinary fluid and dark energy. Shvartsman \cite{1971JETP...33..475S} suggests that stars can carry electric charges also. \par
Hence for this, a model of singularity-free charged dark energy star has been presented here. The charged dark energy star model has increased in astrophysical relevance for a variety of reasons, one of which is its potential replacement for a black hole. Furthermore, it has been discovered that dark energy, combined with anisotropic stress, can help to stabilize stars from gravitational collapse by exerting a repulsive force on the dark energy surrounding them, similar to an electric charge.  
A straightforward numerical technique has been developed to integrate the stellar model equations from the stellar core to its surface. Here, all the essential physical properties for charged dark energy star has been plotted for Rastall coupling parameter values $\alpha = -0.01, -0.02, -0.03, -0.04, -0.05$.  We must compare these findings to the observational data that is now available. The model is horizon free and satisfies all the physical requirements, suggesting that this kind of star could exist in the real world. The following points highlight some intriguing aspects of the study on the proposed dark-energy stellar model that is being presented: \\

\begin{itemize}
\item We found that the chosen metric coefficients are physically well- behaved and remain regular at all interior points of the stellar structure, as shown by their graphical representations in Fig.~(\ref{metric}).
\item Fig. (\ref{rho}) shows that the OBM energy density $\rho(r)$ and pressure $p(r)$ are continuous and positively finite inside the star, with a steady decrease towards the surface for different values of $\alpha$. At $r = \mathcal{R}$, OBM pressure and density diminish. We observe that at the center ($r = 0$), $\rho$ and $p$ get their maximums, which suggests that extremely compact cores are present. In contrast, the intensity of the electric field $E^2$ increases with $r$. It continues to increase monotonically and positively across the fluid sphere.
\item In Fig.~(\ref{grad}), the gradient components of pressure and matter-energy density exhibit a negative trend. Within the stellar configuration, the pressure and density gradients remain negative and diminish at the center ($r=0$). These characteristics acknowledge that this model may produce a new dark energy stellar star model and are in good agreement with previous results.
\item According to Fig.~(\ref{dark}), this model illustrates the repulsive nature of dark energy, which results in radial dark pressure $(p^{DE}_r)$ being negative and energy density $(\rho^{DE})$ being positive. Also, transverse dark pressure $(p^{DE}_t)$ remains negative throughout the stellar region.
\item Fig.~(\ref{mass}) demonstrates that the mass-radius relation and the compactness factor are evolutionary functions of the star radius. The plots of these quantities show positive evolution throughout the composition and maximum at $r \sim \mathcal{R}$. These functions of radius $r$ increase monotonically and can be combined to form a new ultra-relativistic dark energy star model.
\item Furthermore, from Fig.~(\ref{red}), we note the intriguing observation that the surface redshift $(z_s)$ behaves exactly opposite to $z_g$, whereas the gravitational (or internal) redshift $(z_g)$ is maximum at the center and minimum at the surface.
\item The numerically computed values of the dark energy coupling ($\beta$), central density $(\rho_c)$, surface density $(\rho_s)$, central pressure $(p_c)$, compactness ratio $(u(\mathcal{R}))$, surface redshift $(z_s(\mathcal{R}))$, mass ($M_{\odot}$), and $(p_c/\rho_c)$ have been presented for various values of Rastall parameter $\alpha$ in Table~\ref{tab1}.
\item In addition to meet the Causality condition, which states that the square of the sound velocity $V^2$ lies within $(0,1)$ (from Fig.~(\ref{sv})) within the stellar body. Consequently, the proposed model of the charged dark energy star is confirmed to be physically stable and in equilibrium by the causality condition.
\item The effect of the relativistic adiabatic index ($\Gamma$) on the model's stability has been examined graphically in Fig.~(\ref{gam}). This requirement appears to be in good accord and provides compelling evidence for the model's stable equilibrium state. The fact that the adiabatic index is greater than the collapsing condition ($\Gamma < 4/3$) suggests that the stability condition is altered as $\Gamma > 4/3$ when anisotropic pressure increases.
\item For a range of $\alpha$ values, the resulting model satisfies all five energy conditions, namely NEC, DEC, SEC, WEC, and TEC. Fig.~(\ref{ec}) provides a graphic representation of this. Furthermore, these consistently stay positive throughout the region, demonstrating the viability of this model.
\item Fig.~(\ref{tov}) provides a full illustration for the profiles of $F_g$, $F_h$, $F_d$, $F_R$ and $F_e$ for this suggested stellar structure. The setting up of a static equilibrium configuration is possible by combining the effects of all these forces. Notably for $\alpha = 0$, this equilibrium situation can reproduce to a standard GR solution.
\item The EoS parameters are analyzed in Fig.~(\ref{eos}) using the Zel'dovich condition for stability, which states that $\Omega(r)=\frac{p(r)}{\rho(r)}$ should be less than unity. This is confirmed even though the star contains dark energy, OBM, and electric charge. The plot of $p$ w.r.t. $\rho$ has also been given. Thus, it is evident that the matter content is non-exotic in nature and that this proposed model's solution behaves physically well. 
\end{itemize}

In light of all the important findings, we can thus at last say that, by using the dark energy equation of state, we can construct a physically plausible, stable, and singularity-free generalized model for charged strange stars that is appropriate for research on dark energy stars using the Rastall framework. Additionally, the obtained results can be compared to standard GR and other modified gravity theories. It can be anticipated that this model might potentially make a larger-scale contribution to the astrophysical scenario.

\section*{Author contributions}
\textbf{Pramit Rej} contributed to conceptualization, editing, mathematical analysis, designing computer codes for data analysis, revision work, validation, and the draft of this manuscript. 

\section*{Acknowledgements} 
\textbf{Pramit Rej} is thankful to the Inter-University Centre for Astronomy and Astrophysics (IUCAA), Pune, Government of India, for providing a Visiting Associateship.

\section*{Declarations}
\textbf{Funding:} The author did not receive any funding in the form of financial aid or a grant from any institution or organization for the present research work.\par
\textbf{Data Availability Statement:} The results are obtained
via purely theoretical calculations and can be verified analytically;
thus this manuscript has no associated data, or the data will not be deposited. \par
\textbf{Conflicts of Interest:} The author has no known competing financial interests or personal relationships that could have appeared to influence the work reported in this paper.

\bibliographystyle{apsrev4-1}
\bibliography{de_ras}

\begin{thebibliography}{134}%
\makeatletter
\providecommand \@ifxundefined [1]{%
 \@ifx{#1\undefined}
}%
\providecommand \@ifnum [1]{%
 \ifnum #1\expandafter \@firstoftwo
 \else \expandafter \@secondoftwo
 \fi
}%
\providecommand \@ifx [1]{%
 \ifx #1\expandafter \@firstoftwo
 \else \expandafter \@secondoftwo
 \fi
}%
\providecommand \natexlab [1]{#1}%
\providecommand \enquote  [1]{``#1''}%
\providecommand \bibnamefont  [1]{#1}%
\providecommand \bibfnamefont [1]{#1}%
\providecommand \citenamefont [1]{#1}%
\providecommand \href@noop [0]{\@secondoftwo}%
\providecommand \href [0]{\begingroup \@sanitize@url \@href}%
\providecommand \@href[1]{\@@startlink{#1}\@@href}%
\providecommand \@@href[1]{\endgroup#1\@@endlink}%
\providecommand \@sanitize@url [0]{\catcode `\\12\catcode `\$12\catcode `\&12\catcode `\#12\catcode `\^12\catcode `\_12\catcode `\%12\relax}%
\providecommand \@@startlink[1]{}%
\providecommand \@@endlink[0]{}%
\providecommand \url  [0]{\begingroup\@sanitize@url \@url }%
\providecommand \@url [1]{\endgroup\@href {#1}{\urlprefix }}%
\providecommand \urlprefix  [0]{URL }%
\providecommand \Eprint [0]{\href }%
\providecommand \doibase [0]{http://dx.doi.org/}%
\providecommand \selectlanguage [0]{\@gobble}%
\providecommand \bibinfo  [0]{\@secondoftwo}%
\providecommand \bibfield  [0]{\@secondoftwo}%
\providecommand \translation [1]{[#1]}%
\providecommand \BibitemOpen [0]{}%
\providecommand \bibitemStop [0]{}%
\providecommand \bibitemNoStop [0]{.\EOS\space}%
\providecommand \EOS [0]{\spacefactor3000\relax}%
\providecommand \BibitemShut  [1]{\csname bibitem#1\endcsname}%
\let\auto@bib@innerbib\@empty
%</preamble>
\bibitem [{\citenamefont {Maurya}\ and\ \citenamefont {Tello-Ortiz}(2019)}]{Maurya:2019zyc}%
  \BibitemOpen
  \bibfield  {author} {\bibinfo {author} {\bibfnamefont {S.~K.}\ \bibnamefont {Maurya}}\ and\ \bibinfo {author} {\bibfnamefont {F.}~\bibnamefont {Tello-Ortiz}},\ }\href {\doibase 10.1140/epjc/s10052-019-6575-0} {\bibfield  {journal} {\bibinfo  {journal} {Eur. Phys. J. C}\ }\textbf {\bibinfo {volume} {79}},\ \bibinfo {pages} {33} (\bibinfo {year} {2019})}\BibitemShut {NoStop}%
\bibitem [{\citenamefont {Tamta}\ and\ \citenamefont {Fuloria}(2017)}]{tamta2017new}%
  \BibitemOpen
  \bibfield  {author} {\bibinfo {author} {\bibfnamefont {R.}~\bibnamefont {Tamta}}\ and\ \bibinfo {author} {\bibfnamefont {P.}~\bibnamefont {Fuloria}},\ }\href {\doibase 10.4236/jmp.2017.811104} {\bibfield  {journal} {\bibinfo  {journal} {Journal of Modern Physics}\ }\textbf {\bibinfo {volume} {8}},\ \bibinfo {pages} {1762} (\bibinfo {year} {2017})}\BibitemShut {NoStop}%
\bibitem [{\citenamefont {Bondi}(1992)}]{bondi1992anisotropic}%
  \BibitemOpen
  \bibfield  {author} {\bibinfo {author} {\bibfnamefont {H.}~\bibnamefont {Bondi}},\ }\href {\doibase 10.1093/mnras/259.2.365} {\bibfield  {journal} {\bibinfo  {journal} {Monthly Notices of the Royal Astronomical Society}\ }\textbf {\bibinfo {volume} {259}},\ \bibinfo {pages} {365} (\bibinfo {year} {1992})}\BibitemShut {NoStop}%
\bibitem [{\citenamefont {Deb}\ \emph {et~al.}(2017)\citenamefont {Deb}, \citenamefont {Chowdhury}, \citenamefont {Ray}, \citenamefont {Rahaman},\ and\ \citenamefont {Guha}}]{DEB2017239}%
  \BibitemOpen
  \bibfield  {author} {\bibinfo {author} {\bibfnamefont {D.}~\bibnamefont {Deb}}, \bibinfo {author} {\bibfnamefont {S.~R.}\ \bibnamefont {Chowdhury}}, \bibinfo {author} {\bibfnamefont {S.}~\bibnamefont {Ray}}, \bibinfo {author} {\bibfnamefont {F.}~\bibnamefont {Rahaman}}, \ and\ \bibinfo {author} {\bibfnamefont {B.}~\bibnamefont {Guha}},\ }\href {\doibase https://doi.org/10.1016/j.aop.2017.10.010} {\bibfield  {journal} {\bibinfo  {journal} {Annals of Physics}\ }\textbf {\bibinfo {volume} {387}},\ \bibinfo {pages} {239} (\bibinfo {year} {2017})}\BibitemShut {NoStop}%
\bibitem [{\citenamefont {Dev}\ and\ \citenamefont {Gleiser}(2002)}]{dev2002anisotropic}%
  \BibitemOpen
  \bibfield  {author} {\bibinfo {author} {\bibfnamefont {K.}~\bibnamefont {Dev}}\ and\ \bibinfo {author} {\bibfnamefont {M.}~\bibnamefont {Gleiser}},\ }\href {\doibase 10.1023/A:1020707906543} {\bibfield  {journal} {\bibinfo  {journal} {Gen. Rel. Grav.}\ }\textbf {\bibinfo {volume} {34}},\ \bibinfo {pages} {1793} (\bibinfo {year} {2002})},\ \Eprint {http://arxiv.org/abs/astro-ph/0012265} {arXiv:astro-ph/0012265} \BibitemShut {NoStop}%
\bibitem [{\citenamefont {Di~Prisco}\ \emph {et~al.}(1997)\citenamefont {Di~Prisco}, \citenamefont {Herrera},\ and\ \citenamefont {Varela}}]{di1997cracking}%
  \BibitemOpen
  \bibfield  {author} {\bibinfo {author} {\bibfnamefont {A.}~\bibnamefont {Di~Prisco}}, \bibinfo {author} {\bibfnamefont {L.}~\bibnamefont {Herrera}}, \ and\ \bibinfo {author} {\bibfnamefont {V.}~\bibnamefont {Varela}},\ }\href {\doibase 10.1023/A:1018859712881} {\bibfield  {journal} {\bibinfo  {journal} {Gen. Rel. Grav.}\ }\textbf {\bibinfo {volume} {29}},\ \bibinfo {pages} {1239} (\bibinfo {year} {1997})}\BibitemShut {NoStop}%
\bibitem [{\citenamefont {Krori}\ \emph {et~al.}(1984)\citenamefont {Krori}, \citenamefont {Borgohain},\ and\ \citenamefont {Devi}}]{krori1984some}%
  \BibitemOpen
  \bibfield  {author} {\bibinfo {author} {\bibfnamefont {K.}~\bibnamefont {Krori}}, \bibinfo {author} {\bibfnamefont {P.}~\bibnamefont {Borgohain}}, \ and\ \bibinfo {author} {\bibfnamefont {R.}~\bibnamefont {Devi}},\ }\href {\doibase 10.1139/p84-038} {\bibfield  {journal} {\bibinfo  {journal} {Canadian journal of physics}\ }\textbf {\bibinfo {volume} {62}},\ \bibinfo {pages} {239} (\bibinfo {year} {1984})}\BibitemShut {NoStop}%
\bibitem [{\citenamefont {Mak}\ and\ \citenamefont {Harko}(2003)}]{mak2003anisotropic}%
  \BibitemOpen
  \bibfield  {author} {\bibinfo {author} {\bibfnamefont {M.~K.}\ \bibnamefont {Mak}}\ and\ \bibinfo {author} {\bibfnamefont {T.}~\bibnamefont {Harko}},\ }\href {\doibase 10.1098/rspa.2002.1014} {\bibfield  {journal} {\bibinfo  {journal} {Proc. Roy. Soc. Lond. A}\ }\textbf {\bibinfo {volume} {459}},\ \bibinfo {pages} {393} (\bibinfo {year} {2003})},\ \Eprint {http://arxiv.org/abs/gr-qc/0110103} {arXiv:gr-qc/0110103} \BibitemShut {NoStop}%
\bibitem [{\citenamefont {Maurya}\ \emph {et~al.}(2016)\citenamefont {Maurya}, \citenamefont {T.~T.}, \citenamefont {Gupta},\ and\ \citenamefont {Rahaman}}]{maurya2016new}%
  \BibitemOpen
  \bibfield  {author} {\bibinfo {author} {\bibfnamefont {S.~K.}\ \bibnamefont {Maurya}}, \bibinfo {author} {\bibfnamefont {S.}~\bibnamefont {T.~T.}}, \bibinfo {author} {\bibfnamefont {Y.~K.}\ \bibnamefont {Gupta}}, \ and\ \bibinfo {author} {\bibfnamefont {F.}~\bibnamefont {Rahaman}},\ }\href {\doibase 10.1140/epja/i2016-16191-1} {\bibfield  {journal} {\bibinfo  {journal} {Eur. Phys. J. A}\ }\textbf {\bibinfo {volume} {52}},\ \bibinfo {pages} {191} (\bibinfo {year} {2016})},\ \Eprint {http://arxiv.org/abs/1512.01667} {arXiv:1512.01667 [gr-qc]} \BibitemShut {NoStop}%
\bibitem [{\citenamefont {Lobo}(2006)}]{Lobo:2005uf}%
  \BibitemOpen
  \bibfield  {author} {\bibinfo {author} {\bibfnamefont {F.~S.~N.}\ \bibnamefont {Lobo}},\ }\href {\doibase 10.1088/0264-9381/23/5/006} {\bibfield  {journal} {\bibinfo  {journal} {Class. Quant. Grav.}\ }\textbf {\bibinfo {volume} {23}},\ \bibinfo {pages} {1525} (\bibinfo {year} {2006})},\ \Eprint {http://arxiv.org/abs/gr-qc/0508115} {arXiv:gr-qc/0508115} \BibitemShut {NoStop}%
\bibitem [{\citenamefont {Perlmutter}\ \emph {et~al.}(1999)\citenamefont {Perlmutter} \emph {et~al.}}]{SupernovaCosmologyProject:1998vns}%
  \BibitemOpen
  \bibfield  {author} {\bibinfo {author} {\bibfnamefont {S.}~\bibnamefont {Perlmutter}} \emph {et~al.} (\bibinfo {collaboration} {Supernova Cosmology Project}),\ }\href {\doibase 10.1086/307221} {\bibfield  {journal} {\bibinfo  {journal} {Astrophys. J.}\ }\textbf {\bibinfo {volume} {517}},\ \bibinfo {pages} {565} (\bibinfo {year} {1999})},\ \Eprint {http://arxiv.org/abs/astro-ph/9812133} {arXiv:astro-ph/9812133} \BibitemShut {NoStop}%
\bibitem [{\citenamefont {Beltracchi}\ and\ \citenamefont {Gondolo}(2019)}]{beltracchi2019formation}%
  \BibitemOpen
  \bibfield  {author} {\bibinfo {author} {\bibfnamefont {P.}~\bibnamefont {Beltracchi}}\ and\ \bibinfo {author} {\bibfnamefont {P.}~\bibnamefont {Gondolo}},\ }\href {\doibase 10.1103/PhysRevD.99.044037} {\bibfield  {journal} {\bibinfo  {journal} {Phys. Rev. D}\ }\textbf {\bibinfo {volume} {99}},\ \bibinfo {pages} {044037} (\bibinfo {year} {2019})},\ \Eprint {http://arxiv.org/abs/1810.12400} {arXiv:1810.12400 [gr-qc]} \BibitemShut {NoStop}%
\bibitem [{\citenamefont {Chapline}(2004)}]{Chapline:2004jfp}%
  \BibitemOpen
  \bibfield  {author} {\bibinfo {author} {\bibfnamefont {G.}~\bibnamefont {Chapline}},\ }\href@noop {} {\bibfield  {journal} {\bibinfo  {journal} {eConf}\ }\textbf {\bibinfo {volume} {C041213}},\ \bibinfo {pages} {0205} (\bibinfo {year} {2004})},\ \Eprint {http://arxiv.org/abs/astro-ph/0503200} {arXiv:astro-ph/0503200} \BibitemShut {NoStop}%
\bibitem [{\citenamefont {Riess}\ \emph {et~al.}(2019)\citenamefont {Riess}, \citenamefont {Casertano}, \citenamefont {Yuan}, \citenamefont {Macri},\ and\ \citenamefont {Scolnic}}]{Riess:2019cxk}%
  \BibitemOpen
  \bibfield  {author} {\bibinfo {author} {\bibfnamefont {A.~G.}\ \bibnamefont {Riess}}, \bibinfo {author} {\bibfnamefont {S.}~\bibnamefont {Casertano}}, \bibinfo {author} {\bibfnamefont {W.}~\bibnamefont {Yuan}}, \bibinfo {author} {\bibfnamefont {L.~M.}\ \bibnamefont {Macri}}, \ and\ \bibinfo {author} {\bibfnamefont {D.}~\bibnamefont {Scolnic}},\ }\href {\doibase 10.3847/1538-4357/ab1422} {\bibfield  {journal} {\bibinfo  {journal} {Astrophys. J.}\ }\textbf {\bibinfo {volume} {876}},\ \bibinfo {pages} {85} (\bibinfo {year} {2019})},\ \Eprint {http://arxiv.org/abs/1903.07603} {arXiv:1903.07603 [astro-ph.CO]} \BibitemShut {NoStop}%
\bibitem [{\citenamefont {Sushkov}(2005)}]{sushkov2005wormholes}%
  \BibitemOpen
  \bibfield  {author} {\bibinfo {author} {\bibfnamefont {S.~V.}\ \bibnamefont {Sushkov}},\ }\href {\doibase 10.1103/PhysRevD.71.043520} {\bibfield  {journal} {\bibinfo  {journal} {Phys. Rev. D}\ }\textbf {\bibinfo {volume} {71}},\ \bibinfo {pages} {043520} (\bibinfo {year} {2005})},\ \Eprint {http://arxiv.org/abs/gr-qc/0502084} {arXiv:gr-qc/0502084} \BibitemShut {NoStop}%
\bibitem [{\citenamefont {Bibi}\ \emph {et~al.}(2016)\citenamefont {Bibi}, \citenamefont {Feroze},\ and\ \citenamefont {Siddiqui}}]{bibi2016solution}%
  \BibitemOpen
  \bibfield  {author} {\bibinfo {author} {\bibfnamefont {R.}~\bibnamefont {Bibi}}, \bibinfo {author} {\bibfnamefont {T.}~\bibnamefont {Feroze}}, \ and\ \bibinfo {author} {\bibfnamefont {A.~A.}\ \bibnamefont {Siddiqui}},\ }\href {\doibase 10.1139/cjp-2016-0069} {\bibfield  {journal} {\bibinfo  {journal} {Canadian Journal of Physics}\ }\textbf {\bibinfo {volume} {94}},\ \bibinfo {pages} {758} (\bibinfo {year} {2016})}\BibitemShut {NoStop}%
\bibitem [{\citenamefont {Feng}\ \emph {et~al.}(2008)\citenamefont {Feng}, \citenamefont {Wang}, \citenamefont {Abdalla},\ and\ \citenamefont {Su}}]{feng2008observational}%
  \BibitemOpen
  \bibfield  {author} {\bibinfo {author} {\bibfnamefont {C.}~\bibnamefont {Feng}}, \bibinfo {author} {\bibfnamefont {B.}~\bibnamefont {Wang}}, \bibinfo {author} {\bibfnamefont {E.}~\bibnamefont {Abdalla}}, \ and\ \bibinfo {author} {\bibfnamefont {R.-K.}\ \bibnamefont {Su}},\ }\href {\doibase 10.1016/j.physletb.2008.05.066} {\bibfield  {journal} {\bibinfo  {journal} {Phys. Lett. B}\ }\textbf {\bibinfo {volume} {665}},\ \bibinfo {pages} {111} (\bibinfo {year} {2008})},\ \Eprint {http://arxiv.org/abs/0804.0110} {arXiv:0804.0110 [astro-ph]} \BibitemShut {NoStop}%
\bibitem [{\citenamefont {Capozziello}(2002)}]{Capozziello:2002rd}%
  \BibitemOpen
  \bibfield  {author} {\bibinfo {author} {\bibfnamefont {S.}~\bibnamefont {Capozziello}},\ }\href {\doibase 10.1142/S0218271802002025} {\bibfield  {journal} {\bibinfo  {journal} {Int. J. Mod. Phys. D}\ }\textbf {\bibinfo {volume} {11}},\ \bibinfo {pages} {483} (\bibinfo {year} {2002})},\ \Eprint {http://arxiv.org/abs/gr-qc/0201033} {arXiv:gr-qc/0201033} \BibitemShut {NoStop}%
\bibitem [{\citenamefont {Nojiri}\ and\ \citenamefont {Odintsov}(2006)}]{Nojiri:2006ri}%
  \BibitemOpen
  \bibfield  {author} {\bibinfo {author} {\bibfnamefont {S.}~\bibnamefont {Nojiri}}\ and\ \bibinfo {author} {\bibfnamefont {S.~D.}\ \bibnamefont {Odintsov}},\ }\href {\doibase 10.1142/S0219887807001928} {\bibfield  {journal} {\bibinfo  {journal} {eConf}\ }\textbf {\bibinfo {volume} {C0602061}},\ \bibinfo {pages} {06} (\bibinfo {year} {2006})},\ \Eprint {http://arxiv.org/abs/hep-th/0601213} {arXiv:hep-th/0601213} \BibitemShut {NoStop}%
\bibitem [{\citenamefont {Nojiri}\ and\ \citenamefont {Odintsov}(2011)}]{Nojiri:2010wj}%
  \BibitemOpen
  \bibfield  {author} {\bibinfo {author} {\bibfnamefont {S.}~\bibnamefont {Nojiri}}\ and\ \bibinfo {author} {\bibfnamefont {S.~D.}\ \bibnamefont {Odintsov}},\ }\href {\doibase 10.1016/j.physrep.2011.04.001} {\bibfield  {journal} {\bibinfo  {journal} {Phys. Rept.}\ }\textbf {\bibinfo {volume} {505}},\ \bibinfo {pages} {59} (\bibinfo {year} {2011})},\ \Eprint {http://arxiv.org/abs/1011.0544} {arXiv:1011.0544 [gr-qc]} \BibitemShut {NoStop}%
\bibitem [{\citenamefont {Bamba}\ \emph {et~al.}(2012)\citenamefont {Bamba}, \citenamefont {Capozziello}, \citenamefont {Nojiri},\ and\ \citenamefont {Odintsov}}]{Bamba:2012cp}%
  \BibitemOpen
  \bibfield  {author} {\bibinfo {author} {\bibfnamefont {K.}~\bibnamefont {Bamba}}, \bibinfo {author} {\bibfnamefont {S.}~\bibnamefont {Capozziello}}, \bibinfo {author} {\bibfnamefont {S.}~\bibnamefont {Nojiri}}, \ and\ \bibinfo {author} {\bibfnamefont {S.~D.}\ \bibnamefont {Odintsov}},\ }\href {\doibase 10.1007/s10509-012-1181-8} {\bibfield  {journal} {\bibinfo  {journal} {Astrophys. Space Sci.}\ }\textbf {\bibinfo {volume} {342}},\ \bibinfo {pages} {155} (\bibinfo {year} {2012})},\ \Eprint {http://arxiv.org/abs/1205.3421} {arXiv:1205.3421 [gr-qc]} \BibitemShut {NoStop}%
\bibitem [{\citenamefont {Rastall}(1972)}]{Rastall:1972swe}%
  \BibitemOpen
  \bibfield  {author} {\bibinfo {author} {\bibfnamefont {P.}~\bibnamefont {Rastall}},\ }\href {\doibase 10.1103/PhysRevD.6.3357} {\bibfield  {journal} {\bibinfo  {journal} {Phys. Rev. D}\ }\textbf {\bibinfo {volume} {6}},\ \bibinfo {pages} {3357} (\bibinfo {year} {1972})}\BibitemShut {NoStop}%
\bibitem [{\citenamefont {Rastall}(1976)}]{Rastall:1976uh}%
  \BibitemOpen
  \bibfield  {author} {\bibinfo {author} {\bibfnamefont {P.}~\bibnamefont {Rastall}},\ }\href {\doibase 10.1139/p76-008} {\bibfield  {journal} {\bibinfo  {journal} {Can. J. Phys.}\ }\textbf {\bibinfo {volume} {54}},\ \bibinfo {pages} {66} (\bibinfo {year} {1976})}\BibitemShut {NoStop}%
\bibitem [{\citenamefont {Harko}\ \emph {et~al.}(2011)\citenamefont {Harko}, \citenamefont {Lobo}, \citenamefont {Nojiri},\ and\ \citenamefont {Odintsov}}]{Harko:2011kv}%
  \BibitemOpen
  \bibfield  {author} {\bibinfo {author} {\bibfnamefont {T.}~\bibnamefont {Harko}}, \bibinfo {author} {\bibfnamefont {F.~S.~N.}\ \bibnamefont {Lobo}}, \bibinfo {author} {\bibfnamefont {S.}~\bibnamefont {Nojiri}}, \ and\ \bibinfo {author} {\bibfnamefont {S.~D.}\ \bibnamefont {Odintsov}},\ }\href {\doibase 10.1103/PhysRevD.84.024020} {\bibfield  {journal} {\bibinfo  {journal} {Phys. Rev. D}\ }\textbf {\bibinfo {volume} {84}},\ \bibinfo {pages} {024020} (\bibinfo {year} {2011})},\ \Eprint {http://arxiv.org/abs/1104.2669} {arXiv:1104.2669 [gr-qc]} \BibitemShut {NoStop}%
\bibitem [{\citenamefont {Odintsov}\ and\ \citenamefont {S\'aez-G\'omez}(2013)}]{Odintsov:2013iba}%
  \BibitemOpen
  \bibfield  {author} {\bibinfo {author} {\bibfnamefont {S.~D.}\ \bibnamefont {Odintsov}}\ and\ \bibinfo {author} {\bibfnamefont {D.}~\bibnamefont {S\'aez-G\'omez}},\ }\href {\doibase 10.1016/j.physletb.2013.07.026} {\bibfield  {journal} {\bibinfo  {journal} {Phys. Lett. B}\ }\textbf {\bibinfo {volume} {725}},\ \bibinfo {pages} {437} (\bibinfo {year} {2013})},\ \Eprint {http://arxiv.org/abs/1304.5411} {arXiv:1304.5411 [gr-qc]} \BibitemShut {NoStop}%
\bibitem [{\citenamefont {Haghani}\ \emph {et~al.}(2013)\citenamefont {Haghani}, \citenamefont {Harko}, \citenamefont {Lobo}, \citenamefont {Sepangi},\ and\ \citenamefont {Shahidi}}]{Haghani:2013oma}%
  \BibitemOpen
  \bibfield  {author} {\bibinfo {author} {\bibfnamefont {Z.}~\bibnamefont {Haghani}}, \bibinfo {author} {\bibfnamefont {T.}~\bibnamefont {Harko}}, \bibinfo {author} {\bibfnamefont {F.~S.~N.}\ \bibnamefont {Lobo}}, \bibinfo {author} {\bibfnamefont {H.~R.}\ \bibnamefont {Sepangi}}, \ and\ \bibinfo {author} {\bibfnamefont {S.}~\bibnamefont {Shahidi}},\ }\href {\doibase 10.1103/PhysRevD.88.044023} {\bibfield  {journal} {\bibinfo  {journal} {Phys. Rev. D}\ }\textbf {\bibinfo {volume} {88}},\ \bibinfo {pages} {044023} (\bibinfo {year} {2013})},\ \Eprint {http://arxiv.org/abs/1304.5957} {arXiv:1304.5957 [gr-qc]} \BibitemShut {NoStop}%
\bibitem [{\citenamefont {Ayuso}\ \emph {et~al.}(2015)\citenamefont {Ayuso}, \citenamefont {Beltran~Jimenez},\ and\ \citenamefont {de~la Cruz-Dombriz}}]{Ayuso:2014jda}%
  \BibitemOpen
  \bibfield  {author} {\bibinfo {author} {\bibfnamefont {I.}~\bibnamefont {Ayuso}}, \bibinfo {author} {\bibfnamefont {J.}~\bibnamefont {Beltran~Jimenez}}, \ and\ \bibinfo {author} {\bibfnamefont {A.}~\bibnamefont {de~la Cruz-Dombriz}},\ }\href {\doibase 10.1103/PhysRevD.91.104003} {\bibfield  {journal} {\bibinfo  {journal} {Phys. Rev. D}\ }\textbf {\bibinfo {volume} {91}},\ \bibinfo {pages} {104003} (\bibinfo {year} {2015})},\ \bibinfo {note} {[Addendum: Phys.Rev.D 93, 089901 (2016)]},\ \Eprint {http://arxiv.org/abs/1411.1636} {arXiv:1411.1636 [hep-th]} \BibitemShut {NoStop}%
\bibitem [{\citenamefont {Shirafuji}\ \emph {et~al.}(1996)\citenamefont {Shirafuji}, \citenamefont {Nashed},\ and\ \citenamefont {Kobayashi}}]{Shirafuji:1996im}%
  \BibitemOpen
  \bibfield  {author} {\bibinfo {author} {\bibfnamefont {T.}~\bibnamefont {Shirafuji}}, \bibinfo {author} {\bibfnamefont {G.~G.~L.}\ \bibnamefont {Nashed}}, \ and\ \bibinfo {author} {\bibfnamefont {Y.}~\bibnamefont {Kobayashi}},\ }\href {\doibase 10.1143/PTP.96.933} {\bibfield  {journal} {\bibinfo  {journal} {Prog. Theor. Phys.}\ }\textbf {\bibinfo {volume} {96}},\ \bibinfo {pages} {933} (\bibinfo {year} {1996})},\ \Eprint {http://arxiv.org/abs/gr-qc/9609060} {arXiv:gr-qc/9609060} \BibitemShut {NoStop}%
\bibitem [{\citenamefont {Nashed}(2002{\natexlab{a}})}]{Nashed:2001im}%
  \BibitemOpen
  \bibfield  {author} {\bibinfo {author} {\bibfnamefont {G.~G.~L.}\ \bibnamefont {Nashed}},\ }\href@noop {} {\bibfield  {journal} {\bibinfo  {journal} {Nuovo Cim. B}\ }\textbf {\bibinfo {volume} {117}},\ \bibinfo {pages} {521} (\bibinfo {year} {2002}{\natexlab{a}})},\ \Eprint {http://arxiv.org/abs/gr-qc/0109017} {arXiv:gr-qc/0109017} \BibitemShut {NoStop}%
\bibitem [{\citenamefont {Nashed}(2002{\natexlab{b}})}]{Nashed:2001cp}%
  \BibitemOpen
  \bibfield  {author} {\bibinfo {author} {\bibfnamefont {G.~G.~L.}\ \bibnamefont {Nashed}},\ }\href {\doibase 10.1023/A:1016509920499} {\bibfield  {journal} {\bibinfo  {journal} {Gen. Rel. Grav.}\ }\textbf {\bibinfo {volume} {34}},\ \bibinfo {pages} {1047} (\bibinfo {year} {2002}{\natexlab{b}})},\ \Eprint {http://arxiv.org/abs/gr-qc/0109033} {arXiv:gr-qc/0109033} \BibitemShut {NoStop}%
\bibitem [{\citenamefont {Nashed}(2010)}]{Nashed:2010ocg}%
  \BibitemOpen
  \bibfield  {author} {\bibinfo {author} {\bibfnamefont {G.~G.~L.}\ \bibnamefont {Nashed}},\ }\href {\doibase 10.1007/s10509-010-0375-1} {\bibfield  {journal} {\bibinfo  {journal} {Astrophys. Space Sci.}\ }\textbf {\bibinfo {volume} {330}},\ \bibinfo {pages} {173} (\bibinfo {year} {2010})},\ \Eprint {http://arxiv.org/abs/1503.01379} {arXiv:1503.01379 [gr-qc]} \BibitemShut {NoStop}%
\bibitem [{\citenamefont {El~Hanafy}\ and\ \citenamefont {Nashed}(2016)}]{ElHanafy:2014efn}%
  \BibitemOpen
  \bibfield  {author} {\bibinfo {author} {\bibfnamefont {W.}~\bibnamefont {El~Hanafy}}\ and\ \bibinfo {author} {\bibfnamefont {G.~L.}\ \bibnamefont {Nashed}},\ }\href {\doibase 10.1007/s10509-016-2786-0} {\bibfield  {journal} {\bibinfo  {journal} {Astrophys. Space Sci.}\ }\textbf {\bibinfo {volume} {361}},\ \bibinfo {pages} {197} (\bibinfo {year} {2016})},\ \Eprint {http://arxiv.org/abs/1410.2467} {arXiv:1410.2467 [hep-th]} \BibitemShut {NoStop}%
\bibitem [{\citenamefont {Nashed}(2015)}]{Nashed:2015pga}%
  \BibitemOpen
  \bibfield  {author} {\bibinfo {author} {\bibfnamefont {G.~G.~L.}\ \bibnamefont {Nashed}},\ }\href {\doibase 10.1140/epjp/i2015-15124-3} {\bibfield  {journal} {\bibinfo  {journal} {Eur. Phys. J. Plus}\ }\textbf {\bibinfo {volume} {130}},\ \bibinfo {pages} {124} (\bibinfo {year} {2015})}\BibitemShut {NoStop}%
\bibitem [{\citenamefont {Nashed}(2018)}]{Nashed:2018piz}%
  \BibitemOpen
  \bibfield  {author} {\bibinfo {author} {\bibfnamefont {G.~G.~L.}\ \bibnamefont {Nashed}},\ }\href {\doibase 10.1155/2018/7323574} {\bibfield  {journal} {\bibinfo  {journal} {Adv. High Energy Phys.}\ }\textbf {\bibinfo {volume} {2018}},\ \bibinfo {pages} {7323574} (\bibinfo {year} {2018})}\BibitemShut {NoStop}%
\bibitem [{\citenamefont {Nashed}\ \emph {et~al.}(2019)\citenamefont {Nashed}, \citenamefont {El~Hanafy},\ and\ \citenamefont {Bamba}}]{Nashed:2018qag}%
  \BibitemOpen
  \bibfield  {author} {\bibinfo {author} {\bibfnamefont {G.~G.~L.}\ \bibnamefont {Nashed}}, \bibinfo {author} {\bibfnamefont {W.}~\bibnamefont {El~Hanafy}}, \ and\ \bibinfo {author} {\bibfnamefont {K.}~\bibnamefont {Bamba}},\ }\href {\doibase 10.1088/1475-7516/2019/01/058} {\bibfield  {journal} {\bibinfo  {journal} {JCAP}\ }\textbf {\bibinfo {volume} {01}},\ \bibinfo {pages} {058} (\bibinfo {year} {2019})},\ \Eprint {http://arxiv.org/abs/1809.02289} {arXiv:1809.02289 [gr-qc]} \BibitemShut {NoStop}%
\bibitem [{\citenamefont {Nashed}\ \emph {et~al.}(2020)\citenamefont {Nashed}, \citenamefont {El~Hanafy}, \citenamefont {Odintsov},\ and\ \citenamefont {Oikonomou}}]{Nashed:2019yto}%
  \BibitemOpen
  \bibfield  {author} {\bibinfo {author} {\bibfnamefont {G.~G.~L.}\ \bibnamefont {Nashed}}, \bibinfo {author} {\bibfnamefont {W.}~\bibnamefont {El~Hanafy}}, \bibinfo {author} {\bibfnamefont {S.~D.}\ \bibnamefont {Odintsov}}, \ and\ \bibinfo {author} {\bibfnamefont {V.~K.}\ \bibnamefont {Oikonomou}},\ }\href {\doibase 10.1142/S021827182050090X} {\bibfield  {journal} {\bibinfo  {journal} {Int. J. Mod. Phys. D}\ }\textbf {\bibinfo {volume} {29}},\ \bibinfo {pages} {2050090} (\bibinfo {year} {2020})},\ \Eprint {http://arxiv.org/abs/1912.03897} {arXiv:1912.03897 [gr-qc]} \BibitemShut {NoStop}%
\bibitem [{\citenamefont {Nashed}\ and\ \citenamefont {Nojiri}(2020)}]{Nashed:2020mnp}%
  \BibitemOpen
  \bibfield  {author} {\bibinfo {author} {\bibfnamefont {G.~G.~L.}\ \bibnamefont {Nashed}}\ and\ \bibinfo {author} {\bibfnamefont {S.}~\bibnamefont {Nojiri}},\ }\href {\doibase 10.1103/PhysRevD.102.124022} {\bibfield  {journal} {\bibinfo  {journal} {Phys. Rev. D}\ }\textbf {\bibinfo {volume} {102}},\ \bibinfo {pages} {124022} (\bibinfo {year} {2020})},\ \Eprint {http://arxiv.org/abs/2012.05711} {arXiv:2012.05711 [gr-qc]} \BibitemShut {NoStop}%
\bibitem [{\citenamefont {Nashed}\ and\ \citenamefont {El~Hanafy}(2022)}]{Nashed:2022zyi}%
  \BibitemOpen
  \bibfield  {author} {\bibinfo {author} {\bibfnamefont {G.~G.~L.}\ \bibnamefont {Nashed}}\ and\ \bibinfo {author} {\bibfnamefont {W.}~\bibnamefont {El~Hanafy}},\ }\href {\doibase 10.1140/epjc/s10052-022-10634-0} {\bibfield  {journal} {\bibinfo  {journal} {Eur. Phys. J. C}\ }\textbf {\bibinfo {volume} {82}},\ \bibinfo {pages} {679} (\bibinfo {year} {2022})},\ \Eprint {http://arxiv.org/abs/2208.13814} {arXiv:2208.13814 [gr-qc]} \BibitemShut {NoStop}%
\bibitem [{\citenamefont {Darabi}\ \emph {et~al.}(2018)\citenamefont {Darabi}, \citenamefont {Moradpour}, \citenamefont {Licata}, \citenamefont {Heydarzade},\ and\ \citenamefont {Corda}}]{Darabi:2017coc}%
  \BibitemOpen
  \bibfield  {author} {\bibinfo {author} {\bibfnamefont {F.}~\bibnamefont {Darabi}}, \bibinfo {author} {\bibfnamefont {H.}~\bibnamefont {Moradpour}}, \bibinfo {author} {\bibfnamefont {I.}~\bibnamefont {Licata}}, \bibinfo {author} {\bibfnamefont {Y.}~\bibnamefont {Heydarzade}}, \ and\ \bibinfo {author} {\bibfnamefont {C.}~\bibnamefont {Corda}},\ }\href {\doibase 10.1140/epjc/s10052-017-5502-5} {\bibfield  {journal} {\bibinfo  {journal} {Eur. Phys. J. C}\ }\textbf {\bibinfo {volume} {78}},\ \bibinfo {pages} {25} (\bibinfo {year} {2018})},\ \Eprint {http://arxiv.org/abs/1712.09307} {arXiv:1712.09307 [gr-qc]} \BibitemShut {NoStop}%
\bibitem [{\citenamefont {Moradpour}\ and\ \citenamefont {Salako}(2016)}]{Moradpour:2016fur}%
  \BibitemOpen
  \bibfield  {author} {\bibinfo {author} {\bibfnamefont {H.}~\bibnamefont {Moradpour}}\ and\ \bibinfo {author} {\bibfnamefont {I.~G.}\ \bibnamefont {Salako}},\ }\href {\doibase 10.1155/2016/3492796} {\bibfield  {journal} {\bibinfo  {journal} {Adv. High Energy Phys.}\ }\textbf {\bibinfo {volume} {2016}},\ \bibinfo {pages} {3492796} (\bibinfo {year} {2016})},\ \Eprint {http://arxiv.org/abs/1606.06589} {arXiv:1606.06589 [gr-qc]} \BibitemShut {NoStop}%
\bibitem [{\citenamefont {Li}\ \emph {et~al.}(2019)\citenamefont {Li}, \citenamefont {Wang}, \citenamefont {Xu},\ and\ \citenamefont {Guo}}]{Li:2019jkv}%
  \BibitemOpen
  \bibfield  {author} {\bibinfo {author} {\bibfnamefont {R.}~\bibnamefont {Li}}, \bibinfo {author} {\bibfnamefont {J.}~\bibnamefont {Wang}}, \bibinfo {author} {\bibfnamefont {Z.}~\bibnamefont {Xu}}, \ and\ \bibinfo {author} {\bibfnamefont {X.}~\bibnamefont {Guo}},\ }\href {\doibase 10.1093/mnras/stz967} {\bibfield  {journal} {\bibinfo  {journal} {Mon. Not. Roy. Astron. Soc.}\ }\textbf {\bibinfo {volume} {486}},\ \bibinfo {pages} {2407} (\bibinfo {year} {2019})},\ \Eprint {http://arxiv.org/abs/1903.08790} {arXiv:1903.08790} \BibitemShut {NoStop}%
\bibitem [{\citenamefont {Koivisto}(2006)}]{Koivisto:2005yk}%
  \BibitemOpen
  \bibfield  {author} {\bibinfo {author} {\bibfnamefont {T.}~\bibnamefont {Koivisto}},\ }\href {\doibase 10.1088/0264-9381/23/12/N01} {\bibfield  {journal} {\bibinfo  {journal} {Class. Quant. Grav.}\ }\textbf {\bibinfo {volume} {23}},\ \bibinfo {pages} {4289} (\bibinfo {year} {2006})},\ \Eprint {http://arxiv.org/abs/gr-qc/0505128} {arXiv:gr-qc/0505128} \BibitemShut {NoStop}%
\bibitem [{\citenamefont {Bertolami}\ \emph {et~al.}(2007)\citenamefont {Bertolami}, \citenamefont {Boehmer}, \citenamefont {Harko},\ and\ \citenamefont {Lobo}}]{Bertolami:2007gv}%
  \BibitemOpen
  \bibfield  {author} {\bibinfo {author} {\bibfnamefont {O.}~\bibnamefont {Bertolami}}, \bibinfo {author} {\bibfnamefont {C.~G.}\ \bibnamefont {Boehmer}}, \bibinfo {author} {\bibfnamefont {T.}~\bibnamefont {Harko}}, \ and\ \bibinfo {author} {\bibfnamefont {F.~S.~N.}\ \bibnamefont {Lobo}},\ }\href {\doibase 10.1103/PhysRevD.75.104016} {\bibfield  {journal} {\bibinfo  {journal} {Phys. Rev. D}\ }\textbf {\bibinfo {volume} {75}},\ \bibinfo {pages} {104016} (\bibinfo {year} {2007})},\ \Eprint {http://arxiv.org/abs/0704.1733} {arXiv:0704.1733 [gr-qc]} \BibitemShut {NoStop}%
\bibitem [{\citenamefont {Harko}\ and\ \citenamefont {Lobo}(2014)}]{Harko:2014gwa}%
  \BibitemOpen
  \bibfield  {author} {\bibinfo {author} {\bibfnamefont {T.}~\bibnamefont {Harko}}\ and\ \bibinfo {author} {\bibfnamefont {F.~S.~N.}\ \bibnamefont {Lobo}},\ }\href {\doibase 10.3390/galaxies2030410} {\bibfield  {journal} {\bibinfo  {journal} {Galaxies}\ }\textbf {\bibinfo {volume} {2}},\ \bibinfo {pages} {410} (\bibinfo {year} {2014})},\ \Eprint {http://arxiv.org/abs/1407.2013} {arXiv:1407.2013 [gr-qc]} \BibitemShut {NoStop}%
\bibitem [{\citenamefont {Capone}\ \emph {et~al.}(2010)\citenamefont {Capone}, \citenamefont {Cardone},\ and\ \citenamefont {Ruggiero}}]{capone2010possibility}%
  \BibitemOpen
  \bibfield  {author} {\bibinfo {author} {\bibfnamefont {M.}~\bibnamefont {Capone}}, \bibinfo {author} {\bibfnamefont {V.}~\bibnamefont {Cardone}}, \ and\ \bibinfo {author} {\bibfnamefont {M.~L.}\ \bibnamefont {Ruggiero}},\ }in\ \href@noop {} {\emph {\bibinfo {booktitle} {Journal of Physics: Conference Series}}},\ Vol.\ \bibinfo {volume} {222}\ (\bibinfo {organization} {IOP Publishing},\ \bibinfo {year} {2010})\ p.\ \bibinfo {pages} {012012}\BibitemShut {NoStop}%
\bibitem [{\citenamefont {Batista}\ \emph {et~al.}(2012)\citenamefont {Batista}, \citenamefont {Daouda}, \citenamefont {Fabris}, \citenamefont {Piattella},\ and\ \citenamefont {Rodrigues}}]{Batista:2011nu}%
  \BibitemOpen
  \bibfield  {author} {\bibinfo {author} {\bibfnamefont {C.~E.~M.}\ \bibnamefont {Batista}}, \bibinfo {author} {\bibfnamefont {M.~H.}\ \bibnamefont {Daouda}}, \bibinfo {author} {\bibfnamefont {J.~C.}\ \bibnamefont {Fabris}}, \bibinfo {author} {\bibfnamefont {O.~F.}\ \bibnamefont {Piattella}}, \ and\ \bibinfo {author} {\bibfnamefont {D.~C.}\ \bibnamefont {Rodrigues}},\ }\href {\doibase 10.1103/PhysRevD.85.084008} {\bibfield  {journal} {\bibinfo  {journal} {Phys. Rev. D}\ }\textbf {\bibinfo {volume} {85}},\ \bibinfo {pages} {084008} (\bibinfo {year} {2012})},\ \Eprint {http://arxiv.org/abs/1112.4141} {arXiv:1112.4141 [astro-ph.CO]} \BibitemShut {NoStop}%
\bibitem [{\citenamefont {Akarsu}\ \emph {et~al.}(2020)\citenamefont {Akarsu}, \citenamefont {Kat\i{}rc\i{}}, \citenamefont {Kumar}, \citenamefont {Nunes}, \citenamefont {\"Ozt\"urk},\ and\ \citenamefont {Sharma}}]{Akarsu:2020yqa}%
  \BibitemOpen
  \bibfield  {author} {\bibinfo {author} {\bibfnamefont {O.}~\bibnamefont {Akarsu}}, \bibinfo {author} {\bibfnamefont {N.}~\bibnamefont {Kat\i{}rc\i{}}}, \bibinfo {author} {\bibfnamefont {S.}~\bibnamefont {Kumar}}, \bibinfo {author} {\bibfnamefont {R.~C.}\ \bibnamefont {Nunes}}, \bibinfo {author} {\bibfnamefont {B.}~\bibnamefont {\"Ozt\"urk}}, \ and\ \bibinfo {author} {\bibfnamefont {S.}~\bibnamefont {Sharma}},\ }\href {\doibase 10.1140/epjc/s10052-020-08586-4} {\bibfield  {journal} {\bibinfo  {journal} {Eur. Phys. J. C}\ }\textbf {\bibinfo {volume} {80}},\ \bibinfo {pages} {1050} (\bibinfo {year} {2020})},\ \Eprint {http://arxiv.org/abs/2004.04074} {arXiv:2004.04074 [astro-ph.CO]} \BibitemShut {NoStop}%
\bibitem [{\citenamefont {Majernik}\ and\ \citenamefont {Richterek}(2006)}]{Majernik:2006jg}%
  \BibitemOpen
  \bibfield  {author} {\bibinfo {author} {\bibfnamefont {V.}~\bibnamefont {Majernik}}\ and\ \bibinfo {author} {\bibfnamefont {L.}~\bibnamefont {Richterek}},\ }\href@noop {} {\  (\bibinfo {year} {2006})},\ \Eprint {http://arxiv.org/abs/gr-qc/0610070} {arXiv:gr-qc/0610070} \BibitemShut {NoStop}%
\bibitem [{\citenamefont {Fabris}\ \emph {et~al.}(2012{\natexlab{a}})\citenamefont {Fabris}, \citenamefont {Piattella}, \citenamefont {Rodrigues}, \citenamefont {Batista},\ and\ \citenamefont {Daouda}}]{Fabris:2012hw}%
  \BibitemOpen
  \bibfield  {author} {\bibinfo {author} {\bibfnamefont {J.~C.}\ \bibnamefont {Fabris}}, \bibinfo {author} {\bibfnamefont {O.~F.}\ \bibnamefont {Piattella}}, \bibinfo {author} {\bibfnamefont {D.~C.}\ \bibnamefont {Rodrigues}}, \bibinfo {author} {\bibfnamefont {C.~E.~M.}\ \bibnamefont {Batista}}, \ and\ \bibinfo {author} {\bibfnamefont {M.~H.}\ \bibnamefont {Daouda}},\ }\href {\doibase 10.1142/S2010194512008227} {\bibfield  {journal} {\bibinfo  {journal} {Int. J. Mod. Phys. Conf. Ser.}\ }\textbf {\bibinfo {volume} {18}},\ \bibinfo {pages} {67} (\bibinfo {year} {2012}{\natexlab{a}})},\ \Eprint {http://arxiv.org/abs/1205.1198} {arXiv:1205.1198 [astro-ph.CO]} \BibitemShut {NoStop}%
\bibitem [{\citenamefont {Fabris}\ \emph {et~al.}(2012{\natexlab{b}})\citenamefont {Fabris}, \citenamefont {Daouda},\ and\ \citenamefont {Piattella}}]{Fabris:2011wz}%
  \BibitemOpen
  \bibfield  {author} {\bibinfo {author} {\bibfnamefont {J.~C.}\ \bibnamefont {Fabris}}, \bibinfo {author} {\bibfnamefont {M.~H.}\ \bibnamefont {Daouda}}, \ and\ \bibinfo {author} {\bibfnamefont {O.~F.}\ \bibnamefont {Piattella}},\ }\href {\doibase 10.1016/j.physletb.2012.04.020} {\bibfield  {journal} {\bibinfo  {journal} {Phys. Lett. B}\ }\textbf {\bibinfo {volume} {711}},\ \bibinfo {pages} {232} (\bibinfo {year} {2012}{\natexlab{b}})},\ \Eprint {http://arxiv.org/abs/1109.2096} {arXiv:1109.2096 [astro-ph.CO]} \BibitemShut {NoStop}%
\bibitem [{\citenamefont {Batista}\ \emph {et~al.}(2013)\citenamefont {Batista}, \citenamefont {Fabris}, \citenamefont {Piattella},\ and\ \citenamefont {Velasquez-Toribio}}]{Batista:2012hv}%
  \BibitemOpen
  \bibfield  {author} {\bibinfo {author} {\bibfnamefont {C.~E.~M.}\ \bibnamefont {Batista}}, \bibinfo {author} {\bibfnamefont {J.~C.}\ \bibnamefont {Fabris}}, \bibinfo {author} {\bibfnamefont {O.~F.}\ \bibnamefont {Piattella}}, \ and\ \bibinfo {author} {\bibfnamefont {A.~M.}\ \bibnamefont {Velasquez-Toribio}},\ }\href {\doibase 10.1140/epjc/s10052-013-2425-7} {\bibfield  {journal} {\bibinfo  {journal} {Eur. Phys. J. C}\ }\textbf {\bibinfo {volume} {73}},\ \bibinfo {pages} {2425} (\bibinfo {year} {2013})},\ \Eprint {http://arxiv.org/abs/1208.6327} {arXiv:1208.6327 [astro-ph.CO]} \BibitemShut {NoStop}%
\bibitem [{\citenamefont {Campos}\ \emph {et~al.}(2013)\citenamefont {Campos}, \citenamefont {Fabris}, \citenamefont {Perez}, \citenamefont {Piattella},\ and\ \citenamefont {Velten}}]{Campos:2012ez}%
  \BibitemOpen
  \bibfield  {author} {\bibinfo {author} {\bibfnamefont {J.~P.}\ \bibnamefont {Campos}}, \bibinfo {author} {\bibfnamefont {J.~C.}\ \bibnamefont {Fabris}}, \bibinfo {author} {\bibfnamefont {R.}~\bibnamefont {Perez}}, \bibinfo {author} {\bibfnamefont {O.~F.}\ \bibnamefont {Piattella}}, \ and\ \bibinfo {author} {\bibfnamefont {H.}~\bibnamefont {Velten}},\ }\href {\doibase 10.1140/epjc/s10052-013-2357-2} {\bibfield  {journal} {\bibinfo  {journal} {Eur. Phys. J. C}\ }\textbf {\bibinfo {volume} {73}},\ \bibinfo {pages} {2357} (\bibinfo {year} {2013})},\ \Eprint {http://arxiv.org/abs/1212.4136} {arXiv:1212.4136 [astro-ph.CO]} \BibitemShut {NoStop}%
\bibitem [{\citenamefont {Caram\^es}\ \emph {et~al.}(2014)\citenamefont {Caram\^es}, \citenamefont {Daouda}, \citenamefont {Fabris}, \citenamefont {Oliveira}, \citenamefont {Piattella},\ and\ \citenamefont {Strokov}}]{Carames:2014twa}%
  \BibitemOpen
  \bibfield  {author} {\bibinfo {author} {\bibfnamefont {T.~R.~P.}\ \bibnamefont {Caram\^es}}, \bibinfo {author} {\bibfnamefont {M.~H.}\ \bibnamefont {Daouda}}, \bibinfo {author} {\bibfnamefont {J.~C.}\ \bibnamefont {Fabris}}, \bibinfo {author} {\bibfnamefont {A.~M.}\ \bibnamefont {Oliveira}}, \bibinfo {author} {\bibfnamefont {O.~F.}\ \bibnamefont {Piattella}}, \ and\ \bibinfo {author} {\bibfnamefont {V.}~\bibnamefont {Strokov}},\ }\href {\doibase 10.1140/epjc/s10052-014-3145-3} {\bibfield  {journal} {\bibinfo  {journal} {Eur. Phys. J. C}\ }\textbf {\bibinfo {volume} {74}},\ \bibinfo {pages} {3145} (\bibinfo {year} {2014})},\ \Eprint {http://arxiv.org/abs/1409.2322} {arXiv:1409.2322 [gr-qc]} \BibitemShut {NoStop}%
\bibitem [{\citenamefont {Salako}\ \emph {et~al.}(2016)\citenamefont {Salako}, \citenamefont {Houndjo},\ and\ \citenamefont {Jawad}}]{Salako:2016ihq}%
  \BibitemOpen
  \bibfield  {author} {\bibinfo {author} {\bibfnamefont {I.~G.}\ \bibnamefont {Salako}}, \bibinfo {author} {\bibfnamefont {M.~J.~S.}\ \bibnamefont {Houndjo}}, \ and\ \bibinfo {author} {\bibfnamefont {A.}~\bibnamefont {Jawad}},\ }\href {\doibase 10.1142/S0218271816500760} {\bibfield  {journal} {\bibinfo  {journal} {Int. J. Mod. Phys. D}\ }\textbf {\bibinfo {volume} {25}},\ \bibinfo {pages} {1650076} (\bibinfo {year} {2016})},\ \Eprint {http://arxiv.org/abs/1605.07611} {arXiv:1605.07611 [gr-qc]} \BibitemShut {NoStop}%
\bibitem [{\citenamefont {Moradpour}(2016)}]{Moradpour:2015ymo}%
  \BibitemOpen
  \bibfield  {author} {\bibinfo {author} {\bibfnamefont {H.}~\bibnamefont {Moradpour}},\ }\href {\doibase 10.1016/j.physletb.2016.03.072} {\bibfield  {journal} {\bibinfo  {journal} {Phys. Lett. B}\ }\textbf {\bibinfo {volume} {757}},\ \bibinfo {pages} {187} (\bibinfo {year} {2016})},\ \Eprint {http://arxiv.org/abs/1601.04529} {arXiv:1601.04529 [physics.gen-ph]} \BibitemShut {NoStop}%
\bibitem [{\citenamefont {Moradpour}\ \emph {et~al.}(2017{\natexlab{a}})\citenamefont {Moradpour}, \citenamefont {Heydarzade}, \citenamefont {Darabi},\ and\ \citenamefont {Salako}}]{Moradpour:2017shy}%
  \BibitemOpen
  \bibfield  {author} {\bibinfo {author} {\bibfnamefont {H.}~\bibnamefont {Moradpour}}, \bibinfo {author} {\bibfnamefont {Y.}~\bibnamefont {Heydarzade}}, \bibinfo {author} {\bibfnamefont {F.}~\bibnamefont {Darabi}}, \ and\ \bibinfo {author} {\bibfnamefont {I.~G.}\ \bibnamefont {Salako}},\ }\href {\doibase 10.1140/epjc/s10052-017-4811-z} {\bibfield  {journal} {\bibinfo  {journal} {Eur. Phys. J. C}\ }\textbf {\bibinfo {volume} {77}},\ \bibinfo {pages} {259} (\bibinfo {year} {2017}{\natexlab{a}})},\ \Eprint {http://arxiv.org/abs/1704.02458} {arXiv:1704.02458 [gr-qc]} \BibitemShut {NoStop}%
\bibitem [{\citenamefont {Moradpour}\ \emph {et~al.}(2017{\natexlab{b}})\citenamefont {Moradpour}, \citenamefont {Bonilla}, \citenamefont {Abreu},\ and\ \citenamefont {Neto}}]{Moradpour:2017ycq}%
  \BibitemOpen
  \bibfield  {author} {\bibinfo {author} {\bibfnamefont {H.}~\bibnamefont {Moradpour}}, \bibinfo {author} {\bibfnamefont {A.}~\bibnamefont {Bonilla}}, \bibinfo {author} {\bibfnamefont {E.~M.~C.}\ \bibnamefont {Abreu}}, \ and\ \bibinfo {author} {\bibfnamefont {J.~A.}\ \bibnamefont {Neto}},\ }\href {\doibase 10.1103/PhysRevD.96.123504} {\bibfield  {journal} {\bibinfo  {journal} {Phys. Rev. D}\ }\textbf {\bibinfo {volume} {96}},\ \bibinfo {pages} {123504} (\bibinfo {year} {2017}{\natexlab{b}})},\ \Eprint {http://arxiv.org/abs/1711.08338} {arXiv:1711.08338 [physics.gen-ph]} \BibitemShut {NoStop}%
\bibitem [{\citenamefont {Das}\ \emph {et~al.}(2018)\citenamefont {Das}, \citenamefont {Dutta},\ and\ \citenamefont {Chakraborty}}]{Das:2018dzp}%
  \BibitemOpen
  \bibfield  {author} {\bibinfo {author} {\bibfnamefont {D.}~\bibnamefont {Das}}, \bibinfo {author} {\bibfnamefont {S.}~\bibnamefont {Dutta}}, \ and\ \bibinfo {author} {\bibfnamefont {S.}~\bibnamefont {Chakraborty}},\ }\href {\doibase 10.1140/epjc/s10052-018-6293-z} {\bibfield  {journal} {\bibinfo  {journal} {Eur. Phys. J. C}\ }\textbf {\bibinfo {volume} {78}},\ \bibinfo {pages} {810} (\bibinfo {year} {2018})},\ \Eprint {http://arxiv.org/abs/1810.11260} {arXiv:1810.11260 [gr-qc]} \BibitemShut {NoStop}%
\bibitem [{\citenamefont {Ghaffari}\ \emph {et~al.}(2020)\citenamefont {Ghaffari}, \citenamefont {Mamon}, \citenamefont {Moradpour},\ and\ \citenamefont {Ziaie}}]{Ghaffari:2020nnk}%
  \BibitemOpen
  \bibfield  {author} {\bibinfo {author} {\bibfnamefont {S.}~\bibnamefont {Ghaffari}}, \bibinfo {author} {\bibfnamefont {A.~A.}\ \bibnamefont {Mamon}}, \bibinfo {author} {\bibfnamefont {H.}~\bibnamefont {Moradpour}}, \ and\ \bibinfo {author} {\bibfnamefont {A.~H.}\ \bibnamefont {Ziaie}},\ }\href {\doibase 10.1142/S0217732320502764} {\bibfield  {journal} {\bibinfo  {journal} {Mod. Phys. Lett. A}\ }\textbf {\bibinfo {volume} {35}},\ \bibinfo {pages} {2050276} (\bibinfo {year} {2020})},\ \Eprint {http://arxiv.org/abs/2002.04972} {arXiv:2002.04972 [gr-qc]} \BibitemShut {NoStop}%
\bibitem [{\citenamefont {Saleem}\ and\ \citenamefont {Hassan}(2020)}]{Saleem:2020bwa}%
  \BibitemOpen
  \bibfield  {author} {\bibinfo {author} {\bibfnamefont {R.}~\bibnamefont {Saleem}}\ and\ \bibinfo {author} {\bibfnamefont {J.}~\bibnamefont {Hassan}},\ }\href {\doibase 10.1016/j.dark.2020.100515} {\bibfield  {journal} {\bibinfo  {journal} {Phys. Dark Univ.}\ }\textbf {\bibinfo {volume} {28}},\ \bibinfo {pages} {100515} (\bibinfo {year} {2020})}\BibitemShut {NoStop}%
\bibitem [{\citenamefont {Singh}\ and\ \citenamefont {Mishra}(2020)}]{Singh:2020akk}%
  \BibitemOpen
  \bibfield  {author} {\bibinfo {author} {\bibfnamefont {A.}~\bibnamefont {Singh}}\ and\ \bibinfo {author} {\bibfnamefont {K.~C.}\ \bibnamefont {Mishra}},\ }\href {\doibase 10.1140/epjp/s13360-020-00783-0} {\bibfield  {journal} {\bibinfo  {journal} {Eur. Phys. J. Plus}\ }\textbf {\bibinfo {volume} {135}},\ \bibinfo {pages} {752} (\bibinfo {year} {2020})}\BibitemShut {NoStop}%
\bibitem [{\citenamefont {Bishi}\ \emph {et~al.}(2023)\citenamefont {Bishi}, \citenamefont {Lepse},\ and\ \citenamefont {Beesham}}]{Bishi:2023mwv}%
  \BibitemOpen
  \bibfield  {author} {\bibinfo {author} {\bibfnamefont {B.~K.}\ \bibnamefont {Bishi}}, \bibinfo {author} {\bibfnamefont {P.~V.}\ \bibnamefont {Lepse}}, \ and\ \bibinfo {author} {\bibfnamefont {A.}~\bibnamefont {Beesham}},\ }\href {\doibase 10.3390/ECU2023-14057} {\bibfield  {journal} {\bibinfo  {journal} {Phys. Sci. Forum}\ }\textbf {\bibinfo {volume} {7}},\ \bibinfo {pages} {38} (\bibinfo {year} {2023})}\BibitemShut {NoStop}%
\bibitem [{\citenamefont {Gibbons}\ and\ \citenamefont {Hawking}(1977)}]{Gibbons:1977mu}%
  \BibitemOpen
  \bibfield  {author} {\bibinfo {author} {\bibfnamefont {G.~W.}\ \bibnamefont {Gibbons}}\ and\ \bibinfo {author} {\bibfnamefont {S.~W.}\ \bibnamefont {Hawking}},\ }\href {\doibase 10.1103/PhysRevD.15.2738} {\bibfield  {journal} {\bibinfo  {journal} {Phys. Rev. D}\ }\textbf {\bibinfo {volume} {15}},\ \bibinfo {pages} {2738} (\bibinfo {year} {1977})}\BibitemShut {NoStop}%
\bibitem [{\citenamefont {Parker}(1971)}]{Parker:1971pt}%
  \BibitemOpen
  \bibfield  {author} {\bibinfo {author} {\bibfnamefont {L.}~\bibnamefont {Parker}},\ }\href {\doibase 10.1103/PhysRevD.3.346} {\bibfield  {journal} {\bibinfo  {journal} {Phys. Rev. D}\ }\textbf {\bibinfo {volume} {3}},\ \bibinfo {pages} {346} (\bibinfo {year} {1971})},\ \bibinfo {note} {[Erratum: Phys.Rev.D 3, 2546--2546 (1971)]}\BibitemShut {NoStop}%
\bibitem [{\citenamefont {Ford}(1987)}]{Ford:1986sy}%
  \BibitemOpen
  \bibfield  {author} {\bibinfo {author} {\bibfnamefont {L.~H.}\ \bibnamefont {Ford}},\ }\href {\doibase 10.1103/PhysRevD.35.2955} {\bibfield  {journal} {\bibinfo  {journal} {Phys. Rev. D}\ }\textbf {\bibinfo {volume} {35}},\ \bibinfo {pages} {2955} (\bibinfo {year} {1987})}\BibitemShut {NoStop}%
\bibitem [{\citenamefont {Pereira}\ \emph {et~al.}(2010)\citenamefont {Pereira}, \citenamefont {Bessa},\ and\ \citenamefont {Lima}}]{Pereira:2009kv}%
  \BibitemOpen
  \bibfield  {author} {\bibinfo {author} {\bibfnamefont {S.~H.}\ \bibnamefont {Pereira}}, \bibinfo {author} {\bibfnamefont {C.~H.~G.}\ \bibnamefont {Bessa}}, \ and\ \bibinfo {author} {\bibfnamefont {J.~A.~S.}\ \bibnamefont {Lima}},\ }\href {\doibase 10.1016/j.physletb.2010.05.027} {\bibfield  {journal} {\bibinfo  {journal} {Phys. Lett. B}\ }\textbf {\bibinfo {volume} {690}},\ \bibinfo {pages} {103} (\bibinfo {year} {2010})},\ \Eprint {http://arxiv.org/abs/0911.0622} {arXiv:0911.0622 [astro-ph.CO]} \BibitemShut {NoStop}%
\bibitem [{\citenamefont {Abbas}\ and\ \citenamefont {Shahzad}(2019)}]{Abbas:2019evn}%
  \BibitemOpen
  \bibfield  {author} {\bibinfo {author} {\bibfnamefont {G.}~\bibnamefont {Abbas}}\ and\ \bibinfo {author} {\bibfnamefont {M.~R.}\ \bibnamefont {Shahzad}},\ }\href {\doibase 10.1007/s10509-019-3537-9} {\bibfield  {journal} {\bibinfo  {journal} {Astrophys. Space Sci.}\ }\textbf {\bibinfo {volume} {364}},\ \bibinfo {pages} {50} (\bibinfo {year} {2019})}\BibitemShut {NoStop}%
\bibitem [{\citenamefont {Shahzad}\ and\ \citenamefont {Abbas}(2019)}]{Shahzad:2019bqf}%
  \BibitemOpen
  \bibfield  {author} {\bibinfo {author} {\bibfnamefont {M.~R.}\ \bibnamefont {Shahzad}}\ and\ \bibinfo {author} {\bibfnamefont {G.}~\bibnamefont {Abbas}},\ }\href {\doibase 10.1142/S0219887819501329} {\bibfield  {journal} {\bibinfo  {journal} {Int. J. Geom. Meth. Mod. Phys.}\ }\textbf {\bibinfo {volume} {16}},\ \bibinfo {pages} {1950132} (\bibinfo {year} {2019})}\BibitemShut {NoStop}%
\bibitem [{\citenamefont {Shahzad}\ and\ \citenamefont {Abbas}(2020{\natexlab{a}})}]{Shahzad:2020gjj}%
  \BibitemOpen
  \bibfield  {author} {\bibinfo {author} {\bibfnamefont {M.~R.}\ \bibnamefont {Shahzad}}\ and\ \bibinfo {author} {\bibfnamefont {G.}~\bibnamefont {Abbas}},\ }\href {\doibase 10.1140/epjp/s13360-020-00508-3} {\bibfield  {journal} {\bibinfo  {journal} {Eur. Phys. J. Plus}\ }\textbf {\bibinfo {volume} {135}},\ \bibinfo {pages} {502} (\bibinfo {year} {2020}{\natexlab{a}})}\BibitemShut {NoStop}%
\bibitem [{\citenamefont {Mustafa}\ \emph {et~al.}(2021)\citenamefont {Mustafa}, \citenamefont {Tie-Cheng},\ and\ \citenamefont {Shamir}}]{Mustafa:2021fzk}%
  \BibitemOpen
  \bibfield  {author} {\bibinfo {author} {\bibfnamefont {G.}~\bibnamefont {Mustafa}}, \bibinfo {author} {\bibfnamefont {X.}~\bibnamefont {Tie-Cheng}}, \ and\ \bibinfo {author} {\bibfnamefont {M.~F.}\ \bibnamefont {Shamir}},\ }\href {\doibase 10.1088/1402-4896/ac0ee6} {\bibfield  {journal} {\bibinfo  {journal} {Phys. Scripta}\ }\textbf {\bibinfo {volume} {96}},\ \bibinfo {pages} {105008} (\bibinfo {year} {2021})}\BibitemShut {NoStop}%
\bibitem [{\citenamefont {Hansraj}\ and\ \citenamefont {Banerjee}(2020)}]{Hansraj:2020clg}%
  \BibitemOpen
  \bibfield  {author} {\bibinfo {author} {\bibfnamefont {S.}~\bibnamefont {Hansraj}}\ and\ \bibinfo {author} {\bibfnamefont {A.}~\bibnamefont {Banerjee}},\ }\href {\doibase 10.1142/S0217732320501059} {\bibfield  {journal} {\bibinfo  {journal} {Mod. Phys. Lett. A}\ }\textbf {\bibinfo {volume} {35}},\ \bibinfo {pages} {2050105} (\bibinfo {year} {2020})}\BibitemShut {NoStop}%
\bibitem [{\citenamefont {Moradpour}\ \emph {et~al.}(2017{\natexlab{c}})\citenamefont {Moradpour}, \citenamefont {Sadeghnezhad},\ and\ \citenamefont {Hendi}}]{Moradpour:2016ubd}%
  \BibitemOpen
  \bibfield  {author} {\bibinfo {author} {\bibfnamefont {H.}~\bibnamefont {Moradpour}}, \bibinfo {author} {\bibfnamefont {N.}~\bibnamefont {Sadeghnezhad}}, \ and\ \bibinfo {author} {\bibfnamefont {S.~H.}\ \bibnamefont {Hendi}},\ }\href {\doibase 10.1139/cjp-2017-0040} {\bibfield  {journal} {\bibinfo  {journal} {Can. J. Phys.}\ }\textbf {\bibinfo {volume} {95}},\ \bibinfo {pages} {1257} (\bibinfo {year} {2017}{\natexlab{c}})},\ \Eprint {http://arxiv.org/abs/1606.00846} {arXiv:1606.00846 [gr-qc]} \BibitemShut {NoStop}%
\bibitem [{\citenamefont {Shahzad}\ and\ \citenamefont {Abbas}(2020{\natexlab{b}})}]{Shahzad:2020bwr}%
  \BibitemOpen
  \bibfield  {author} {\bibinfo {author} {\bibfnamefont {M.~R.}\ \bibnamefont {Shahzad}}\ and\ \bibinfo {author} {\bibfnamefont {G.}~\bibnamefont {Abbas}},\ }\href {\doibase 10.1007/s10509-020-03861-y} {\bibfield  {journal} {\bibinfo  {journal} {Astrophys. Space Sci.}\ }\textbf {\bibinfo {volume} {365}},\ \bibinfo {pages} {147} (\bibinfo {year} {2020}{\natexlab{b}})}\BibitemShut {NoStop}%
\bibitem [{\citenamefont {Saleem}\ \emph {et~al.}(2024)\citenamefont {Saleem}, \citenamefont {Aslam},\ and\ \citenamefont {Shahid}}]{Saleem:2023ukn}%
  \BibitemOpen
  \bibfield  {author} {\bibinfo {author} {\bibfnamefont {R.}~\bibnamefont {Saleem}}, \bibinfo {author} {\bibfnamefont {M.~I.}\ \bibnamefont {Aslam}}, \ and\ \bibinfo {author} {\bibfnamefont {S.}~\bibnamefont {Shahid}},\ }\href {\doibase 10.1142/S0219887824501068} {\bibfield  {journal} {\bibinfo  {journal} {Int. J. Geom. Meth. Mod. Phys.}\ }\textbf {\bibinfo {volume} {21}},\ \bibinfo {pages} {2450106} (\bibinfo {year} {2024})}\BibitemShut {NoStop}%
\bibitem [{\citenamefont {Ashraf}\ \emph {et~al.}(2024)\citenamefont {Ashraf}, \citenamefont {Shahzad},\ and\ \citenamefont {Ma}}]{ashraf2024possible}%
  \BibitemOpen
  \bibfield  {author} {\bibinfo {author} {\bibfnamefont {A.}~\bibnamefont {Ashraf}}, \bibinfo {author} {\bibfnamefont {M.}~\bibnamefont {Shahzad}}, \ and\ \bibinfo {author} {\bibfnamefont {W.-X.}\ \bibnamefont {Ma}},\ }\href@noop {} {\bibfield  {journal} {\bibinfo  {journal} {International Journal of Geometric Methods in Modern Physics}\ ,\ \bibinfo {pages} {2450200}} (\bibinfo {year} {2024})}\BibitemShut {NoStop}%
\bibitem [{\citenamefont {El~Hanafy}(2022)}]{ElHanafy:2022kjl}%
  \BibitemOpen
  \bibfield  {author} {\bibinfo {author} {\bibfnamefont {W.}~\bibnamefont {El~Hanafy}},\ }\href {\doibase 10.3847/1538-4357/ac9410} {\bibfield  {journal} {\bibinfo  {journal} {Astrophys. J.}\ }\textbf {\bibinfo {volume} {940}},\ \bibinfo {pages} {51} (\bibinfo {year} {2022})},\ \Eprint {http://arxiv.org/abs/2209.10287} {arXiv:2209.10287 [astro-ph.HE]} \BibitemShut {NoStop}%
\bibitem [{\citenamefont {Waseem}(2024)}]{waseem2024isotropic}%
  \BibitemOpen
  \bibfield  {author} {\bibinfo {author} {\bibfnamefont {A.}~\bibnamefont {Waseem}},\ }\href@noop {} {\bibfield  {journal} {\bibinfo  {journal} {International Journal of Geometric Methods in Modern Physics}\ ,\ \bibinfo {pages} {2450194}} (\bibinfo {year} {2024})}\BibitemShut {NoStop}%
\bibitem [{\citenamefont {Ghosh}\ \emph {et~al.}(2021)\citenamefont {Ghosh}, \citenamefont {Dey}, \citenamefont {Das}, \citenamefont {Chanda},\ and\ \citenamefont {Paul}}]{Ghosh:2021byh}%
  \BibitemOpen
  \bibfield  {author} {\bibinfo {author} {\bibfnamefont {S.}~\bibnamefont {Ghosh}}, \bibinfo {author} {\bibfnamefont {S.}~\bibnamefont {Dey}}, \bibinfo {author} {\bibfnamefont {A.}~\bibnamefont {Das}}, \bibinfo {author} {\bibfnamefont {A.}~\bibnamefont {Chanda}}, \ and\ \bibinfo {author} {\bibfnamefont {B.~C.}\ \bibnamefont {Paul}},\ }\href {\doibase 10.1088/1475-7516/2021/07/004} {\bibfield  {journal} {\bibinfo  {journal} {JCAP}\ }\textbf {\bibinfo {volume} {07}},\ \bibinfo {pages} {004} (\bibinfo {year} {2021})},\ \Eprint {http://arxiv.org/abs/2102.01524} {arXiv:2102.01524 [gr-qc]} \BibitemShut {NoStop}%
\bibitem [{\citenamefont {Majeed}\ and\ \citenamefont {Abbas}(2022)}]{Majeed:2022uyz}%
  \BibitemOpen
  \bibfield  {author} {\bibinfo {author} {\bibfnamefont {K.}~\bibnamefont {Majeed}}\ and\ \bibinfo {author} {\bibfnamefont {G.}~\bibnamefont {Abbas}},\ }\href {\doibase 10.1088/2399-6528/ac65fa} {\bibfield  {journal} {\bibinfo  {journal} {J. Phys. Comm.}\ }\textbf {\bibinfo {volume} {6}},\ \bibinfo {pages} {045005} (\bibinfo {year} {2022})}\BibitemShut {NoStop}%
\bibitem [{\citenamefont {Abbas}\ and\ \citenamefont {Majeed}(2020)}]{Abbas:2020kju}%
  \BibitemOpen
  \bibfield  {author} {\bibinfo {author} {\bibfnamefont {G.}~\bibnamefont {Abbas}}\ and\ \bibinfo {author} {\bibfnamefont {K.}~\bibnamefont {Majeed}},\ }\href {\doibase 10.1155/2020/8861168} {\bibfield  {journal} {\bibinfo  {journal} {Adv. Astron.}\ }\textbf {\bibinfo {volume} {2020}},\ \bibinfo {pages} {8861168} (\bibinfo {year} {2020})}\BibitemShut {NoStop}%
\bibitem [{\citenamefont {Ghezzi}(2011)}]{Ghezzi:2009ct}%
  \BibitemOpen
  \bibfield  {author} {\bibinfo {author} {\bibfnamefont {C.~R.}\ \bibnamefont {Ghezzi}},\ }\href {\doibase 10.1007/s10509-011-0663-4} {\bibfield  {journal} {\bibinfo  {journal} {Astrophys. Space Sci.}\ }\textbf {\bibinfo {volume} {333}},\ \bibinfo {pages} {437} (\bibinfo {year} {2011})},\ \Eprint {http://arxiv.org/abs/0908.0779} {arXiv:0908.0779 [gr-qc]} \BibitemShut {NoStop}%
\bibitem [{\citenamefont {Tolman}(1939)}]{Tolman:1939jz}%
  \BibitemOpen
  \bibfield  {author} {\bibinfo {author} {\bibfnamefont {R.~C.}\ \bibnamefont {Tolman}},\ }\href {\doibase 10.1103/PhysRev.55.364} {\bibfield  {journal} {\bibinfo  {journal} {Phys. Rev.}\ }\textbf {\bibinfo {volume} {55}},\ \bibinfo {pages} {364} (\bibinfo {year} {1939})}\BibitemShut {NoStop}%
\bibitem [{\citenamefont {Kuchowicz}(1968)}]{osti_4507306}%
  \BibitemOpen
  \bibfield  {author} {\bibinfo {author} {\bibfnamefont {B.}~\bibnamefont {Kuchowicz}},\ }\href@noop {} {\bibfield  {journal} {\bibinfo  {journal} {Acta Phys. Pol., 33: 541-63}\ } (\bibinfo {year} {1968})}\BibitemShut {NoStop}%
\bibitem [{\citenamefont {Ghezzi}(2005)}]{Ghezzi:2005iy}%
  \BibitemOpen
  \bibfield  {author} {\bibinfo {author} {\bibfnamefont {C.~R.}\ \bibnamefont {Ghezzi}},\ }\href {\doibase 10.1103/PhysRevD.72.104017} {\bibfield  {journal} {\bibinfo  {journal} {Phys. Rev. D}\ }\textbf {\bibinfo {volume} {72}},\ \bibinfo {pages} {104017} (\bibinfo {year} {2005})},\ \Eprint {http://arxiv.org/abs/gr-qc/0510106} {arXiv:gr-qc/0510106} \BibitemShut {NoStop}%
\bibitem [{\citenamefont {Barreto}\ \emph {et~al.}(2007)\citenamefont {Barreto}, \citenamefont {Rodriguez}, \citenamefont {Rosales},\ and\ \citenamefont {Serrano}}]{Barreto:2006cr}%
  \BibitemOpen
  \bibfield  {author} {\bibinfo {author} {\bibfnamefont {W.}~\bibnamefont {Barreto}}, \bibinfo {author} {\bibfnamefont {B.}~\bibnamefont {Rodriguez}}, \bibinfo {author} {\bibfnamefont {L.}~\bibnamefont {Rosales}}, \ and\ \bibinfo {author} {\bibfnamefont {O.}~\bibnamefont {Serrano}},\ }\href {\doibase 10.1007/s10714-006-0365-3} {\bibfield  {journal} {\bibinfo  {journal} {Gen. Rel. Grav.}\ }\textbf {\bibinfo {volume} {39}},\ \bibinfo {pages} {23} (\bibinfo {year} {2007})},\ \bibinfo {note} {[Erratum: Gen.Rel.Grav. 39, 537--538 (2007)]},\ \Eprint {http://arxiv.org/abs/gr-qc/0611089} {arXiv:gr-qc/0611089} \BibitemShut {NoStop}%
\bibitem [{\citenamefont {Das}\ and\ \citenamefont {Ali}(2015)}]{das2015anisotropic}%
  \BibitemOpen
  \bibfield  {author} {\bibinfo {author} {\bibfnamefont {K.}~\bibnamefont {Das}}\ and\ \bibinfo {author} {\bibfnamefont {N.}~\bibnamefont {Ali}},\ }\href@noop {} {\bibfield  {journal} {\bibinfo  {journal} {Astrophysics and Space Science}\ }\textbf {\bibinfo {volume} {356}},\ \bibinfo {pages} {57} (\bibinfo {year} {2015})}\BibitemShut {NoStop}%
\bibitem [{\citenamefont {Reissner}(1916)}]{reissner1916eigengravitation}%
  \BibitemOpen
  \bibfield  {author} {\bibinfo {author} {\bibfnamefont {H.}~\bibnamefont {Reissner}},\ }\href@noop {} {\bibfield  {journal} {\bibinfo  {journal} {Annalen der Physik}\ }\textbf {\bibinfo {volume} {355}},\ \bibinfo {pages} {106} (\bibinfo {year} {1916})}\BibitemShut {NoStop}%
\bibitem [{\citenamefont {Nordstr{\"o}m}(1918)}]{nordstrom1918energy}%
  \BibitemOpen
  \bibfield  {author} {\bibinfo {author} {\bibfnamefont {G.}~\bibnamefont {Nordstr{\"o}m}},\ }\href@noop {} {\bibfield  {journal} {\bibinfo  {journal} {Koninklijke Nederlandse Akademie van Wetenschappen Proceedings Series B Physical Sciences}\ }\textbf {\bibinfo {volume} {20}},\ \bibinfo {pages} {1238} (\bibinfo {year} {1918})}\BibitemShut {NoStop}%
\bibitem [{\citenamefont {Abubekerov}\ \emph {et~al.}(2008)\citenamefont {Abubekerov}, \citenamefont {Antokhina}, \citenamefont {Cherepashchuk},\ and\ \citenamefont {Shimanskii}}]{Abubekerov:2008inw}%
  \BibitemOpen
  \bibfield  {author} {\bibinfo {author} {\bibfnamefont {M.~K.}\ \bibnamefont {Abubekerov}}, \bibinfo {author} {\bibfnamefont {E.~A.}\ \bibnamefont {Antokhina}}, \bibinfo {author} {\bibfnamefont {A.~M.}\ \bibnamefont {Cherepashchuk}}, \ and\ \bibinfo {author} {\bibfnamefont {V.~V.}\ \bibnamefont {Shimanskii}},\ }\href {\doibase 10.1134/S1063772908050041} {\bibfield  {journal} {\bibinfo  {journal} {Astron. Rep.}\ }\textbf {\bibinfo {volume} {52}},\ \bibinfo {pages} {379} (\bibinfo {year} {2008})},\ \Eprint {http://arxiv.org/abs/1201.5519} {arXiv:1201.5519 [astro-ph.SR]} \BibitemShut {NoStop}%
\bibitem [{\citenamefont {Delgaty}\ and\ \citenamefont {Lake}(1998)}]{Delgaty:1998uy}%
  \BibitemOpen
  \bibfield  {author} {\bibinfo {author} {\bibfnamefont {M.~S.~R.}\ \bibnamefont {Delgaty}}\ and\ \bibinfo {author} {\bibfnamefont {K.}~\bibnamefont {Lake}},\ }\href {\doibase 10.1016/S0010-4655(98)00130-1} {\bibfield  {journal} {\bibinfo  {journal} {Comput. Phys. Commun.}\ }\textbf {\bibinfo {volume} {115}},\ \bibinfo {pages} {395} (\bibinfo {year} {1998})},\ \Eprint {http://arxiv.org/abs/gr-qc/9809013} {arXiv:gr-qc/9809013} \BibitemShut {NoStop}%
\bibitem [{\citenamefont {Pant}(2010)}]{Pant:2010iub}%
  \BibitemOpen
  \bibfield  {author} {\bibinfo {author} {\bibfnamefont {N.}~\bibnamefont {Pant}},\ }\href {\doibase 10.1007/s10509-010-0453-4} {\bibfield  {journal} {\bibinfo  {journal} {Astrophys. Space Sci.}\ }\textbf {\bibinfo {volume} {331}},\ \bibinfo {pages} {633} (\bibinfo {year} {2010})}\BibitemShut {NoStop}%
\bibitem [{\citenamefont {Chu}\ and\ \citenamefont {Tan}(2022)}]{Chu:2021uec}%
  \BibitemOpen
  \bibfield  {author} {\bibinfo {author} {\bibfnamefont {C.-S.}\ \bibnamefont {Chu}}\ and\ \bibinfo {author} {\bibfnamefont {H.~S.}\ \bibnamefont {Tan}},\ }\href {\doibase 10.3390/universe8050250} {\bibfield  {journal} {\bibinfo  {journal} {Universe}\ }\textbf {\bibinfo {volume} {8}},\ \bibinfo {pages} {250} (\bibinfo {year} {2022})},\ \Eprint {http://arxiv.org/abs/2103.06314} {arXiv:2103.06314 [hep-th]} \BibitemShut {NoStop}%
\bibitem [{\citenamefont {Darmois}(1927)}]{darmois1927equations}%
  \BibitemOpen
  \bibfield  {author} {\bibinfo {author} {\bibfnamefont {G.}~\bibnamefont {Darmois}},\ }\href@noop {} {\bibfield  {journal} {\bibinfo  {journal} {Paris France}\ } (\bibinfo {year} {1927})}\BibitemShut {NoStop}%
\bibitem [{\citenamefont {Israel}(1966)}]{Israel:1966rt}%
  \BibitemOpen
  \bibfield  {author} {\bibinfo {author} {\bibfnamefont {W.}~\bibnamefont {Israel}},\ }\href {\doibase 10.1007/BF02710419} {\bibfield  {journal} {\bibinfo  {journal} {Nuovo Cim. B}\ }\textbf {\bibinfo {volume} {44S10}},\ \bibinfo {pages} {1} (\bibinfo {year} {1966})},\ \bibinfo {note} {[Erratum: Nuovo Cim.B 48, 463 (1967)]}\BibitemShut {NoStop}%
\bibitem [{\citenamefont {Chandrasekhar}(1984)}]{chandrasekhar1984stars}%
  \BibitemOpen
  \bibfield  {author} {\bibinfo {author} {\bibfnamefont {S.}~\bibnamefont {Chandrasekhar}},\ }\href@noop {} {\bibfield  {journal} {\bibinfo  {journal} {Science}\ }\textbf {\bibinfo {volume} {226}},\ \bibinfo {pages} {497} (\bibinfo {year} {1984})}\BibitemShut {NoStop}%
\bibitem [{\citenamefont {Afshar}\ \emph {et~al.}(2023)\citenamefont {Afshar}, \citenamefont {Moradpour},\ and\ \citenamefont {Shabani}}]{Afshar:2023uyw}%
  \BibitemOpen
  \bibfield  {author} {\bibinfo {author} {\bibfnamefont {B.}~\bibnamefont {Afshar}}, \bibinfo {author} {\bibfnamefont {H.}~\bibnamefont {Moradpour}}, \ and\ \bibinfo {author} {\bibfnamefont {H.}~\bibnamefont {Shabani}},\ }\href {\doibase 10.1016/j.dark.2023.101357} {\bibfield  {journal} {\bibinfo  {journal} {Phys. Dark Univ.}\ }\textbf {\bibinfo {volume} {42}},\ \bibinfo {pages} {101357} (\bibinfo {year} {2023})}\BibitemShut {NoStop}%
\bibitem [{\citenamefont {Peebles}\ and\ \citenamefont {Ratra}(2003)}]{Peebles:2002gy}%
  \BibitemOpen
  \bibfield  {author} {\bibinfo {author} {\bibfnamefont {P.~J.~E.}\ \bibnamefont {Peebles}}\ and\ \bibinfo {author} {\bibfnamefont {B.}~\bibnamefont {Ratra}},\ }\href {\doibase 10.1103/RevModPhys.75.559} {\bibfield  {journal} {\bibinfo  {journal} {Rev. Mod. Phys.}\ }\textbf {\bibinfo {volume} {75}},\ \bibinfo {pages} {559} (\bibinfo {year} {2003})},\ \Eprint {http://arxiv.org/abs/astro-ph/0207347} {arXiv:astro-ph/0207347} \BibitemShut {NoStop}%
\bibitem [{\citenamefont {Baum}\ and\ \citenamefont {Frampton}(2007)}]{Baum:2006ee}%
  \BibitemOpen
  \bibfield  {author} {\bibinfo {author} {\bibfnamefont {L.}~\bibnamefont {Baum}}\ and\ \bibinfo {author} {\bibfnamefont {P.~H.}\ \bibnamefont {Frampton}},\ }\href {\doibase 10.1103/PhysRevLett.98.071301} {\bibfield  {journal} {\bibinfo  {journal} {Phys. Rev. Lett.}\ }\textbf {\bibinfo {volume} {98}},\ \bibinfo {pages} {071301} (\bibinfo {year} {2007})},\ \Eprint {http://arxiv.org/abs/hep-th/0610213} {arXiv:hep-th/0610213} \BibitemShut {NoStop}%
\bibitem [{\citenamefont {Buchdahl}(1959)}]{buchdahl1959general}%
  \BibitemOpen
  \bibfield  {author} {\bibinfo {author} {\bibfnamefont {H.~A.}\ \bibnamefont {Buchdahl}},\ }\href@noop {} {\bibfield  {journal} {\bibinfo  {journal} {Physical Review}\ }\textbf {\bibinfo {volume} {116}},\ \bibinfo {pages} {1027} (\bibinfo {year} {1959})}\BibitemShut {NoStop}%
\bibitem [{\citenamefont {Glendenning}(2012)}]{glendenning2012compact}%
  \BibitemOpen
  \bibfield  {author} {\bibinfo {author} {\bibfnamefont {N.~K.}\ \bibnamefont {Glendenning}},\ }\href@noop {} {\emph {\bibinfo {title} {Compact stars: Nuclear physics, particle physics and general relativity}}}\ (\bibinfo  {publisher} {Springer Science \& Business Media},\ \bibinfo {year} {2012})\BibitemShut {NoStop}%
\bibitem [{\citenamefont {Florides}(1983)}]{florides1983complete}%
  \BibitemOpen
  \bibfield  {author} {\bibinfo {author} {\bibfnamefont {P.~S.}\ \bibnamefont {Florides}},\ }\href@noop {} {\bibfield  {journal} {\bibinfo  {journal} {Journal of Physics A: Mathematical and General}\ }\textbf {\bibinfo {volume} {16}},\ \bibinfo {pages} {1419} (\bibinfo {year} {1983})}\BibitemShut {NoStop}%
\bibitem [{\citenamefont {Kumar}\ and\ \citenamefont {Bharti}(2022)}]{kumar2022isotropic}%
  \BibitemOpen
  \bibfield  {author} {\bibinfo {author} {\bibfnamefont {J.}~\bibnamefont {Kumar}}\ and\ \bibinfo {author} {\bibfnamefont {P.}~\bibnamefont {Bharti}},\ }\href@noop {} {\bibfield  {journal} {\bibinfo  {journal} {The European Physical Journal Plus}\ }\textbf {\bibinfo {volume} {137}},\ \bibinfo {pages} {330} (\bibinfo {year} {2022})}\BibitemShut {NoStop}%
\bibitem [{\citenamefont {Mak}\ \emph {et~al.}(2001)\citenamefont {Mak}, \citenamefont {Dobson Peter~N.},\ and\ \citenamefont {Harko}}]{Mak:2001ie}%
  \BibitemOpen
  \bibfield  {author} {\bibinfo {author} {\bibfnamefont {M.~K.}\ \bibnamefont {Mak}}, \bibinfo {author} {\bibfnamefont {J.}~\bibnamefont {Dobson Peter~N.}}, \ and\ \bibinfo {author} {\bibfnamefont {T.}~\bibnamefont {Harko}},\ }\href {\doibase 10.1209/epl/i2001-00416-x} {\bibfield  {journal} {\bibinfo  {journal} {EPL}\ }\textbf {\bibinfo {volume} {55}},\ \bibinfo {pages} {310} (\bibinfo {year} {2001})},\ \Eprint {http://arxiv.org/abs/gr-qc/0107011} {arXiv:gr-qc/0107011} \BibitemShut {NoStop}%
\bibitem [{\citenamefont {Jotania}\ and\ \citenamefont {Tikekar}(2006)}]{Jotania:2006zq}%
  \BibitemOpen
  \bibfield  {author} {\bibinfo {author} {\bibfnamefont {K.}~\bibnamefont {Jotania}}\ and\ \bibinfo {author} {\bibfnamefont {R.}~\bibnamefont {Tikekar}},\ }\href {\doibase 10.1142/S021827180600884X} {\bibfield  {journal} {\bibinfo  {journal} {Int. J. Mod. Phys. D}\ }\textbf {\bibinfo {volume} {15}},\ \bibinfo {pages} {1175} (\bibinfo {year} {2006})}\BibitemShut {NoStop}%
\bibitem [{\citenamefont {Barraco}\ and\ \citenamefont {Hamity}(2002)}]{Barraco:2002ds}%
  \BibitemOpen
  \bibfield  {author} {\bibinfo {author} {\bibfnamefont {D.}~\bibnamefont {Barraco}}\ and\ \bibinfo {author} {\bibfnamefont {V.~H.}\ \bibnamefont {Hamity}},\ }\href {\doibase 10.1103/PhysRevD.65.124028} {\bibfield  {journal} {\bibinfo  {journal} {Phys. Rev. D}\ }\textbf {\bibinfo {volume} {65}},\ \bibinfo {pages} {124028} (\bibinfo {year} {2002})}\BibitemShut {NoStop}%
\bibitem [{\citenamefont {Boehmer}\ and\ \citenamefont {Harko}(2006)}]{Boehmer:2006ye}%
  \BibitemOpen
  \bibfield  {author} {\bibinfo {author} {\bibfnamefont {C.~G.}\ \bibnamefont {Boehmer}}\ and\ \bibinfo {author} {\bibfnamefont {T.}~\bibnamefont {Harko}},\ }\href {\doibase 10.1088/0264-9381/23/22/023} {\bibfield  {journal} {\bibinfo  {journal} {Class. Quant. Grav.}\ }\textbf {\bibinfo {volume} {23}},\ \bibinfo {pages} {6479} (\bibinfo {year} {2006})},\ \Eprint {http://arxiv.org/abs/gr-qc/0609061} {arXiv:gr-qc/0609061} \BibitemShut {NoStop}%
\bibitem [{\citenamefont {Ivanov}(2002)}]{Ivanov:2002xf}%
  \BibitemOpen
  \bibfield  {author} {\bibinfo {author} {\bibfnamefont {B.~V.}\ \bibnamefont {Ivanov}},\ }\href {\doibase 10.1103/PhysRevD.65.104011} {\bibfield  {journal} {\bibinfo  {journal} {Phys. Rev. D}\ }\textbf {\bibinfo {volume} {65}},\ \bibinfo {pages} {104011} (\bibinfo {year} {2002})},\ \Eprint {http://arxiv.org/abs/gr-qc/0201090} {arXiv:gr-qc/0201090} \BibitemShut {NoStop}%
\bibitem [{\citenamefont {Arba\~nil}\ \emph {et~al.}(2014)\citenamefont {Arba\~nil}, \citenamefont {Lemos},\ and\ \citenamefont {Zanchin}}]{Arbanil:2014usa}%
  \BibitemOpen
  \bibfield  {author} {\bibinfo {author} {\bibfnamefont {J.~D.~V.}\ \bibnamefont {Arba\~nil}}, \bibinfo {author} {\bibfnamefont {J.~P.~S.}\ \bibnamefont {Lemos}}, \ and\ \bibinfo {author} {\bibfnamefont {V.~T.}\ \bibnamefont {Zanchin}},\ }\href {\doibase 10.1103/PhysRevD.89.104054} {\bibfield  {journal} {\bibinfo  {journal} {Phys. Rev. D}\ }\textbf {\bibinfo {volume} {89}},\ \bibinfo {pages} {104054} (\bibinfo {year} {2014})},\ \Eprint {http://arxiv.org/abs/1404.7177} {arXiv:1404.7177 [gr-qc]} \BibitemShut {NoStop}%
\bibitem [{\citenamefont {Herrera}(1992)}]{Herrera:1992lwz}%
  \BibitemOpen
  \bibfield  {author} {\bibinfo {author} {\bibfnamefont {L.}~\bibnamefont {Herrera}},\ }\href {\doibase 10.1016/0375-9601(92)90036-L} {\bibfield  {journal} {\bibinfo  {journal} {Phys. Lett. A}\ }\textbf {\bibinfo {volume} {165}},\ \bibinfo {pages} {206} (\bibinfo {year} {1992})}\BibitemShut {NoStop}%
\bibitem [{\citenamefont {Abreu}\ \emph {et~al.}(2007)\citenamefont {Abreu}, \citenamefont {Hernandez},\ and\ \citenamefont {Nunez}}]{Abreu:2007ew}%
  \BibitemOpen
  \bibfield  {author} {\bibinfo {author} {\bibfnamefont {H.}~\bibnamefont {Abreu}}, \bibinfo {author} {\bibfnamefont {H.}~\bibnamefont {Hernandez}}, \ and\ \bibinfo {author} {\bibfnamefont {L.~A.}\ \bibnamefont {Nunez}},\ }\href {\doibase 10.1088/0264-9381/24/18/005} {\bibfield  {journal} {\bibinfo  {journal} {Class. Quant. Grav.}\ }\textbf {\bibinfo {volume} {24}},\ \bibinfo {pages} {4631} (\bibinfo {year} {2007})},\ \Eprint {http://arxiv.org/abs/0706.3452} {arXiv:0706.3452 [gr-qc]} \BibitemShut {NoStop}%
\bibitem [{\citenamefont {{Hillebrandt}}\ and\ \citenamefont {{Steinmetz}}(1976)}]{1976A&A....53..283H}%
  \BibitemOpen
  \bibfield  {author} {\bibinfo {author} {\bibfnamefont {W.}~\bibnamefont {{Hillebrandt}}}\ and\ \bibinfo {author} {\bibfnamefont {K.~O.}\ \bibnamefont {{Steinmetz}}},\ }\href@noop {} {\bibfield  {journal} {\bibinfo  {journal} {Astron. Astrophys.}\ }\textbf {\bibinfo {volume} {53}},\ \bibinfo {pages} {283} (\bibinfo {year} {1976})}\BibitemShut {NoStop}%
\bibitem [{\citenamefont {Chandrasekhar}(1964)}]{Chandrasekhar:1964zz}%
  \BibitemOpen
  \bibfield  {author} {\bibinfo {author} {\bibfnamefont {S.}~\bibnamefont {Chandrasekhar}},\ }\href {\doibase 10.1086/147938} {\bibfield  {journal} {\bibinfo  {journal} {Astrophys. J.}\ }\textbf {\bibinfo {volume} {140}},\ \bibinfo {pages} {417} (\bibinfo {year} {1964})},\ \bibinfo {note} {[Erratum: Astrophys.J. 140, 1342 (1964)]}\BibitemShut {NoStop}%
\bibitem [{\citenamefont {Zubair}\ \emph {et~al.}(2021)\citenamefont {Zubair}, \citenamefont {Waheed}, \citenamefont {Jamal},\ and\ \citenamefont {Mustafa}}]{Zubair:2021dyg}%
  \BibitemOpen
  \bibfield  {author} {\bibinfo {author} {\bibfnamefont {M.}~\bibnamefont {Zubair}}, \bibinfo {author} {\bibfnamefont {S.}~\bibnamefont {Waheed}}, \bibinfo {author} {\bibfnamefont {M.~F.}\ \bibnamefont {Jamal}}, \ and\ \bibinfo {author} {\bibfnamefont {G.}~\bibnamefont {Mustafa}},\ }\href {\doibase 10.1016/j.rinp.2021.104787} {\bibfield  {journal} {\bibinfo  {journal} {Results Phys.}\ }\textbf {\bibinfo {volume} {29}},\ \bibinfo {pages} {104787} (\bibinfo {year} {2021})}\BibitemShut {NoStop}%
\bibitem [{\citenamefont {Mustafa}\ \emph {et~al.}(2022)\citenamefont {Mustafa}, \citenamefont {Errehymy}, \citenamefont {Ditta},\ and\ \citenamefont {Daoud}}]{Mustafa:2022jso}%
  \BibitemOpen
  \bibfield  {author} {\bibinfo {author} {\bibfnamefont {G.}~\bibnamefont {Mustafa}}, \bibinfo {author} {\bibfnamefont {A.}~\bibnamefont {Errehymy}}, \bibinfo {author} {\bibfnamefont {A.}~\bibnamefont {Ditta}}, \ and\ \bibinfo {author} {\bibfnamefont {M.}~\bibnamefont {Daoud}},\ }\href {\doibase 10.1016/j.cjph.2022.05.010} {\bibfield  {journal} {\bibinfo  {journal} {Chin. J. Phys.}\ }\textbf {\bibinfo {volume} {77}},\ \bibinfo {pages} {2781} (\bibinfo {year} {2022})}\BibitemShut {NoStop}%
\bibitem [{\citenamefont {Bondi}(1947)}]{bondi1947spherically}%
  \BibitemOpen
  \bibfield  {author} {\bibinfo {author} {\bibfnamefont {H.}~\bibnamefont {Bondi}},\ }\href@noop {} {\bibfield  {journal} {\bibinfo  {journal} {Monthly Notices of the Royal Astronomical Society}\ }\textbf {\bibinfo {volume} {107}},\ \bibinfo {pages} {410} (\bibinfo {year} {1947})}\BibitemShut {NoStop}%
\bibitem [{\citenamefont {Witten}(1981)}]{witten1981new}%
  \BibitemOpen
  \bibfield  {author} {\bibinfo {author} {\bibfnamefont {E.}~\bibnamefont {Witten}},\ }\href@noop {} {\bibfield  {journal} {\bibinfo  {journal} {Communications in Mathematical Physics}\ }\textbf {\bibinfo {volume} {80}},\ \bibinfo {pages} {381} (\bibinfo {year} {1981})}\BibitemShut {NoStop}%
\bibitem [{\citenamefont {Visser}(1997)}]{visser1997energy}%
  \BibitemOpen
  \bibfield  {author} {\bibinfo {author} {\bibfnamefont {M.}~\bibnamefont {Visser}},\ }\href@noop {} {\bibfield  {journal} {\bibinfo  {journal} {Science}\ }\textbf {\bibinfo {volume} {276}},\ \bibinfo {pages} {88} (\bibinfo {year} {1997})}\BibitemShut {NoStop}%
\bibitem [{\citenamefont {Andreasson}(2009)}]{Andreasson:2008xw}%
  \BibitemOpen
  \bibfield  {author} {\bibinfo {author} {\bibfnamefont {H.}~\bibnamefont {Andreasson}},\ }\href {\doibase 10.1007/s00220-008-0690-3} {\bibfield  {journal} {\bibinfo  {journal} {Commun. Math. Phys.}\ }\textbf {\bibinfo {volume} {288}},\ \bibinfo {pages} {715} (\bibinfo {year} {2009})},\ \Eprint {http://arxiv.org/abs/0804.1882} {arXiv:0804.1882 [gr-qc]} \BibitemShut {NoStop}%
\bibitem [{\citenamefont {Garcia}\ \emph {et~al.}(2011)\citenamefont {Garcia}, \citenamefont {Harko}, \citenamefont {Lobo},\ and\ \citenamefont {Mimoso}}]{garcia2011energy}%
  \BibitemOpen
  \bibfield  {author} {\bibinfo {author} {\bibfnamefont {N.~M.}\ \bibnamefont {Garcia}}, \bibinfo {author} {\bibfnamefont {T.}~\bibnamefont {Harko}}, \bibinfo {author} {\bibfnamefont {F.~S.}\ \bibnamefont {Lobo}}, \ and\ \bibinfo {author} {\bibfnamefont {J.~P.}\ \bibnamefont {Mimoso}},\ }\href@noop {} {\bibfield  {journal} {\bibinfo  {journal} {Physical Review D}\ }\textbf {\bibinfo {volume} {83}},\ \bibinfo {pages} {104032} (\bibinfo {year} {2011})}\BibitemShut {NoStop}%
\bibitem [{\citenamefont {Oppenheimer}\ and\ \citenamefont {Volkoff}(1939)}]{Oppenheimer:1939ne}%
  \BibitemOpen
  \bibfield  {author} {\bibinfo {author} {\bibfnamefont {J.~R.}\ \bibnamefont {Oppenheimer}}\ and\ \bibinfo {author} {\bibfnamefont {G.~M.}\ \bibnamefont {Volkoff}},\ }\href {\doibase 10.1103/PhysRev.55.374} {\bibfield  {journal} {\bibinfo  {journal} {Phys. Rev.}\ }\textbf {\bibinfo {volume} {55}},\ \bibinfo {pages} {374} (\bibinfo {year} {1939})}\BibitemShut {NoStop}%
\bibitem [{\citenamefont {Ponce~de Leon}(1987)}]{ponce1987general}%
  \BibitemOpen
  \bibfield  {author} {\bibinfo {author} {\bibfnamefont {J.}~\bibnamefont {Ponce~de Leon}},\ }\href {\doibase 10.1007/BF00768215} {\bibfield  {journal} {\bibinfo  {journal} {General relativity and gravitation}\ }\textbf {\bibinfo {volume} {19}},\ \bibinfo {pages} {797} (\bibinfo {year} {1987})}\BibitemShut {NoStop}%
\bibitem [{\citenamefont {Ponce~de Leon}(1993)}]{ponce1993limiting}%
  \BibitemOpen
  \bibfield  {author} {\bibinfo {author} {\bibfnamefont {J.}~\bibnamefont {Ponce~de Leon}},\ }\href@noop {} {\bibfield  {journal} {\bibinfo  {journal} {General relativity and gravitation}\ }\textbf {\bibinfo {volume} {25}},\ \bibinfo {pages} {1123} (\bibinfo {year} {1993})}\BibitemShut {NoStop}%
\bibitem [{\citenamefont {Tolman}(1930)}]{PhysRev.35.875}%
  \BibitemOpen
  \bibfield  {author} {\bibinfo {author} {\bibfnamefont {R.~C.}\ \bibnamefont {Tolman}},\ }\href {\doibase 10.1103/PhysRev.35.875} {\bibfield  {journal} {\bibinfo  {journal} {Phys. Rev.}\ }\textbf {\bibinfo {volume} {35}},\ \bibinfo {pages} {875} (\bibinfo {year} {1930})}\BibitemShut {NoStop}%
\bibitem [{\citenamefont {Shapiro}\ and\ \citenamefont {Teukolsky}(2008)}]{shapiro2008black}%
  \BibitemOpen
  \bibfield  {author} {\bibinfo {author} {\bibfnamefont {S.~L.}\ \bibnamefont {Shapiro}}\ and\ \bibinfo {author} {\bibfnamefont {S.~A.}\ \bibnamefont {Teukolsky}},\ }\href@noop {} {\emph {\bibinfo {title} {Black holes, white dwarfs, and neutron stars: The physics of compact objects}}}\ (\bibinfo  {publisher} {John Wiley \& Sons},\ \bibinfo {year} {2008})\BibitemShut {NoStop}%
\bibitem [{\citenamefont {Zeldovich}\ and\ \citenamefont {Novikov}(1971)}]{zeldovich1971relativistic}%
  \BibitemOpen
  \bibfield  {author} {\bibinfo {author} {\bibfnamefont {Y.~B.}\ \bibnamefont {Zeldovich}}\ and\ \bibinfo {author} {\bibfnamefont {I.~D.}\ \bibnamefont {Novikov}},\ }\href@noop {} {\bibfield  {journal} {\bibinfo  {journal} {Chicago: University of Chicago Press}\ } (\bibinfo {year} {1971})}\BibitemShut {NoStop}%
\bibitem [{\citenamefont {Zel'dovich}\ \emph {et~al.}(1972)\citenamefont {Zel'dovich}, \citenamefont {Novikov},\ and\ \citenamefont {Silk}}]{zel1972relativistic}%
  \BibitemOpen
  \bibfield  {author} {\bibinfo {author} {\bibfnamefont {Y.~B.}\ \bibnamefont {Zel'dovich}}, \bibinfo {author} {\bibfnamefont {I.~D.}\ \bibnamefont {Novikov}}, \ and\ \bibinfo {author} {\bibfnamefont {J.}~\bibnamefont {Silk}},\ }\href@noop {} {\bibfield  {journal} {\bibinfo  {journal} {Physics Today}\ }\textbf {\bibinfo {volume} {25}},\ \bibinfo {pages} {63} (\bibinfo {year} {1972})}\BibitemShut {NoStop}%
\bibitem [{\citenamefont {L'DOVICH}(1962)}]{l1962equation}%
  \BibitemOpen
  \bibfield  {author} {\bibinfo {author} {\bibfnamefont {Y.}~\bibnamefont {L'DOVICH}},\ }\href@noop {} {\bibfield  {journal} {\bibinfo  {journal} {Sov. Phys. JETP}\ }\textbf {\bibinfo {volume} {14}},\ \bibinfo {pages} {1609} (\bibinfo {year} {1962})}\BibitemShut {NoStop}%
\bibitem [{\citenamefont {Zhuravlev}\ and\ \citenamefont {Chervon}(2000)}]{Zhuravlev:2000wg}%
  \BibitemOpen
  \bibfield  {author} {\bibinfo {author} {\bibfnamefont {V.~M.}\ \bibnamefont {Zhuravlev}}\ and\ \bibinfo {author} {\bibfnamefont {S.~V.}\ \bibnamefont {Chervon}},\ }\href {\doibase 10.1134/1.1311981} {\bibfield  {journal} {\bibinfo  {journal} {J. Exp. Theor. Phys.}\ }\textbf {\bibinfo {volume} {91}},\ \bibinfo {pages} {227} (\bibinfo {year} {2000})}\BibitemShut {NoStop}%
\bibitem [{\citenamefont {Usmani}\ \emph {et~al.}(2008)\citenamefont {Usmani}, \citenamefont {Ghosh}, \citenamefont {Mukhopadhyay}, \citenamefont {Ray},\ and\ \citenamefont {Ray}}]{Usmani:2008ce}%
  \BibitemOpen
  \bibfield  {author} {\bibinfo {author} {\bibfnamefont {A.~A.}\ \bibnamefont {Usmani}}, \bibinfo {author} {\bibfnamefont {P.~P.}\ \bibnamefont {Ghosh}}, \bibinfo {author} {\bibfnamefont {U.}~\bibnamefont {Mukhopadhyay}}, \bibinfo {author} {\bibfnamefont {P.~C.}\ \bibnamefont {Ray}}, \ and\ \bibinfo {author} {\bibfnamefont {S.}~\bibnamefont {Ray}},\ }\href {\doibase 10.1111/j.1745-3933.2008.00468.x} {\bibfield  {journal} {\bibinfo  {journal} {Mon. Not. Roy. Astron. Soc.}\ }\textbf {\bibinfo {volume} {386}},\ \bibinfo {pages} {L92} (\bibinfo {year} {2008})},\ \Eprint {http://arxiv.org/abs/0801.4529} {arXiv:0801.4529 [gr-qc]} \BibitemShut {NoStop}%
\bibitem [{\citenamefont {Shee}\ \emph {et~al.}(2016)\citenamefont {Shee}, \citenamefont {Rahaman}, \citenamefont {Guha},\ and\ \citenamefont {Ray}}]{Shee:2015kqa}%
  \BibitemOpen
  \bibfield  {author} {\bibinfo {author} {\bibfnamefont {D.}~\bibnamefont {Shee}}, \bibinfo {author} {\bibfnamefont {F.}~\bibnamefont {Rahaman}}, \bibinfo {author} {\bibfnamefont {B.~K.}\ \bibnamefont {Guha}}, \ and\ \bibinfo {author} {\bibfnamefont {S.}~\bibnamefont {Ray}},\ }\href {\doibase 10.1007/s10509-016-2753-9} {\bibfield  {journal} {\bibinfo  {journal} {Astrophys. Space Sci.}\ }\textbf {\bibinfo {volume} {361}},\ \bibinfo {pages} {167} (\bibinfo {year} {2016})},\ \Eprint {http://arxiv.org/abs/1505.03034} {arXiv:1505.03034 [gr-qc]} \BibitemShut {NoStop}%
\bibitem [{\citenamefont {Biswas}\ \emph {et~al.}(2019)\citenamefont {Biswas}, \citenamefont {Shee}, \citenamefont {Ray}, \citenamefont {Rahaman},\ and\ \citenamefont {Guha}}]{Biswas:2019doe}%
  \BibitemOpen
  \bibfield  {author} {\bibinfo {author} {\bibfnamefont {S.}~\bibnamefont {Biswas}}, \bibinfo {author} {\bibfnamefont {D.}~\bibnamefont {Shee}}, \bibinfo {author} {\bibfnamefont {S.}~\bibnamefont {Ray}}, \bibinfo {author} {\bibfnamefont {F.}~\bibnamefont {Rahaman}}, \ and\ \bibinfo {author} {\bibfnamefont {B.~K.}\ \bibnamefont {Guha}},\ }\href {\doibase 10.1016/j.aop.2019.05.004} {\bibfield  {journal} {\bibinfo  {journal} {Annals Phys.}\ }\textbf {\bibinfo {volume} {409}},\ \bibinfo {pages} {167905} (\bibinfo {year} {2019})},\ \Eprint {http://arxiv.org/abs/1910.00427} {arXiv:1910.00427 [gr-qc]} \BibitemShut {NoStop}%
\bibitem [{\citenamefont {Staykov}\ \emph {et~al.}(2014)\citenamefont {Staykov}, \citenamefont {Doneva}, \citenamefont {Yazadjiev},\ and\ \citenamefont {Kokkotas}}]{Staykov:2014mwa}%
  \BibitemOpen
  \bibfield  {author} {\bibinfo {author} {\bibfnamefont {K.~V.}\ \bibnamefont {Staykov}}, \bibinfo {author} {\bibfnamefont {D.~D.}\ \bibnamefont {Doneva}}, \bibinfo {author} {\bibfnamefont {S.~S.}\ \bibnamefont {Yazadjiev}}, \ and\ \bibinfo {author} {\bibfnamefont {K.~D.}\ \bibnamefont {Kokkotas}},\ }\href {\doibase 10.1088/1475-7516/2014/10/006} {\bibfield  {journal} {\bibinfo  {journal} {JCAP}\ }\textbf {\bibinfo {volume} {10}},\ \bibinfo {pages} {006} (\bibinfo {year} {2014})},\ \Eprint {http://arxiv.org/abs/1407.2180} {arXiv:1407.2180 [gr-qc]} \BibitemShut {NoStop}%
\bibitem [{\citenamefont {Ngubelanga}\ \emph {et~al.}(2015)\citenamefont {Ngubelanga}, \citenamefont {Maharaj},\ and\ \citenamefont {Ray}}]{Ngubelanga:2015jne}%
  \BibitemOpen
  \bibfield  {author} {\bibinfo {author} {\bibfnamefont {S.~A.}\ \bibnamefont {Ngubelanga}}, \bibinfo {author} {\bibfnamefont {S.~D.}\ \bibnamefont {Maharaj}}, \ and\ \bibinfo {author} {\bibfnamefont {S.}~\bibnamefont {Ray}},\ }\href {\doibase 10.1007/s10509-015-2247-1} {\bibfield  {journal} {\bibinfo  {journal} {Astrophys. Space Sci.}\ }\textbf {\bibinfo {volume} {357}},\ \bibinfo {pages} {74} (\bibinfo {year} {2015})},\ \Eprint {http://arxiv.org/abs/1512.08994} {arXiv:1512.08994 [gr-qc]} \BibitemShut {NoStop}%
\bibitem [{\citenamefont {{Shvartsman}}(1971)}]{1971JETP...33..475S}%
  \BibitemOpen
  \bibfield  {author} {\bibinfo {author} {\bibfnamefont {V.~F.}\ \bibnamefont {{Shvartsman}}},\ }\href@noop {} {\bibfield  {journal} {\bibinfo  {journal} {Soviet Journal of Experimental and Theoretical Physics}\ }\textbf {\bibinfo {volume} {33}},\ \bibinfo {pages} {475} (\bibinfo {year} {1971})}\BibitemShut {NoStop}%
\end{thebibliography}%

\end{document}